\documentclass[aps,twocolumn,prc,superscriptaddress,noshowpacs,nofootinbib,noshowkeys,floatfix]{revtex4}
\usepackage[dvips]{graphics,graphicx}
\usepackage[colorlinks=true,linktocpage=true,linkcolor=blue,citecolor=blue]{hyperref}
\usepackage[usenames,dvipsnames]{color}
\usepackage{amsmath, amssymb}
\usepackage{multirow}
\usepackage{longtable}
\usepackage{color}
\usepackage{cleveref}
\usepackage[normalem]{ulem}  

\renewcommand\sout{\bgroup \color{blue} \ULdepth=-.5ex \ULset}

\begin{document}

\title{{\Large Anisotropic flow of charged and identified hadrons at FAIR energies and its dependence on the nuclear equation of state}}

\author{Sudhir Pandurang Rode}
\affiliation{Discipline of Physics, School of Basic Sciences, Indian Institute of Technology Indore, Indore 453552 India}
\author{Partha Pratim Bhaduri}
\affiliation{Variable Energy Cyclotron Centre, HBNI, 1/AF Bidhan Nagar, Kolkata 700 064, India}
\author{Ankhi Roy}
\affiliation{Discipline of Physics, School of Basic Sciences, Indian Institute of Technology Indore, Indore 453552 India}
\date{\today}

\begin{abstract}
In this article, we examine the equation of state (EoS) dependence of the anisotropic flow parameters ($v_{1}$, $v_{2}$ and $v_{4}$) of charged and identified hadrons, as a function of transverse momentum ($p_{\rm T}$), rapidity ($y_{c.m.}$) and the incident beam energy ($\rm E_{\rm Lab}$) in mid-central Au +  Au collisions in the energy range $\rm E_{\rm Lab} = 6 -25$ A GeV. Simulations are carried out by employing different variants of the Ultra-relativistic Quantum Molecular Dynamics (UrQMD) model, namely the pure transport (cascade) mode and the hybrid mode. In the hybrid mode, transport calculations are coupled with the ideal hydrodynamical evolution. Within the hydrodynamic scenario, two different equations of state (EoS) viz. Hadron gas and Chiral + deconfinement EoS have been employed separately to possibly mimic the hadronic and partonic scenarios, respectively. It is observed that the flow parameters are sensitive to the onset of hydrodynamic expansion of the fireball in comparison to the pure transport approach. 
The results would be useful as predictions for the upcoming low energy experiments at Facility for Antiproton and Ion Research (FAIR) and Nuclotron-based Ion Collider fAcility (NICA).
\end{abstract}

\maketitle

\section*{I Introduction}
A central goal of the relativistic heavy-ion collision experiments is the quantitive mapping of the QCD phase diagram from low to high baryon densities \cite{a,c}. The major drive behind such investigations is to search for a first and/or second order phase transition together with the existence of critical end point (CEP) of QCD matter at non-zero baryon chemical potentials, as predicted by several effective QCD models. Exploration of the different phases of strongly interacting matter in the full range of temperatures and baryon densities necessitates the simultaneous measurements of various observables over a wide range of beam energies. Over the past two decades, the region of high temperature and vanishing baryon densities of the QCD phase diagram has been extensively studied in experiments carried out at RHIC ~\cite{rhic1,rhic2} and LHC~\cite{lhc1,lhc2,lhc3}. Compared to that our understanding of the QCD equation of state at non-zero baryon densities is rather limited. Hence, the ongoing beam energy scan (BES) program at RHIC \cite{Okorokov:2017npg,Odyniec:2013aaa} and the upcoming heavy-ion collision experiments at the Nuclotron-based Ion Collider fAcility (NICA)\cite{Kekelidze:2016wkp} at the Joint Institute for Nuclear Research (Dubna) and at the Facility for Antiproton and Ion Research (FAIR, Germany)\cite{Ablyazimov:2017guv,Sturm:2010yit} aim at probing the moderate temperature and high baryonic chemical potential regime of the QCD phase diagram. 

The azimuthal anisotropy of the final-state hadrons produced in the heavy ion collisions has long been considered as a deterministic probe to investigate collective effects in multi-particle production\cite{Aamodt:2010pa,Adcox:2004mh,Adams:2005dq}. As per traditional wisdom, at relatively higher energies, the collective transverse flow in nuclear collisions is driven by the pressure gradients in the early thermalized stages of the reaction and hence encodes the information about the underlying QCD equation of state (EoS). For non-central nucleus-nucleus collisions, the asymmetry of the momentum distributions of hadrons are quantified in terms of coefficients of Fourier expansion of the azimuthal distribution of the emitted particles as
\begin{align*}
v_{n} = <\cos[n(\phi - \Psi)]>
\end{align*} 
Where $\phi$ and $\Psi$ denote the azimuthal angle of the particle and reaction plane angle respectively. The first Fourier coefficient of the above azimuthal distribution is called directed flow $v_{1}$, whereas the second Fourier coefficient $v_{2}$ is known as elliptic flow and is dominant among all the coefficients at midrapidity. Similarly, the $v_{4}$ is the fourth Fourier coefficient of the azimuthal anisotropy.

Directed flow ($v_{1}$) is sensitive to the longitudinal dynamics of the produced medium. 
$v_{1}$ can be very useful to probe the early stages of the collision because it is expected to be built even earlier than elliptic flow \cite{Nara:2016phs,Nara:1999dz,Konchakovski:2014gda}. In the vicinity of a first order phase transition directed flow of hadrons is believed to drop and even vanish due to softening of the underlying EoS, making $v_{1}$ an interesting observable to be studied at RHIC-BES, NICA and FAIR. 
The directed flow which is observed at AGS energies \cite{Liu:2000am,Chung:2000ny,Chung:2001je} and below show linearity as a function of the rapidity with the slope quantifying the strength of the signal. Above SPS energies \cite{Appelshauser:1997dg,Adams:2004bi,Back:2005pc}, the slope of directed flow in the mid-rapidity region is different compared to the slope in the beam rapidity region which makes the structure of $v_{1}(y)$ more complex.

On the other hand, the elliptic flow ($v_2$) of identified hadrons directly reflects the rescattering among the produced particles and hence has been studied intensively to look for thermalization of the produced medium in different experiments \cite{Alver:2006wh,Aamodt:2010pa} at various energies. For non-central collisions, the azimuthal anisotropy of the transverse momentum ($p_T$) distribution is believed to be sensitive to the early evolution of the fireball. A bulk of previous studies have been devoted to investigate the elliptic flow in low energy collisions relevant for FAIR~\cite{Bhaduri:2010wi,Sarkar:2017fuy} and RHIC BES~\cite{Auvinen:2013sba} program using transport models like UrQMD \cite{Bass:1998ca,Bleicher:1999xi} with different configurations, and AMPT \cite{Lin:2004en,Chen:2004vha}.


Apart from $v_{1}$ and $v_{2}$, there is another harmonic of the azimuthal distribution which needs due attention is $v_{4}$. The energy dependence of $v_{4}$ is sensitive to the nuclear equations of state (EoS). Calculations based on \cite{Borghini:2005kd,Gombeaud:2009ye,Luzum:2010ae} indicates that, the $v_{2}$ has influence on the generation of $v_{4}$. However, we refrain from the investigation of triangular flow ($v_{3}$) that originates from initial state fluctuations and is not believed to bear any sensitivity to the underlying EoS.
According to recent hydrodynamical predictions \cite{Luzum:2010ae}, $v_{4}$ encodes an important information on the underlying collision dynamics. Recently \cite{Nara:2018ijw}, the beam energy dependence of $v_{4}$ was studied using microscopic transport model JAM \cite{Nara:1999dz}.

In this article, we have studied directed flow ($v_{1}$) as a function of rapidity, beam energy ($\rm E_{\rm Lab}$), elliptic flow ($v_{2}$) with respect to transverse momentum, rapidity and incident beam energy ($\rm E_{\rm Lab}$) and also, $4^{th}$ order Fourier coefficient ($v_{4}$) as a function of beam energy ($\rm E_{\rm Lab}$) for charged and identified hadrons, in the range $E_{\rm Lab} = 6 - 25$ A GeV, relevant for FAIR. The publicly available version 3.4 of the UrQMD model is employed for this purpose. Within this model, the impact parameter vector is aligned along the X-axis, and the reaction plane angle ($\Psi$) is zero. Our calculations are performed with different variants of the code, namely the pure transport (cascade) mode and hybrid mode. In the hybrid mode, two different nuclear equation of states (EoS) viz. Hadron Gas (HG) and Chiral EoS are used separately, in the intermediate hydrodynamic stage, to replicate the effects of hadronic and partonic scenarios, respectively. It is important to note that UrQMD model has been widely used earlier to the flow co-efficients in low energy nuclear collisions. The rapidity and transverse momentum dependence of $v_1$ and $v_2$ in Pb+Pb recations at $40 A$ and $160 A $ GeV beam energies were calculated in~\cite{Petersen:2006vm} using UrQMD model in cascade mode (v2.2) and contrasted with the data avaiable from NA49 experiment, for three different centrality bins. In addition the energy excitation functions of $v_1$ and $v_2$ are estimated in the energy range of $\rm E_{\rm Lab} = 90 A$ MeV to  $\rm E_{\rm cm} = 200 A$ GeV and contrasted with the available data. The hybrid UrQMD approach with HG EoS has also been employed earlier~\cite{Petersen:2009vx} to calculate the beam energy dependence of $v_2$ for heavy ion reactions from GSI-SIS to the highest CERN-SPS energies. With Chiral EoS, the hybrid UrQMD model has been used to study the collision energy dependence of elliptic flow $v_2$ and triangular flow $v_3$ parameters in Au+Au collisions in the energy range $\sqrt{s_{NN}} = 5 - 200$ GeV~\cite{Auvinen:2013sba}. The hybrid model has also been applied examine the collision energy dependence of $v_1$ for heavy-ion collisions over the range $\sqrt{s_{NN}}= 3 -20$ GeV~\cite{Steinheimer:2014pfa}. In the present paper, we have performed a systematic study of the different flow parameters $v_1$, $v_2$ and $v_4$, with the hope to acess the the impact of the EoS of the strongly-interacting matter on these observables.

The organization of the article is as follows. The basic features of the different variants of the UrQMD model are briefly discussed in Section~2. Section~3 presents the results of our studies on the dependence of anisotropic flow parameters ($v_{1}$, $v_{2}$, $v_{4}$) on different kinematic variables over a range of bombarding energies. Conclusions are drawn in section~4.

\section*{II UrQMD Model}
The Ultra-relativistic Quantum Molecular Dynamics (UrQMD) model is an event generator designed to simulate high energy nucleus-nucleus collisions. The target and projectile nuclei are initialized according to the Woods-Saxon profile in the coordinate space, and the Fermi gas model in the momentum space. The initial momentum for each nucleon in the rest frame of the corresponding nucleus is thus assigned randomly between zero and local Thomas Fermi momentum. The interaction is described via multiple interactions of the incident and newly produced hadrons,  the formation and decay of hadronic resonances and the excitation and formation of color strings~\cite{Bleicher:1999xi}. UrQMD employs longitudinal excitation of the strings stretched uniformly between quark, diquark and their anti states, which subsequently break into hadrons following Lund string fragmentation models~\cite{Lund}. The model incorporates available experimental information like hadronic cross sections, resonance decay widths and decay modes. Propagation of particles between subsequent collisions occurs in straight line trajectories with their velocities ($\rm p/\rm E$, with p is the momentum and E is the energy). 

Pure hadronic transport models have been found to underestimate the large $v_2$ values measured above 40 $\rm A GeV$ beam energy upto RHIC energy $\sqrt{s_{NN}}$ $=$ 200 GeV ~\cite{Petersen:2006vm,Burau:2005}. The failure is attributed to the too low-pressure gradients in the early phase of the collisions to generate enough collectivity. With an aim to capture the entire evolution dynamics of the fireball, the so-called hybrid UrQMD model has been developed where the pure transport approach is embedded with a $3-$D ideal relativistic one fluid evolution for the intermediate hot and dense stage of the reaction. Within this integrated approach, the initial conditions and the final hadronic freeze out are calculated from UrQMD on an event-by-event basis, for proper incorporation of the non-equilibrium dynamics. The hydrodynamical evolution is switched on when the two Lorentz-contracted nuclei have crossed each other~\cite{Petersen:2008dd}. Here the participants are mapped to the hydrodynamic grid, and spectators continue to propagate in the cascade. The primary collisions and fragmentations of strings in the microscopic UrQMD model generate the event wise initial conditions and thus incorporate the event-by-event fluctuations. 

Even though hydrodynamics was found very successful to describe the high $v_2$ values measured at RHIC at $\sqrt{s_{NN}}$ $=$ 130 and 200 GeV \cite{Kolb:2000fha,Back:2004dw}, it's application in low energy nuclear collisions below $25 A$ GeV might demand some justification. Pure transport models aim at description of heavy ion reactions on the basis of an effective solution of the relativistic Boltzmann equation. In most transport approaches, the collision kernel is usually restricted to the level of binary collisions and $2 \rightarrow n$ scattering processes to keep the calculation numerically tractable. In the restriction to binary collisions, mean free paths of the particles are assumed to be large, which becomes questionable in the hot and very dense stage of heavy ion collisions. At FAIR energies, a dense baryonic medium is anticipated, where many body collisions migh not be negligible. At finite baryo-chemical potential, the heated
and compressed nuclear matter might also undergo a phase transition, signatures of which might be imprinted in the flow observables.  Within a purely microscopic approach it is difficult to account for the hadronization and the phase transition between the hadronic and the partonic phase. Within hydrodynamical approach, in contrast, one can explicitly incorporate phase transitions by
changing the EoS. Hydrodynamics is thus accepted as an effective tool to describe collective expansion of the intermediate hot and dense stage of the heavy-ion colliisons. However application of hydrodynamics demands local thermalization and the results depend strongly on the initial and final boundary conditions. Hence a more realistic picture of the whole dynamics of heavy ion reactions can be achieved by so-called microscopic plus macroscopic (micro+macro) hybrid models. Such an approach allows to reduce the parameters for the initial conditions and provides a consistent freeze-out description and allows a direct comparison of the different collision dynamics - ideal fluid dynamics vs. non-equilibrium transport scenario.

In the hybrid model, the hydrodynamic evolution is stopped if the energy density $\epsilon$ drops below the five times the ground state energy density $\epsilon_0$ in all cells~\cite{Petersen:2008dd}. After the hydrodynamical evolution, the Cooper-Frye prescription \cite{Cooper:1974mv} is employed to map the hydrodynamical fields to the hadrons, which evolve further via hadronic cascade through rescatterings and final state decays until all interactions cease and the system decouples. 

Within hybrid mode of UrQMD, different choices of the underlying EoS are available for the intermediate hydrodynamic phase. In the present work, we opt for two different EoS, namely the hadron gas (HG) EoS and the Chiral EoS to mimic the hadronic and partonic scenarios, respectively.

The Hadron gas EoS \cite{Zschiesche:2002zr} consists of a grand canonical description of a free and non-interacting gas of hadrons. The underlying hadronic degrees of freedom involved in this EoS are all the reliably known baryons and mesons with masses up to $2$ GeV and thus in line with the degrees of freedom included in the pure UrQMD model. Note that this EoS does not include any type of phase transition. It gives us the privilege of a direct comparison of the hydrodynamic scenario with the transport simulation.

\begin{figure}[h]
  \includegraphics[scale=0.2]{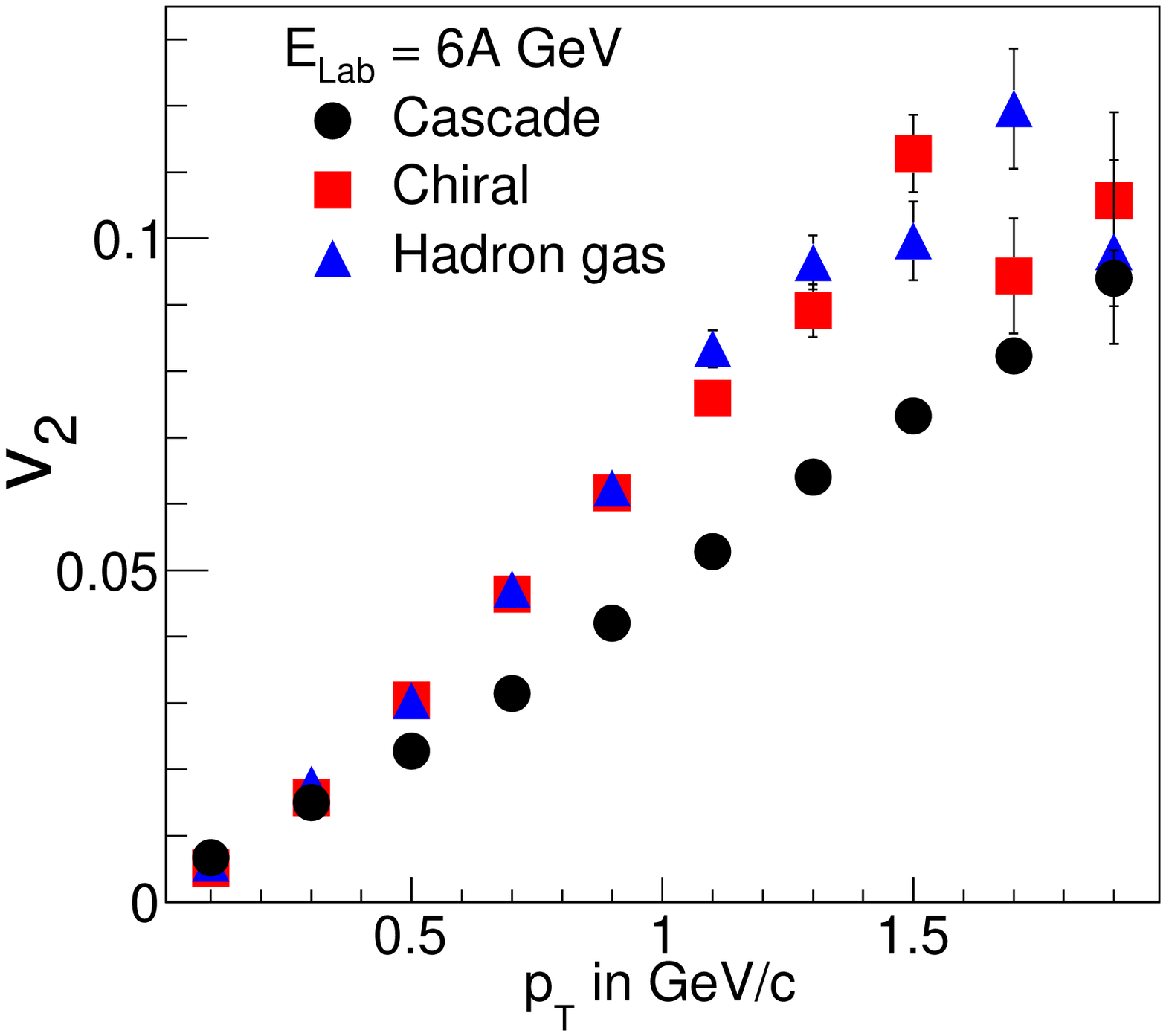}
  \includegraphics[scale=0.2]{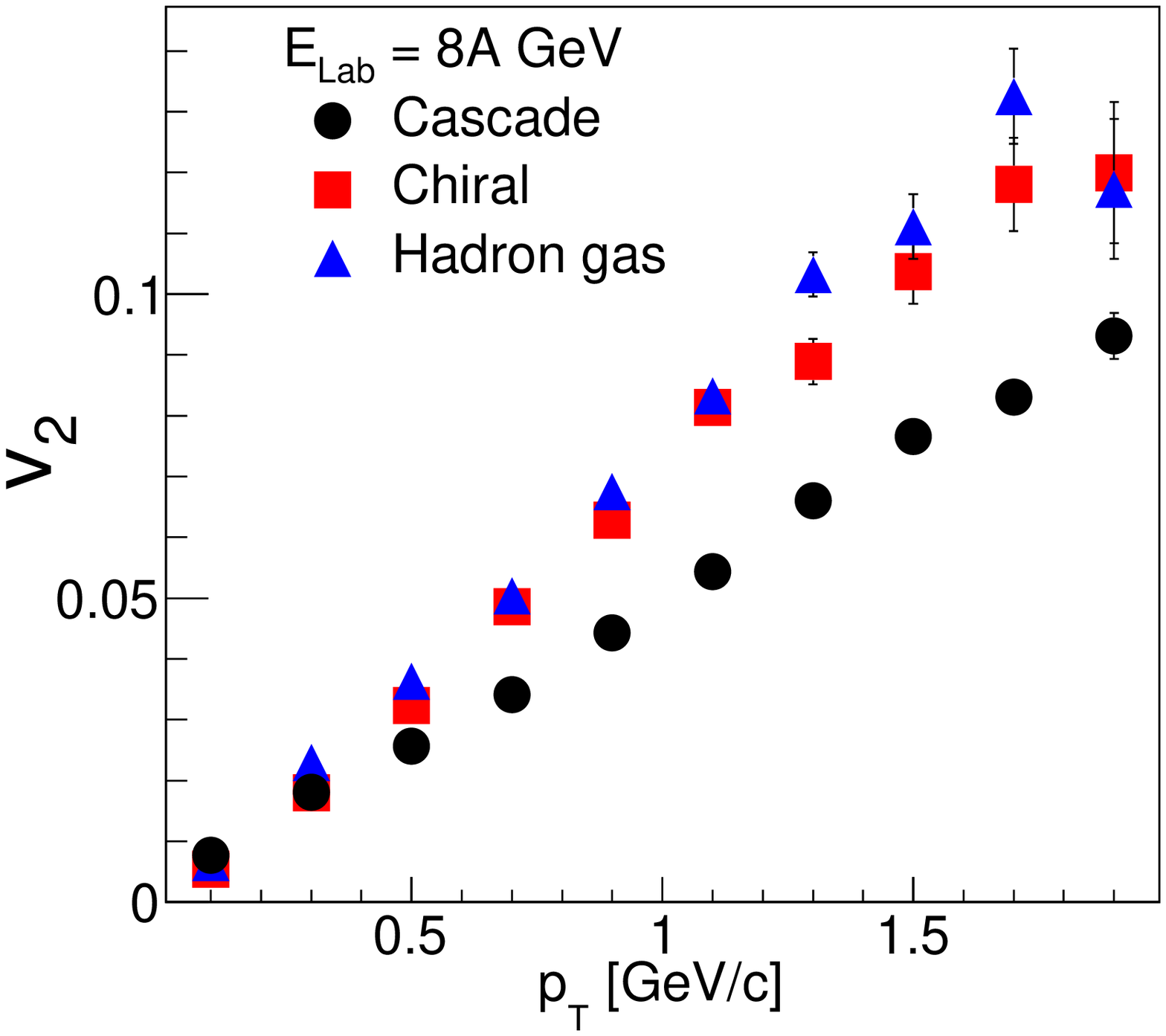}\\
  \includegraphics[scale=0.2]{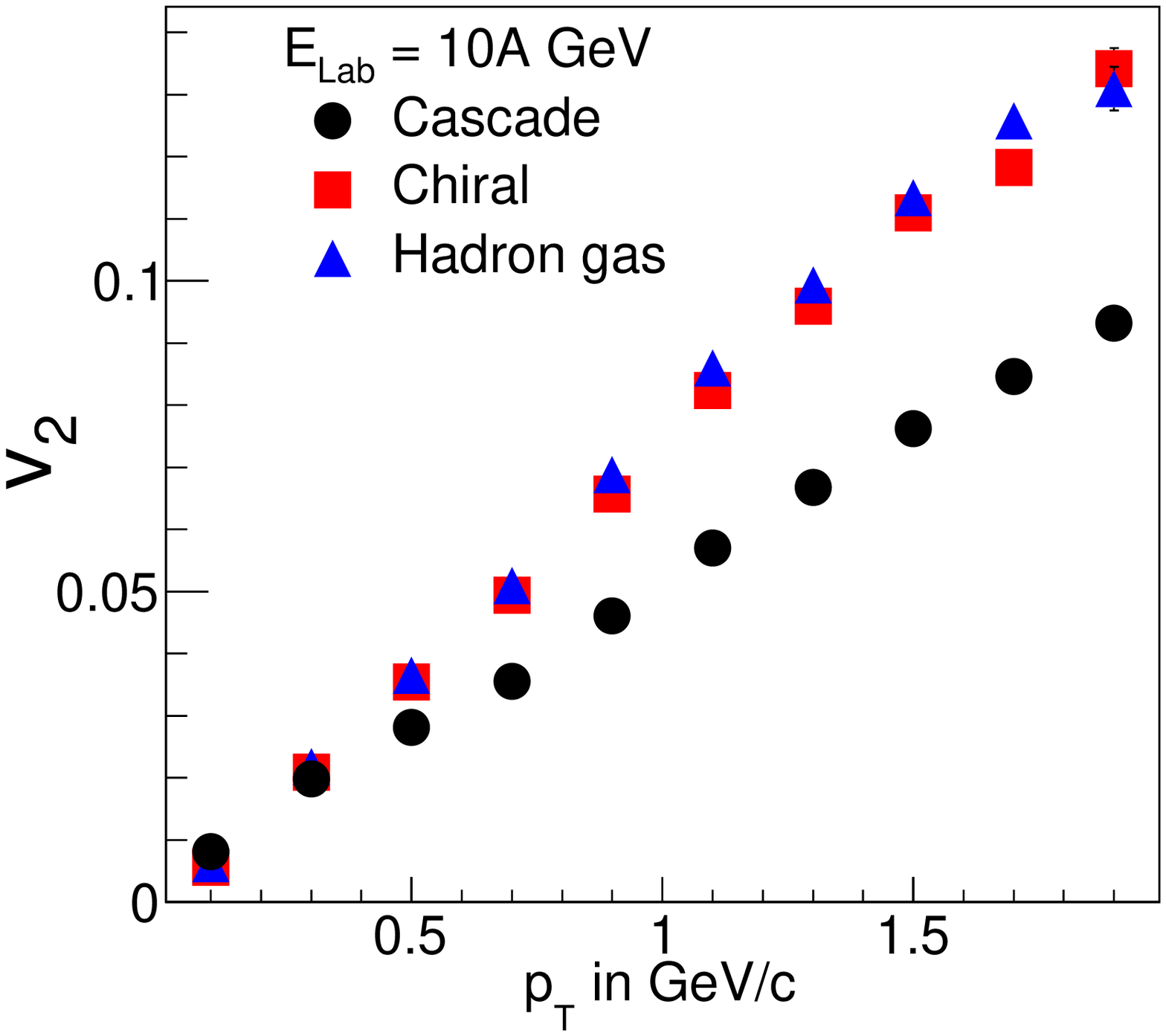}
  \includegraphics[scale=0.2]{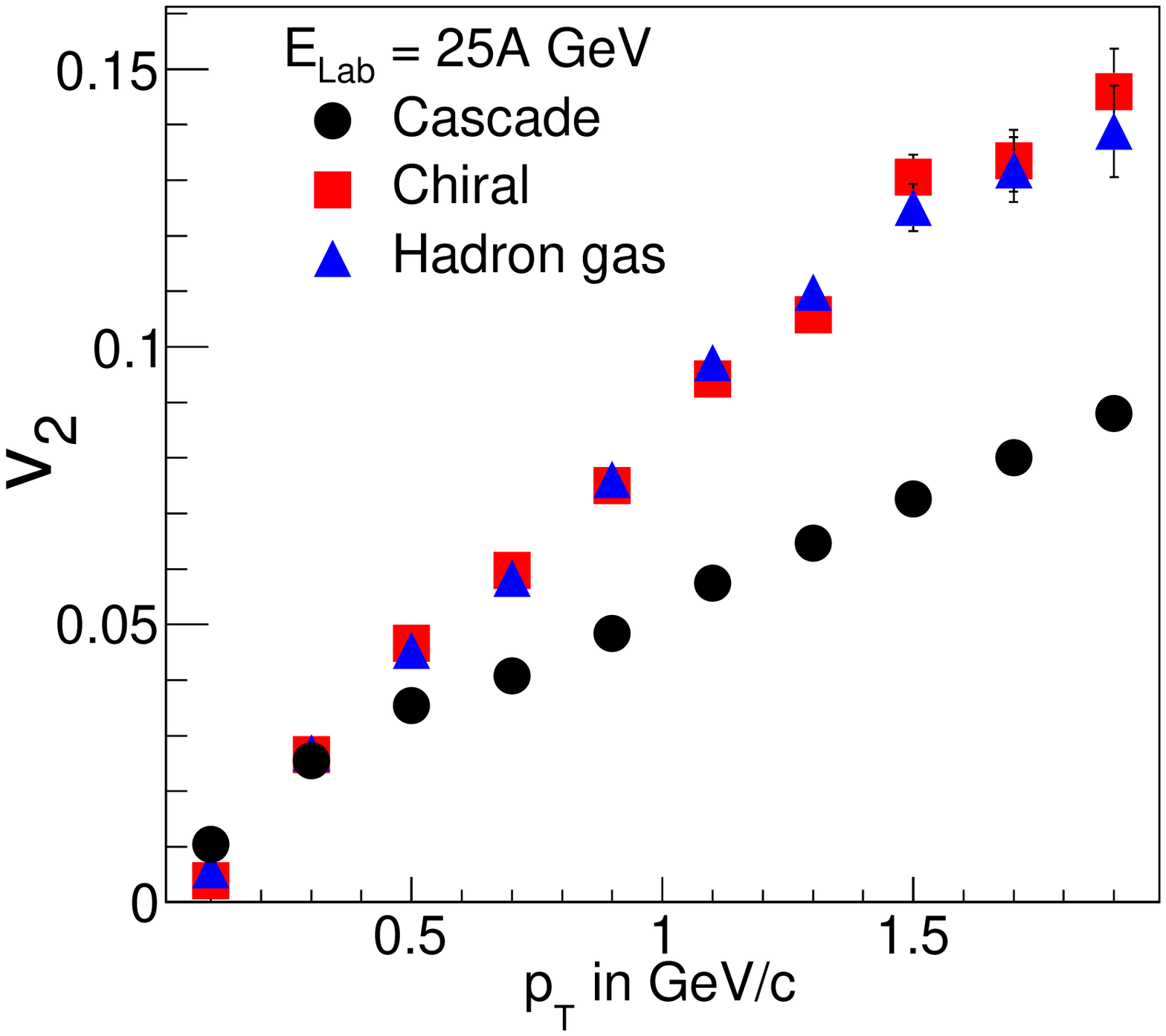}
  
    \caption{$v_{2}$ vs $p_{\rm T}$ for charged hadrons using UrQMD for different EoS for 6A, 8A, 10A \cite{Rode:2017bkd} and 25A GeV }
     \label{v2pt_all_ch}
\end{figure}

\begin{figure}[t]
  \includegraphics[scale=0.2]{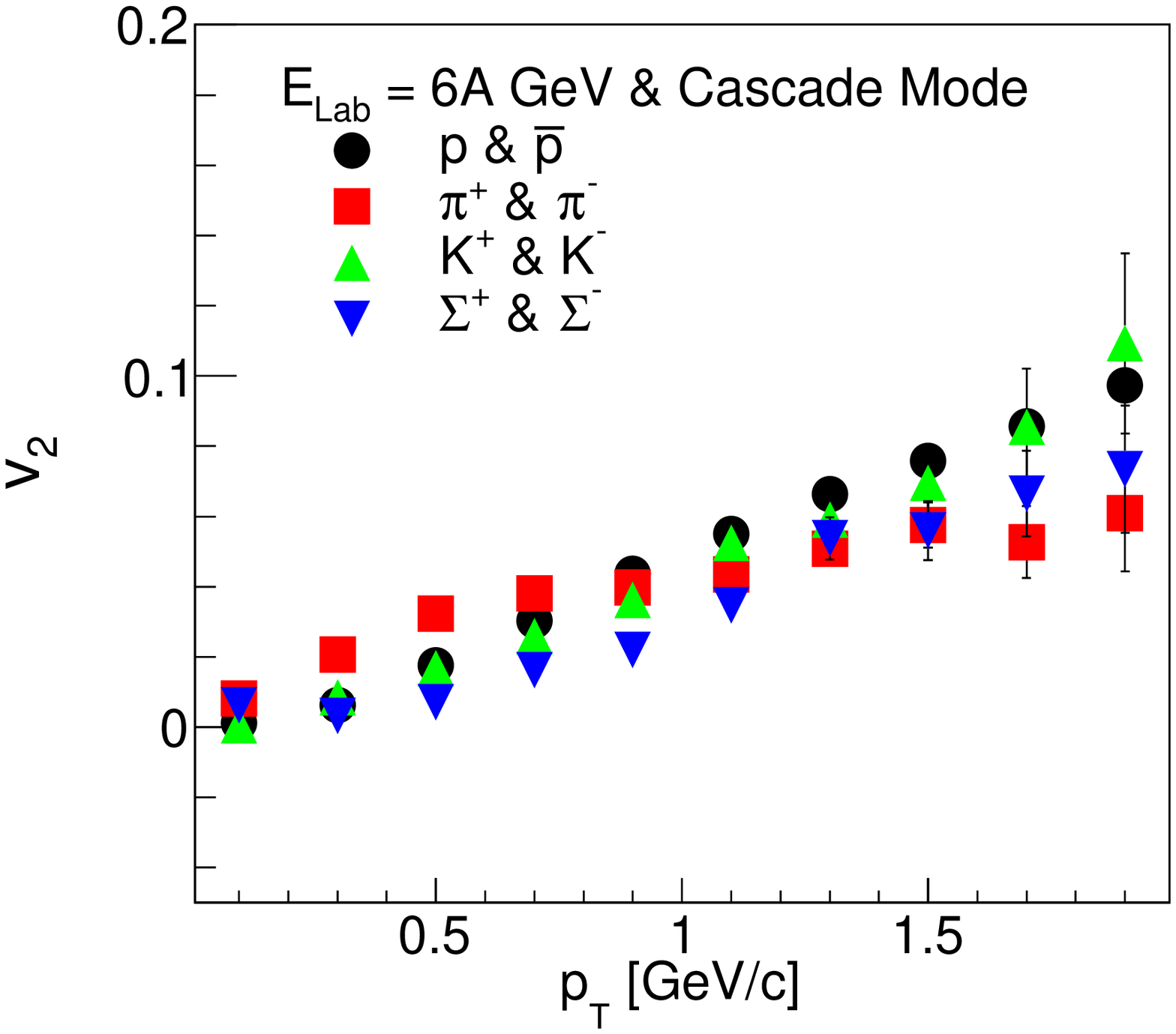}
  \includegraphics[scale=0.2]{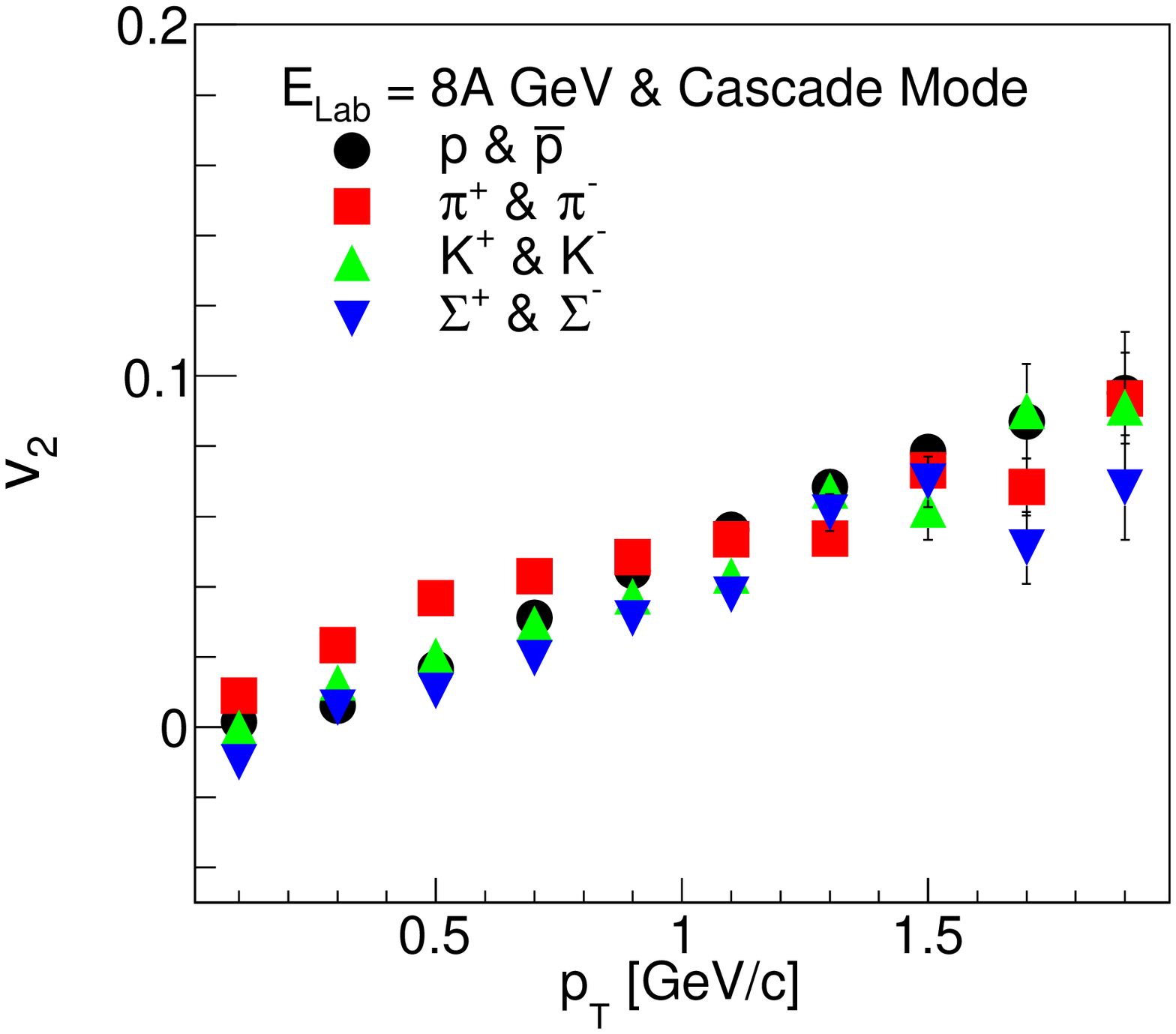}\\
  \includegraphics[scale=0.2]{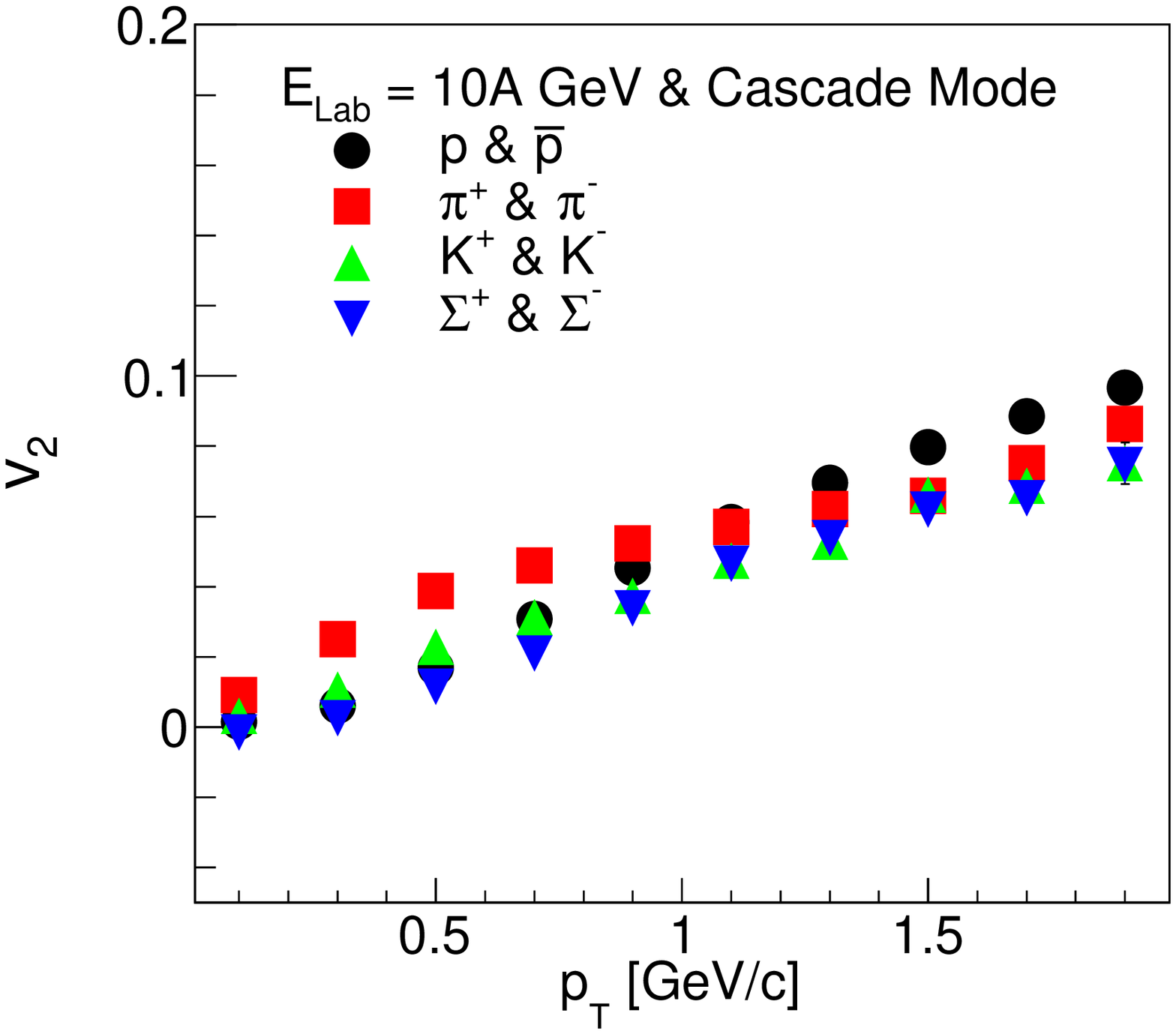}
  \includegraphics[scale=0.2]{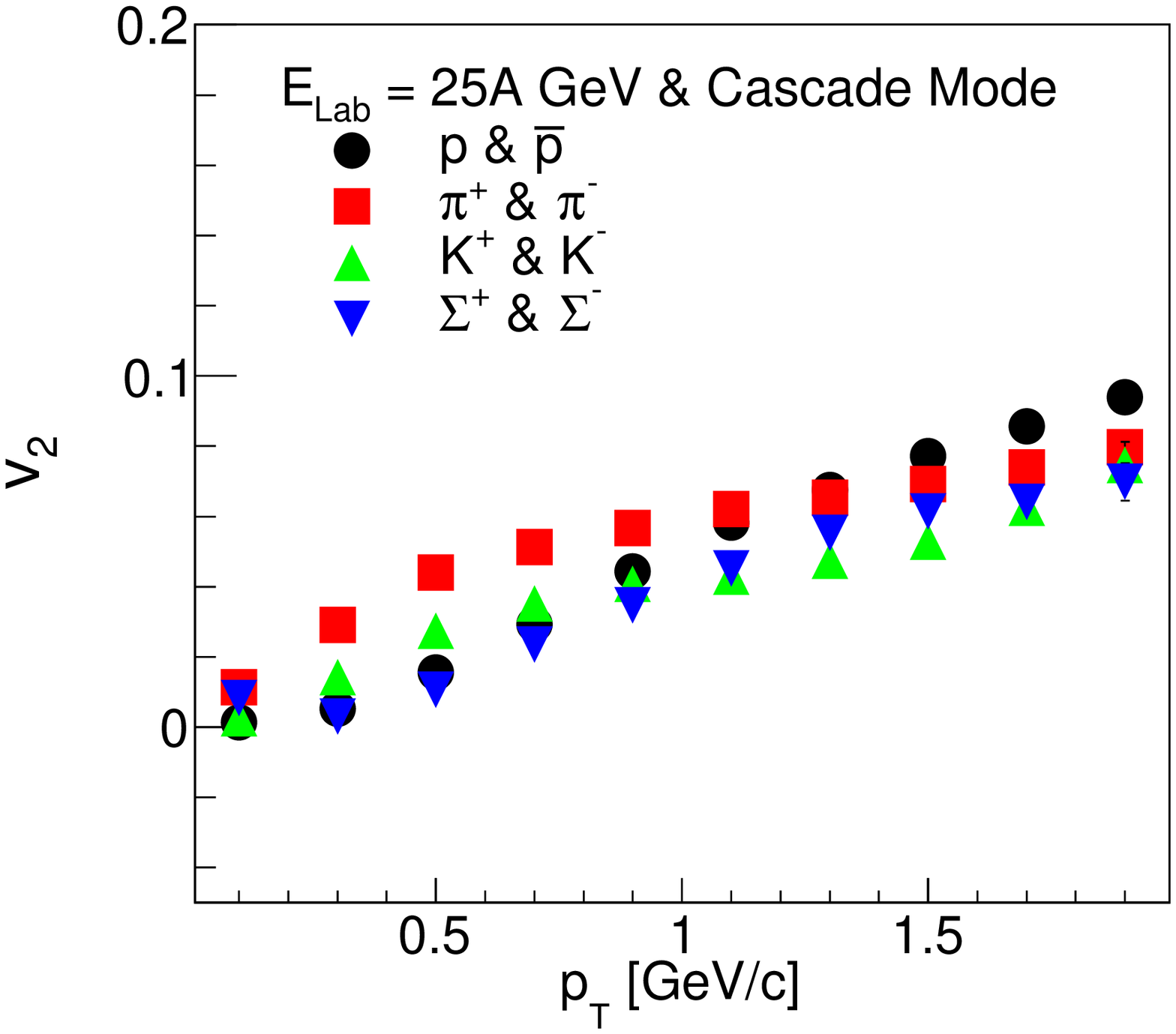}

  \caption{$v_{2}$ vs $p_{\rm T}$ of identified hadrons (p, $\bar{\rm p}$, $\pi^{\pm}$, $\rm K^{\pm}$ and $\Sigma^{\pm}$) using UrQMD in cascade mode for 6A, 8A, 10A and 25A GeV}
   \label{v2pt_cas_id}
\end{figure} 
 
 \begin{figure}[h]
  \includegraphics[scale=0.2]{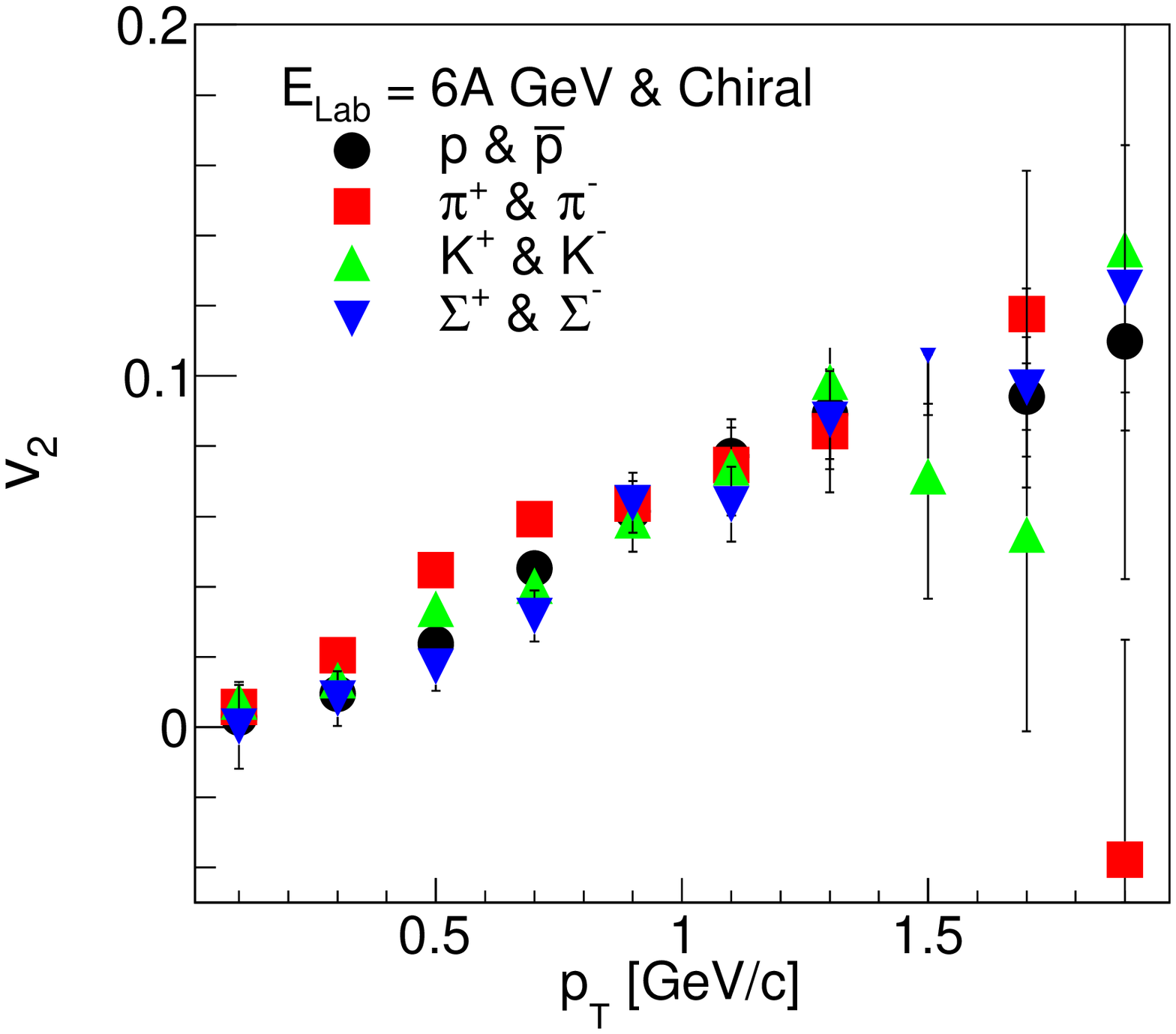}
  \includegraphics[scale=0.2]{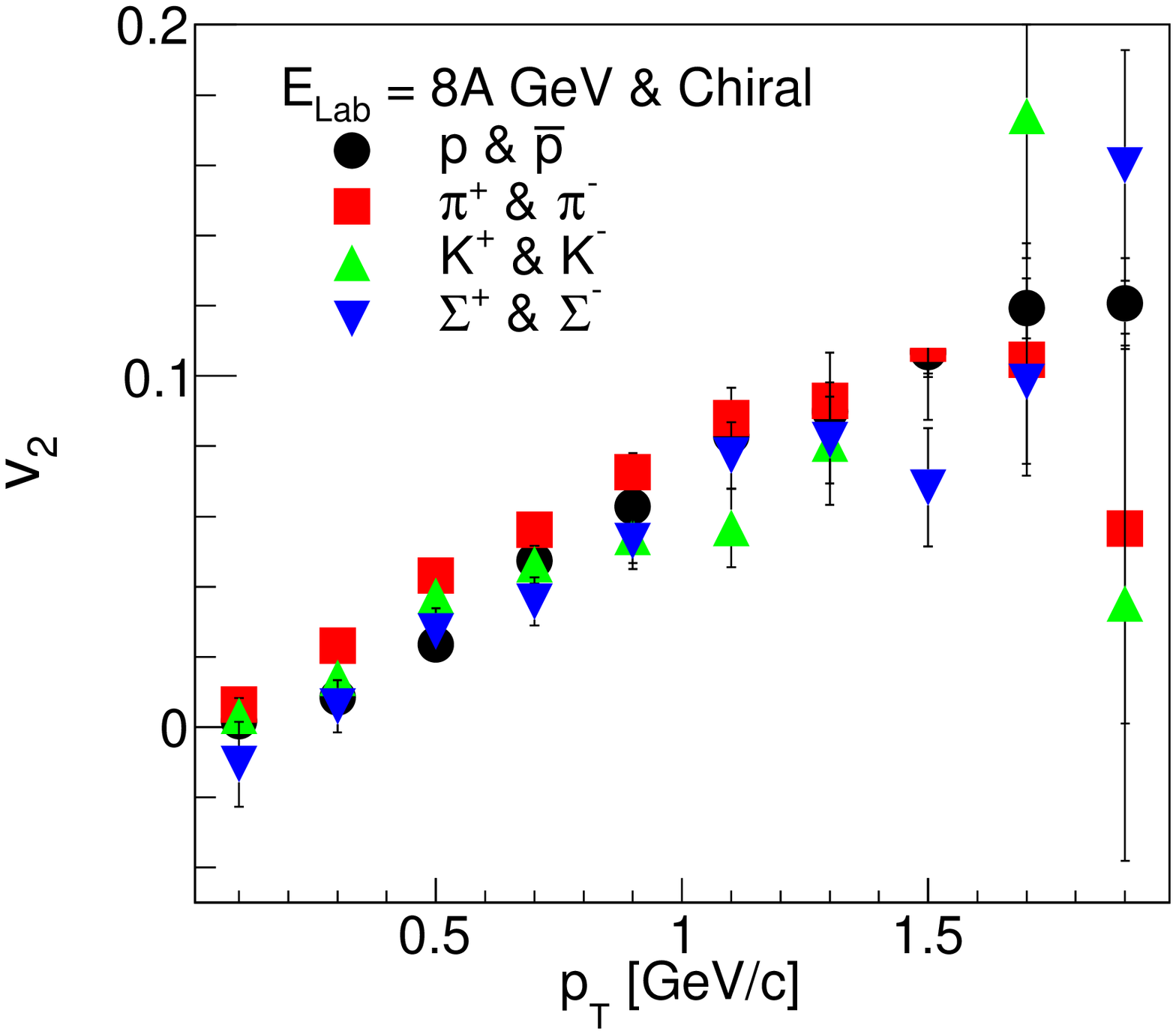}\\
  \includegraphics[scale=0.2]{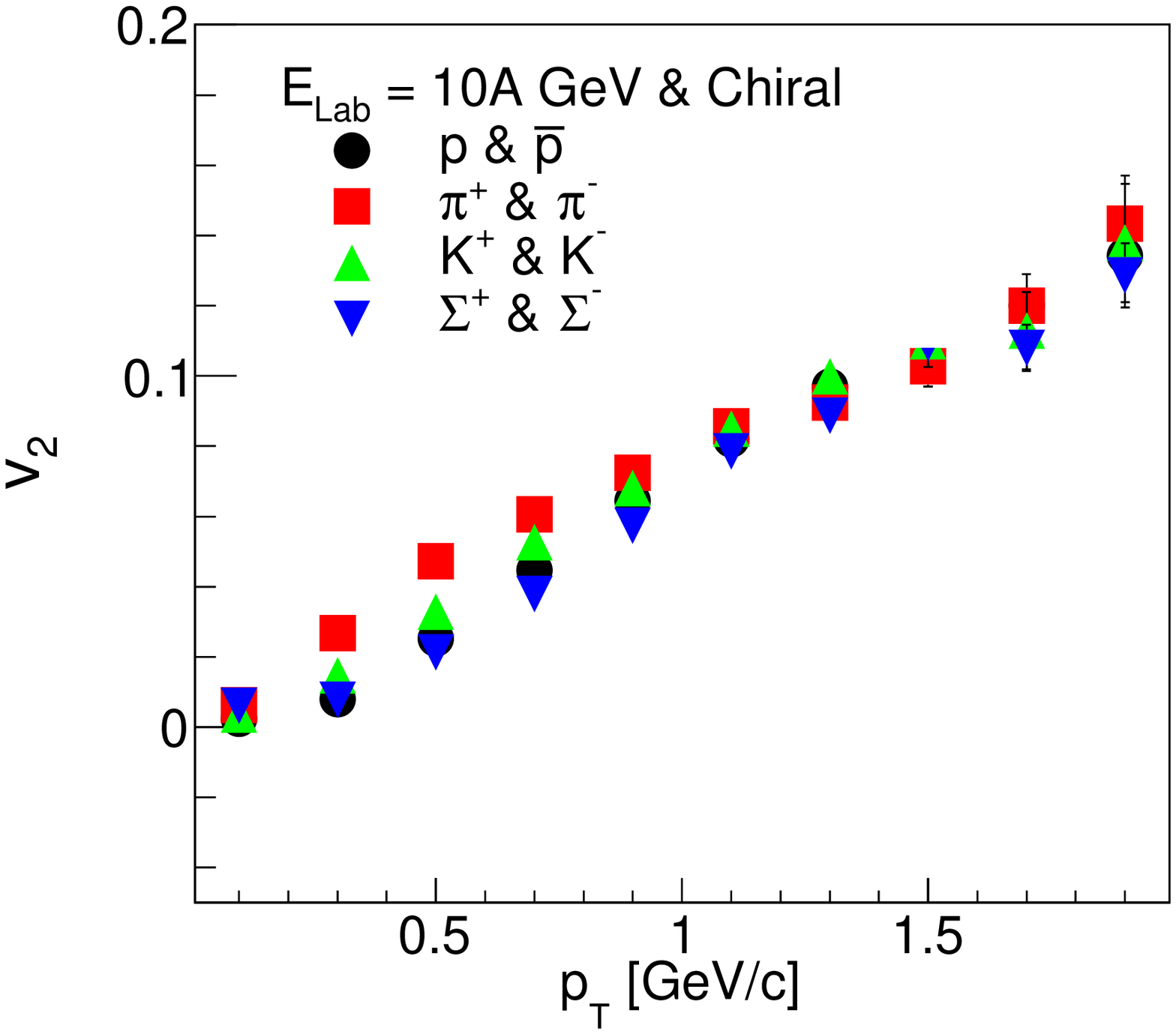}
  \includegraphics[scale=0.2]{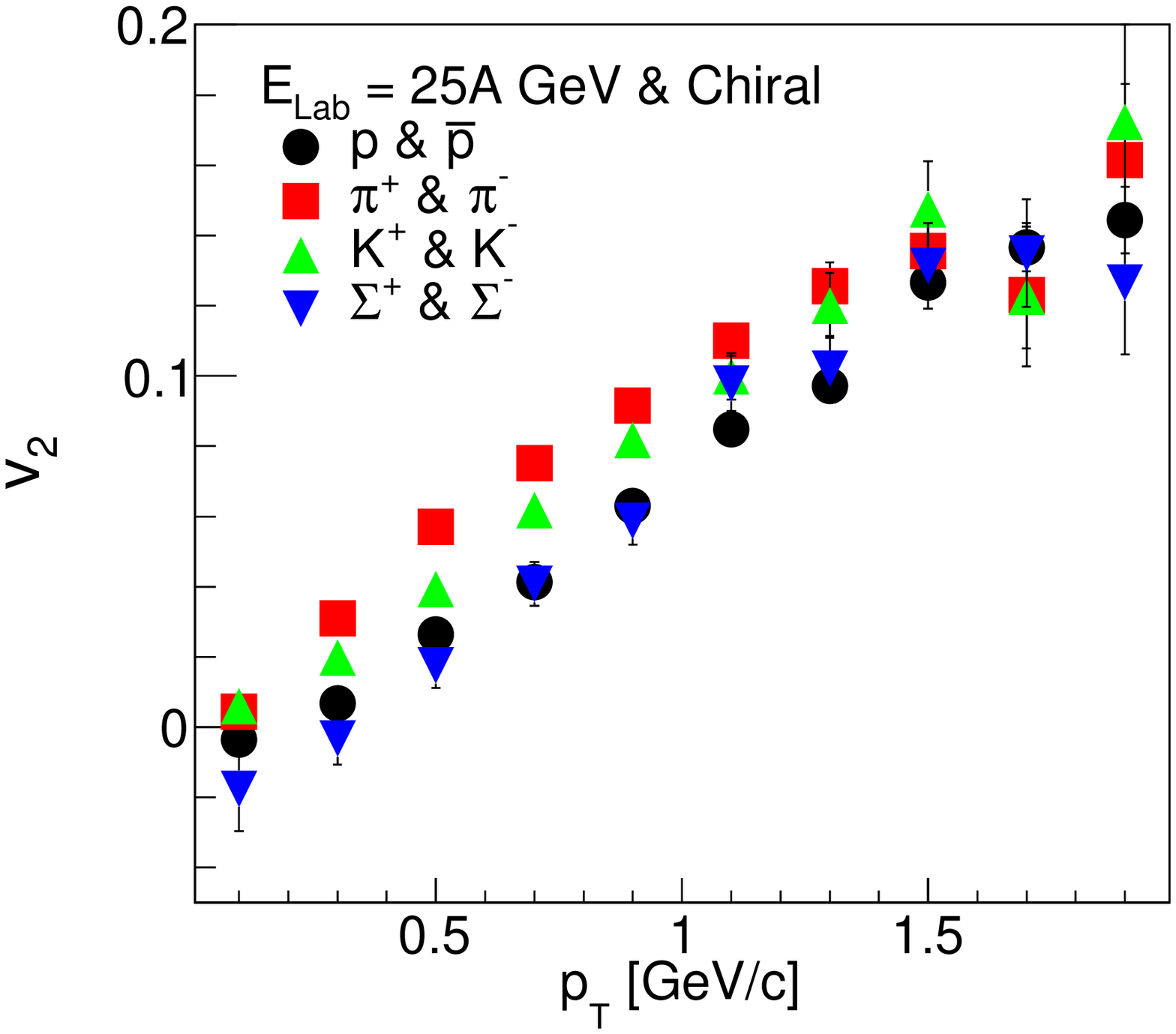}

  \caption{$v_{2}$ vs $p_{\rm T}$ of identified hadrons (p, $\bar{\rm p}$, $\pi^{\pm}$, $\rm K^{\pm}$ and $\Sigma^{\pm}$) using UrQMD in hydro mode using Chiral EoS for 6A, 8A, 10A and 25A GeV}
   \label{v2pt_ch_id}
\end{figure}

\begin{figure}
  \includegraphics[scale=0.2]{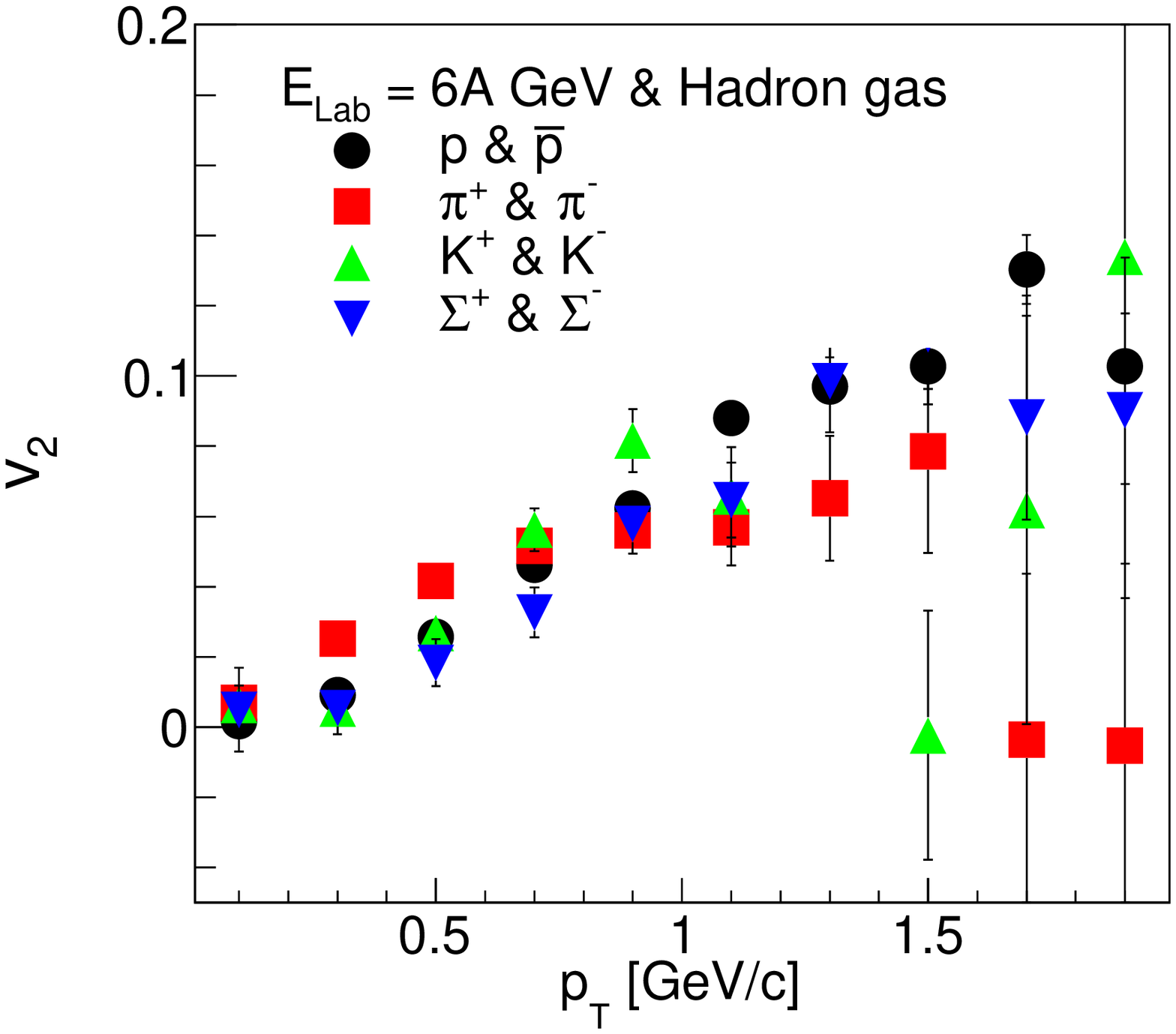}
  \includegraphics[scale=0.2]{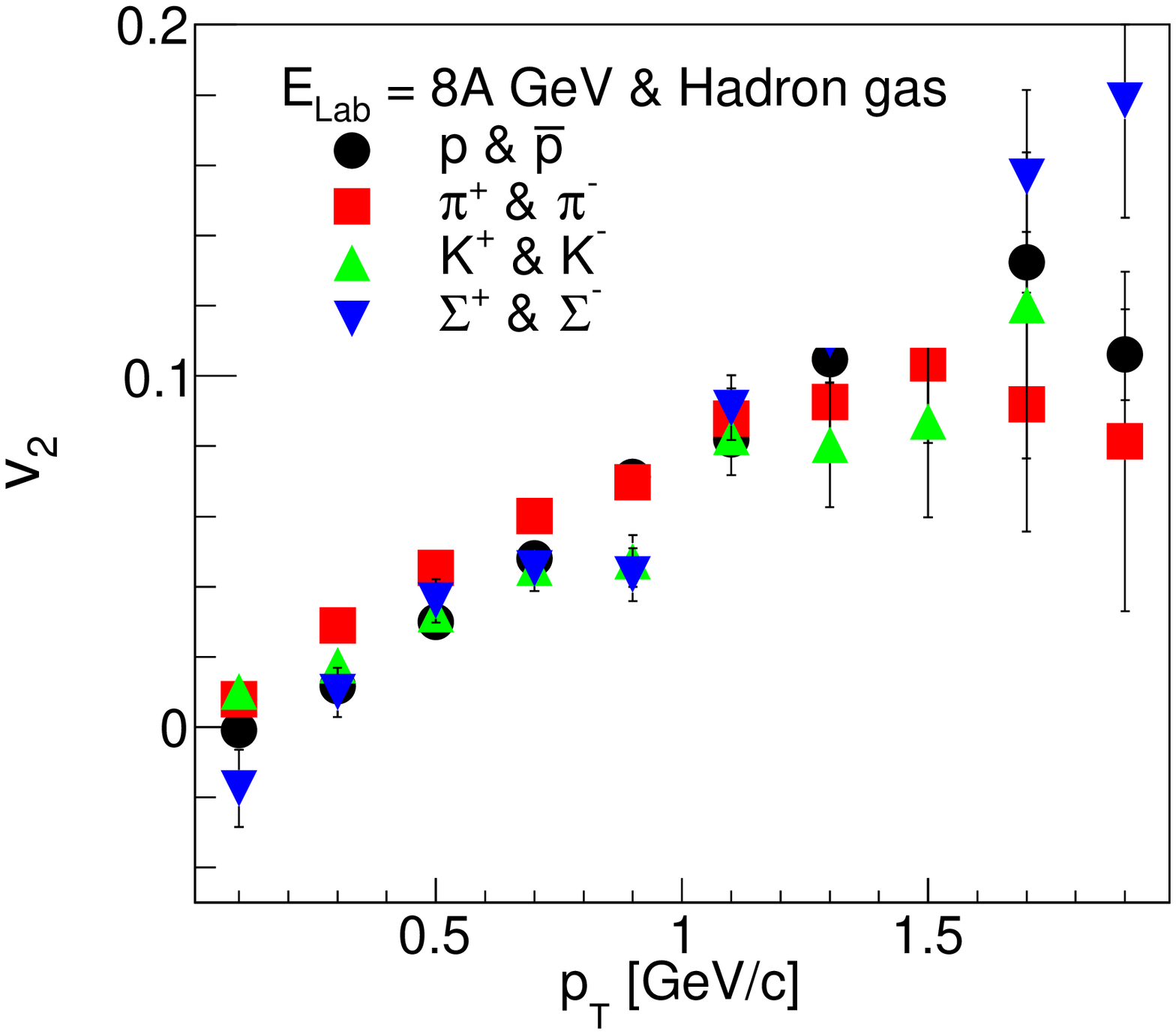}\\
  \includegraphics[scale=0.2]{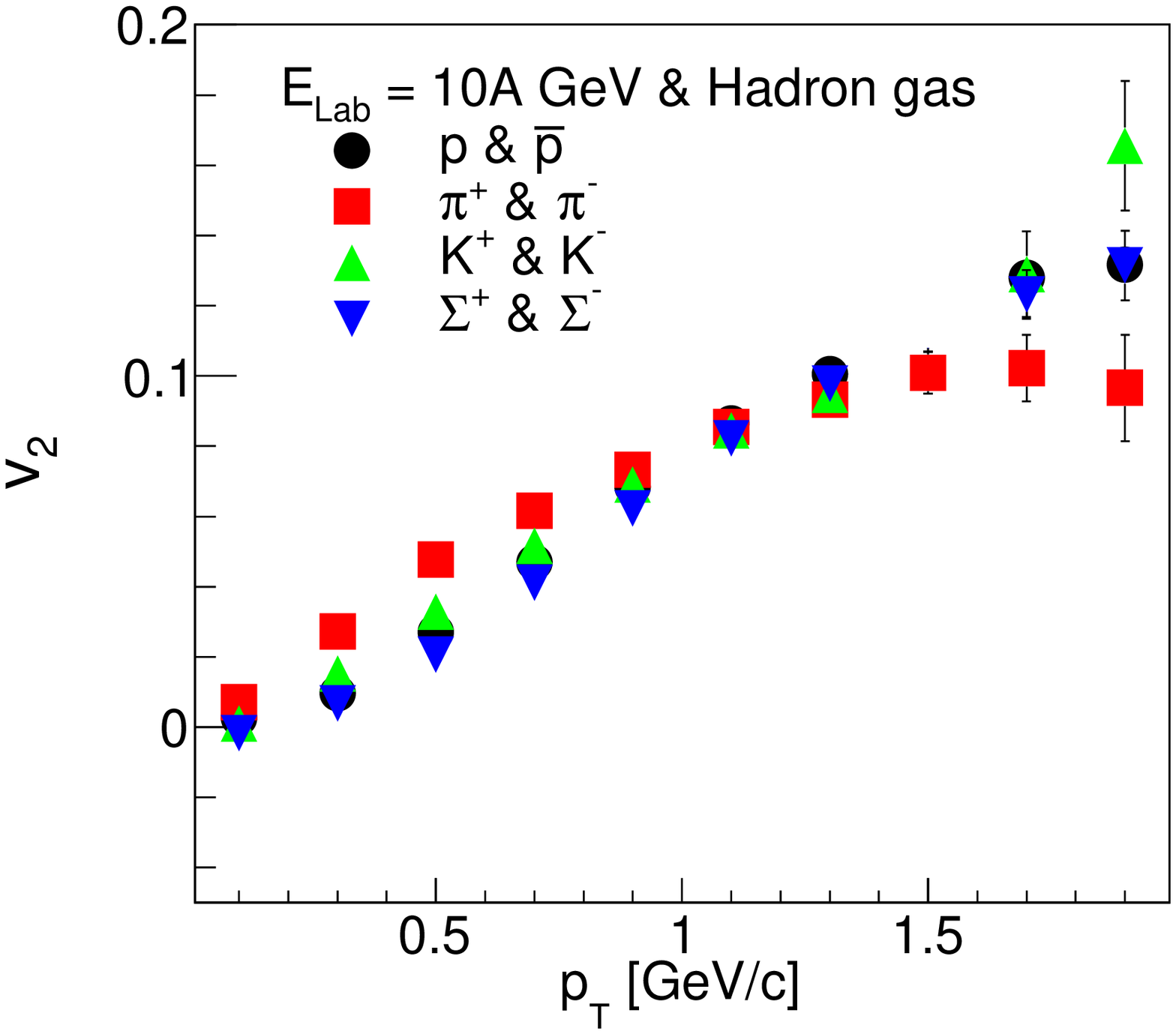}
  \includegraphics[scale=0.2]{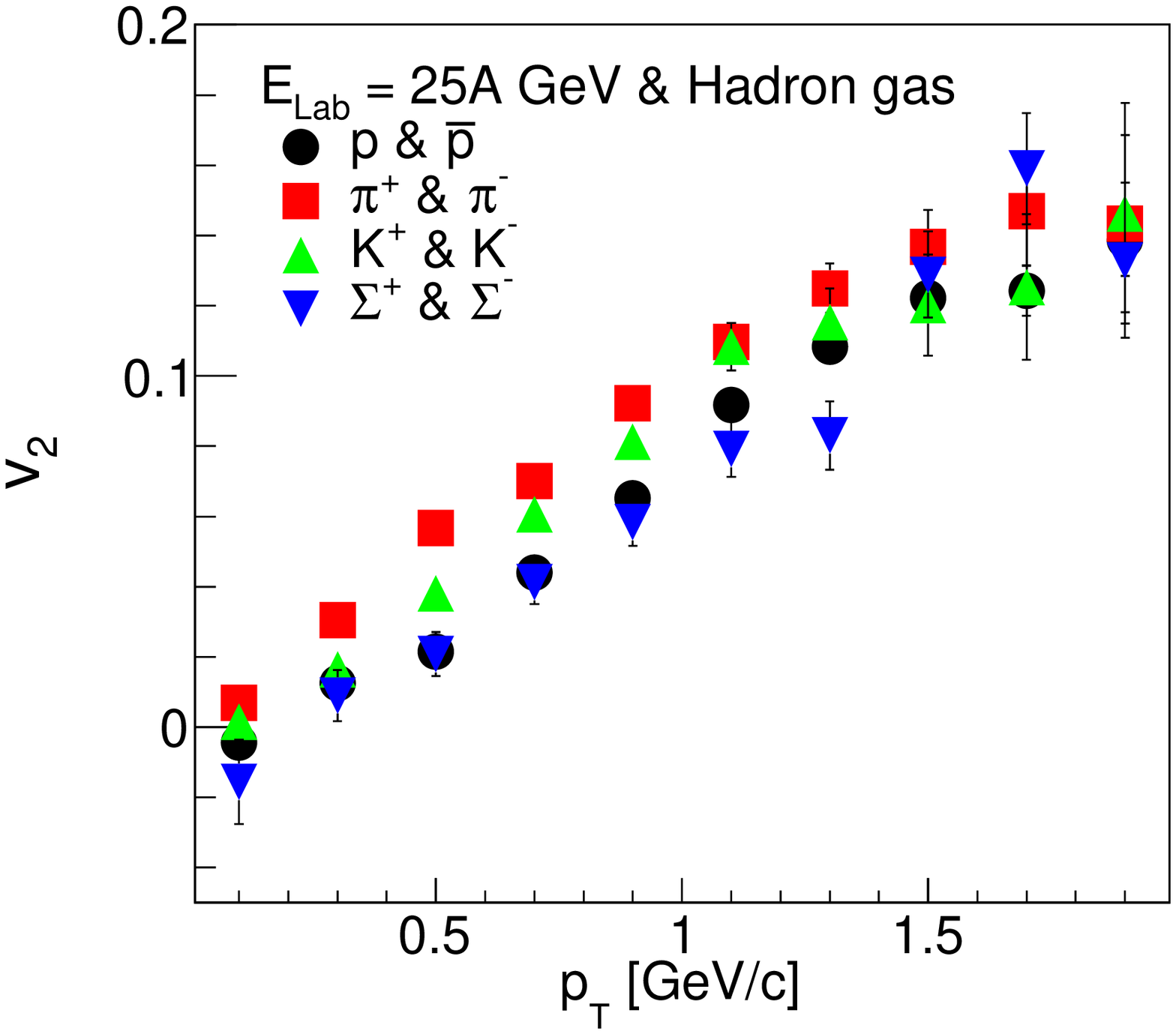}

  \caption{$v_{2}$ vs $p_{\rm T}$ of identified hadrons (p, $\bar{\rm p}$, $\pi^{\pm}$, $\rm K^{\pm}$ and $\Sigma^{\pm}$) using UrQMD in hydro mode using Hadron gas EoS for 6A, 8A, 10A and 25A GeV}
  \label{v2pt_hg_id}
\end{figure}

Another EoS used in this article for the dynamical evolution of the produced medium is Chiral + deconfinement EoS which is taken from the hadronic SU(3) parity doublet model in which quark degrees of freedom are included \cite{Steinheimer:2011ea}. It incorporates chiral as well as deconfinement phase transition. At vanishing baryon chemical potential, this EoS agrees with the lattice QCD results qualitatively. Particularly for this investigation, it is important to note that this EoS is conjectured to be applicable at non-zero baryon chemical potentials. For all values of $\mu_{B}$, this EoS describes the deconfinement transition as a continuous crossover. In this EoS, the deconfinement transition mainly governed by quarks and Polyakov potential and the chiral phase transition by hadronic interactions. Hadrons disappear only at higher temperatures, and quark degrees of freedom becomes dominant. For a more detailed explanation, the reader is referred to Ref~\cite{Steinheimer:2011ea}. 

\section*{III Results and Discussion}

In this section, we discuss the transverse momentum dependence of the elliptic flow, the rapidity dependence of directed flow and the elliptic flow of both the identified and charged hadrons for mid-central ($b= 5-9$ fm) Au-Au collisions at bombarding energies 6A, 8A, 10A and 25A GeV. In the hybrid mode, calculations are performed for two different nuclear EoS, as mentioned before. 
 We also compare proton $v_{1}$ with the measured data from the E895 Collaboration at the AGS \cite{Liu:2000am}, at 6A and 8A GeV beam energies. In addition the constituent quark number scaling of $v_{2}(p_{T})$ is also studied. Finally, we look at the beam energy ($\rm E_{\rm Lab}$) dependence of the slope of the directed flow ($\frac{dv_{1}}{dy}$), elliptic flow, $v_{4}$ and ratio $v_{4}/(v_{2})^{2}$ in the midrapidity region (-0.75 $\leqslant y_{c.m.} \leqslant$ 0.75).

\begin{figure}
  \includegraphics[scale=0.2]{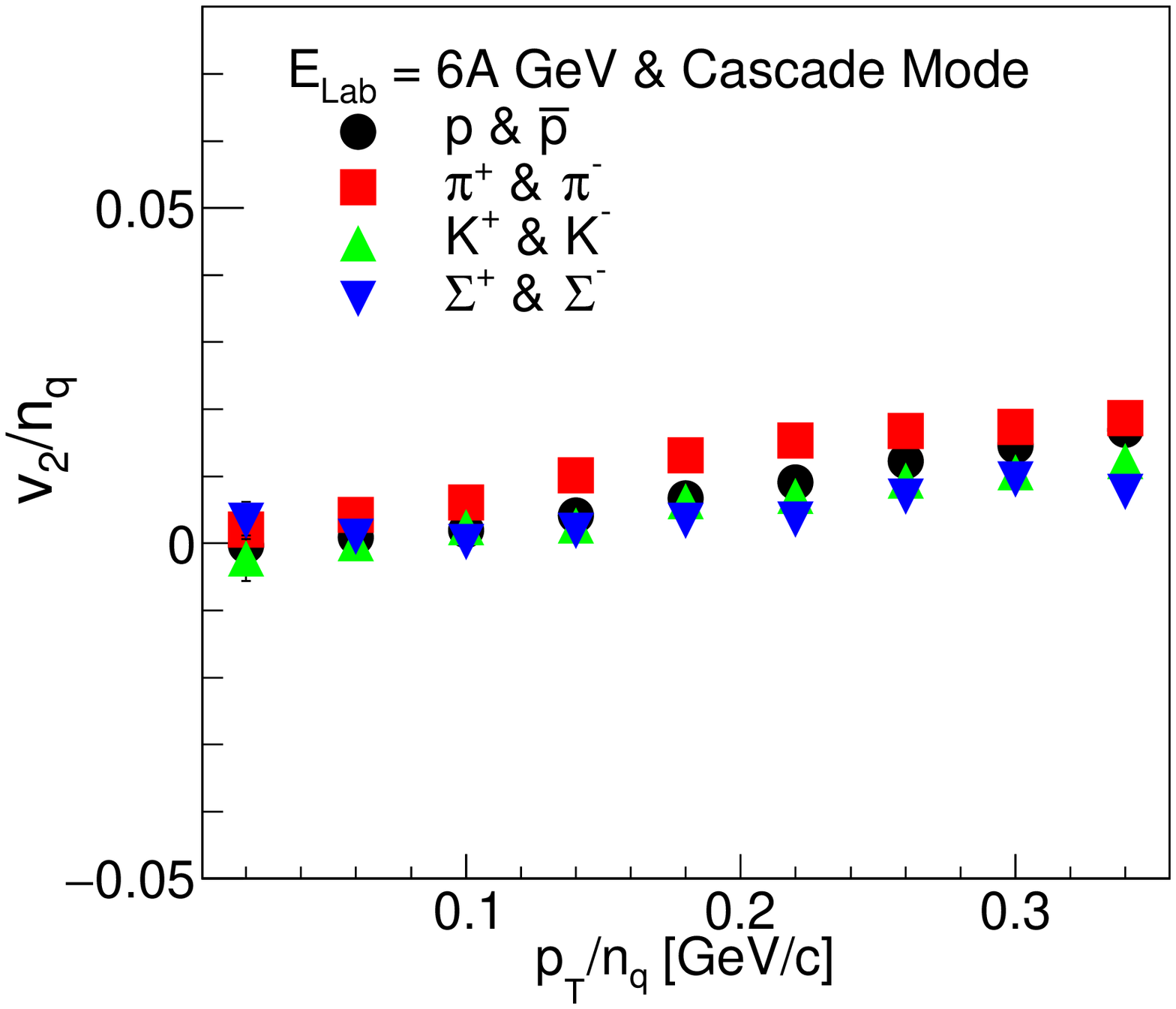}
  \includegraphics[scale=0.2]{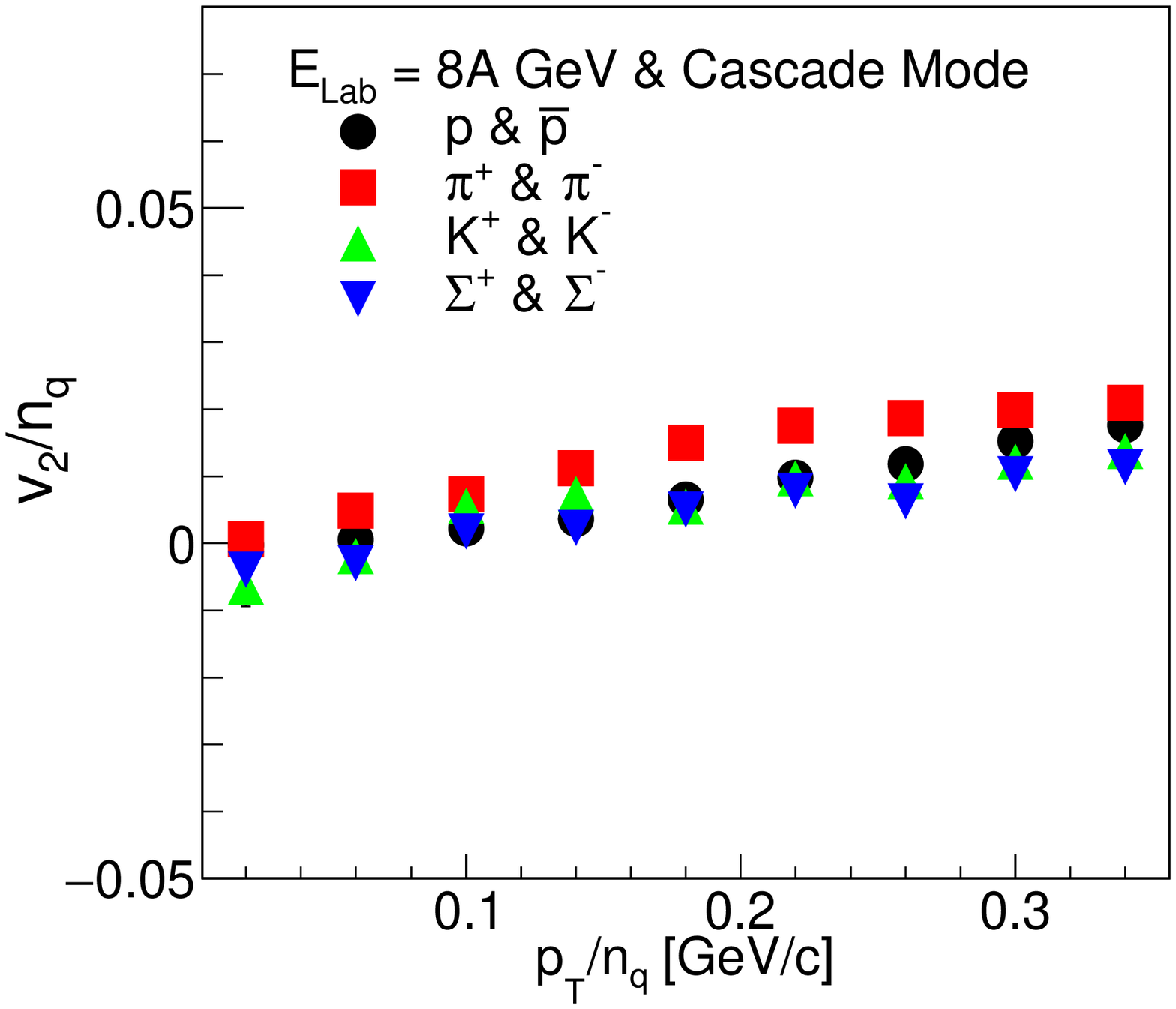}\\
  \includegraphics[scale=0.2]{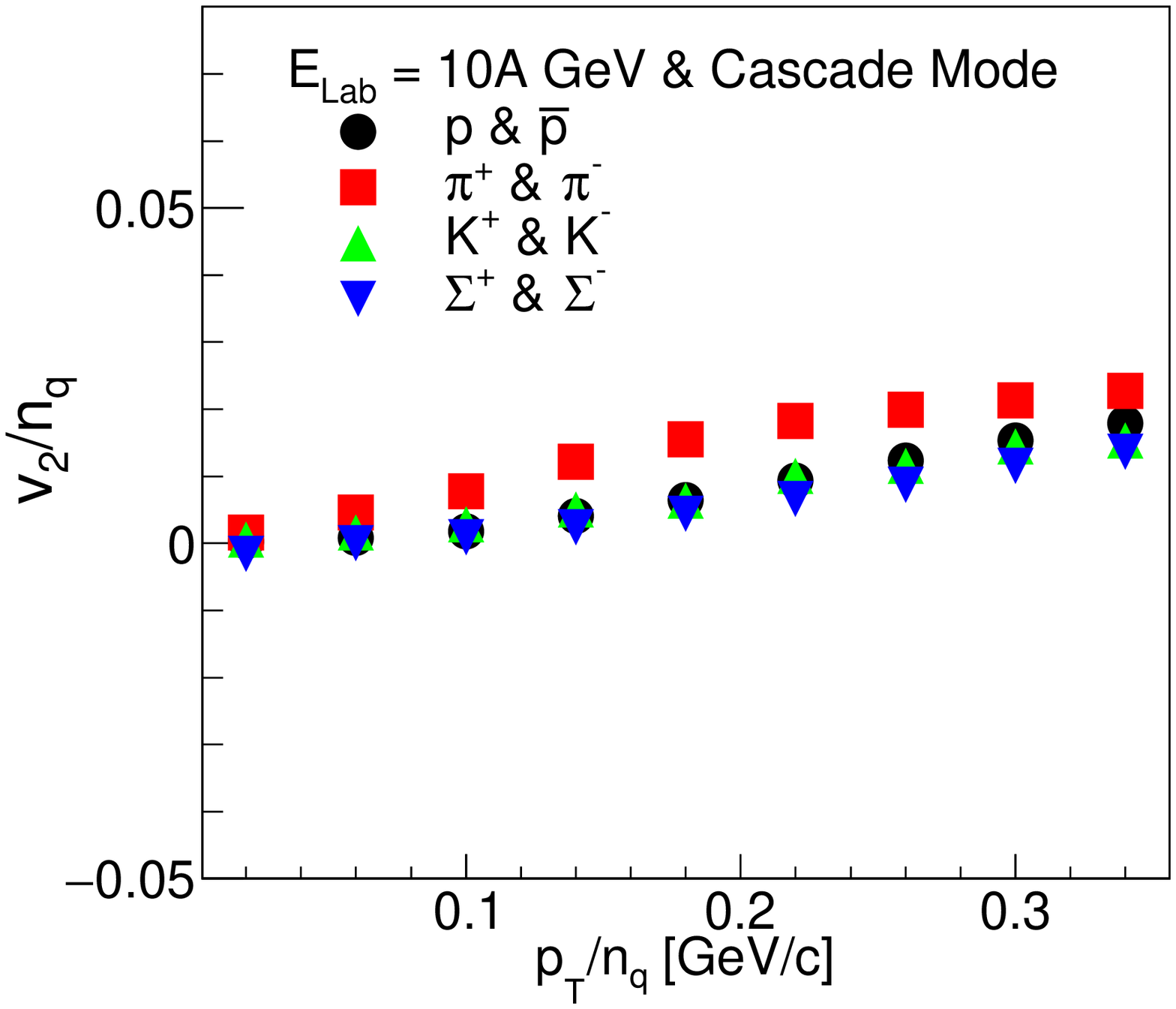}
  \includegraphics[scale=0.2]{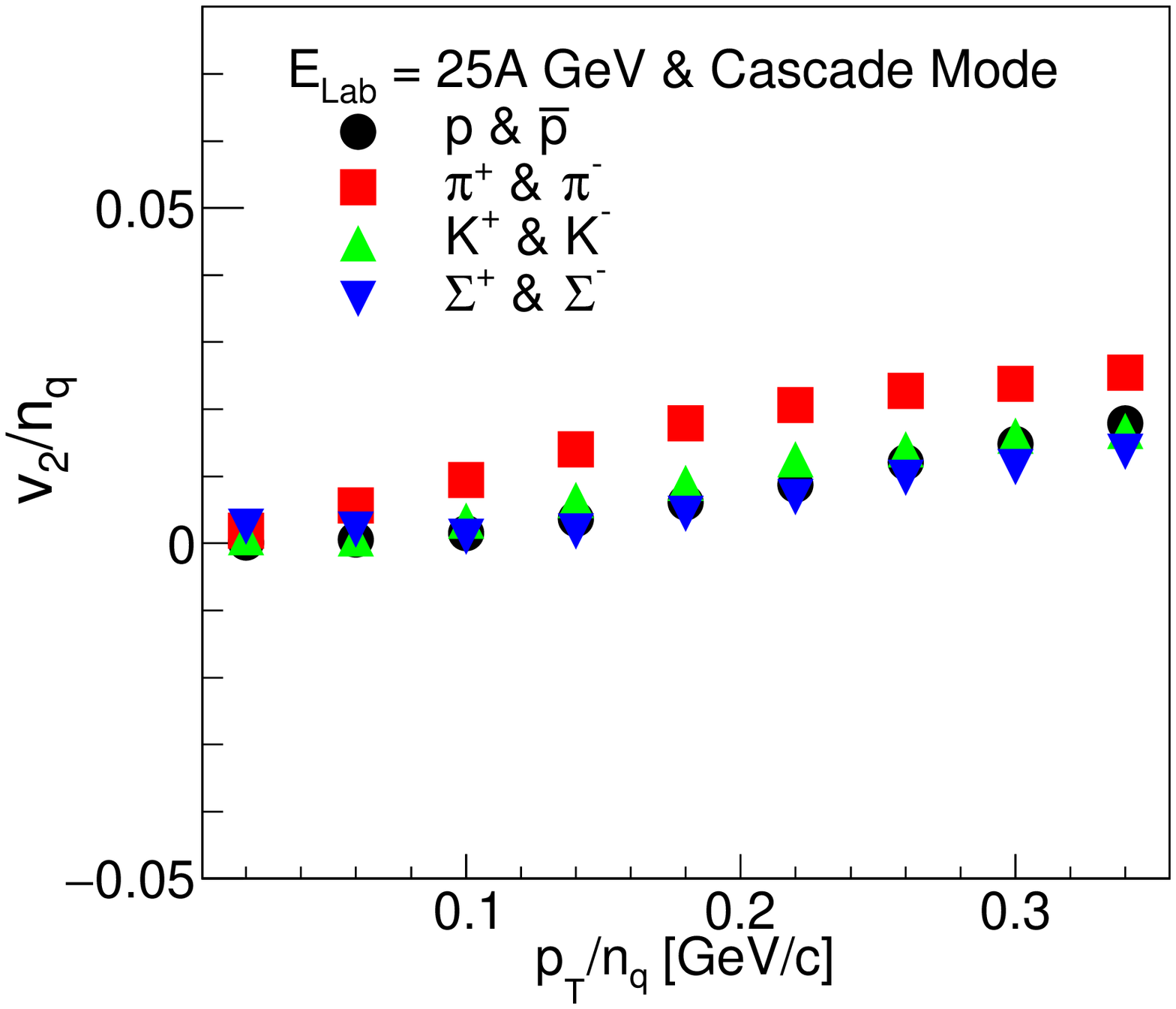}

  \caption{$v_{2}/n_{q}$ vs ${p_{\rm T}}/n_{q}$ of identified hadrons (p, $\bar{\rm p}$, $\pi^{\pm}$, $\rm K^{\pm}$ and $\Sigma^{\pm}$) using UrQMD in cascade mode for 6A, 8A, 10A and 25A GeV}
 \label{scale_cas}
\end{figure}

\subsection*{A. $p_{\rm T}$ dependence}

\begin{figure}[h]
  \includegraphics[scale=0.2]{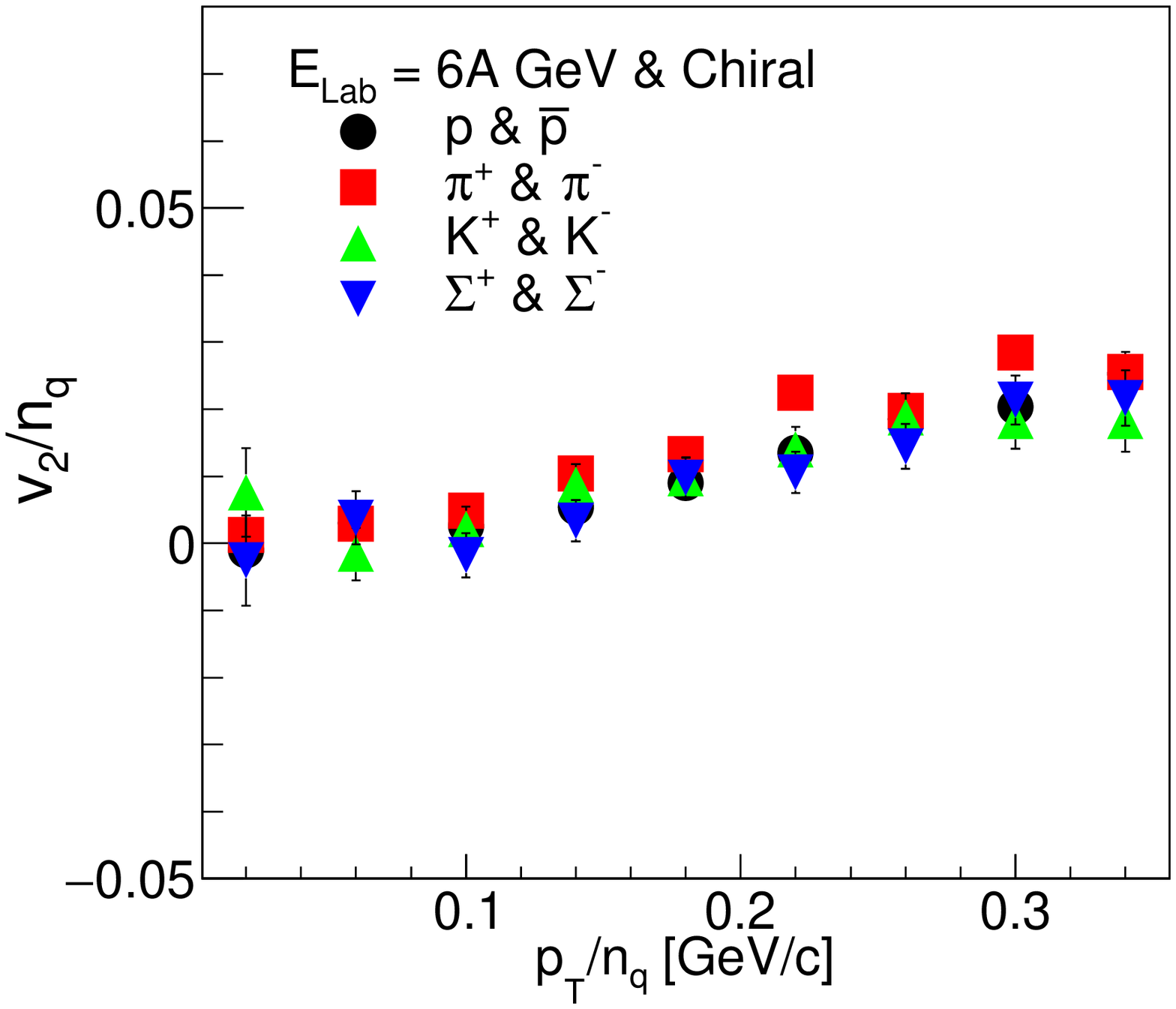}
  \includegraphics[scale=0.2]{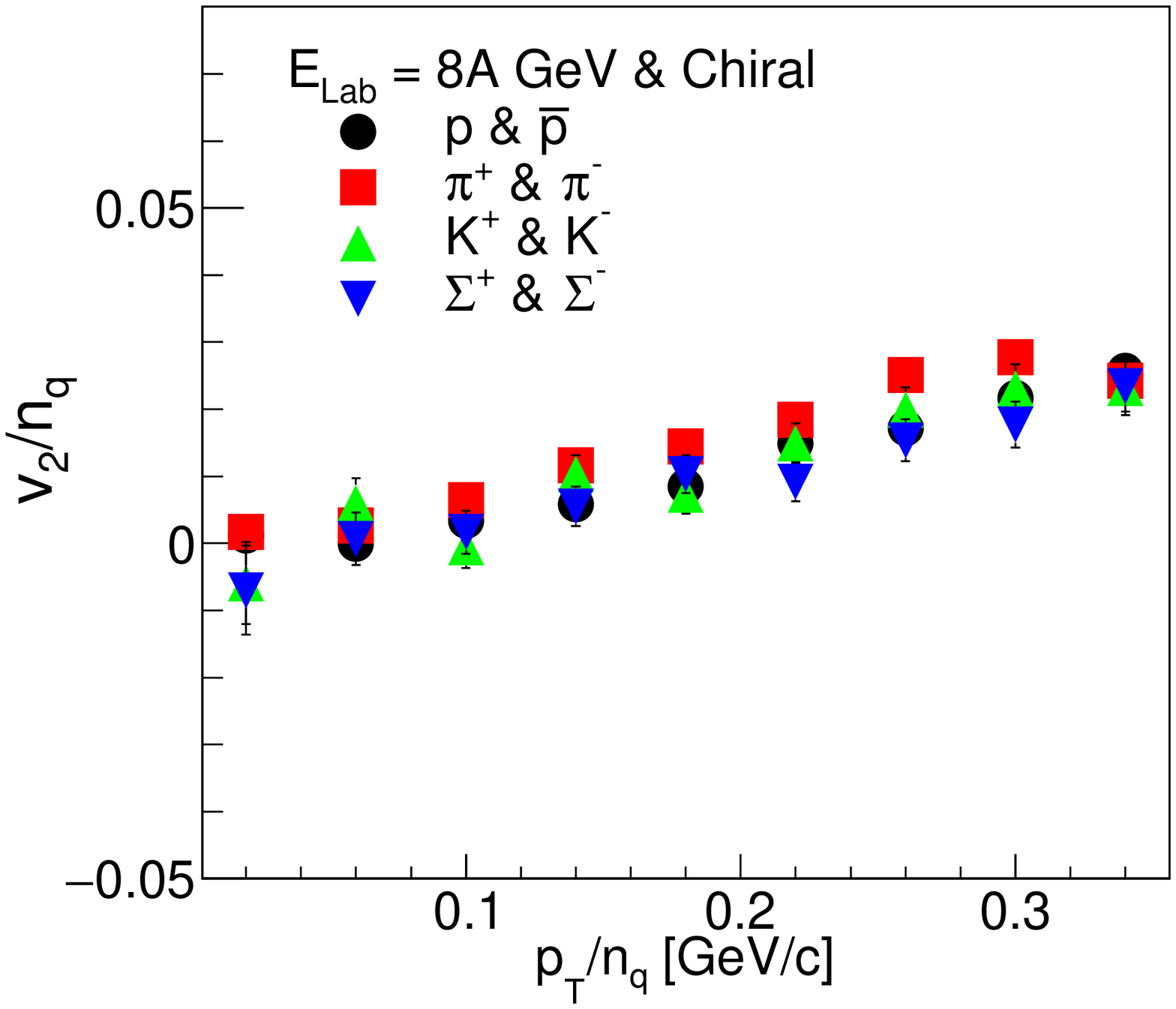}\\
  \includegraphics[scale=0.2]{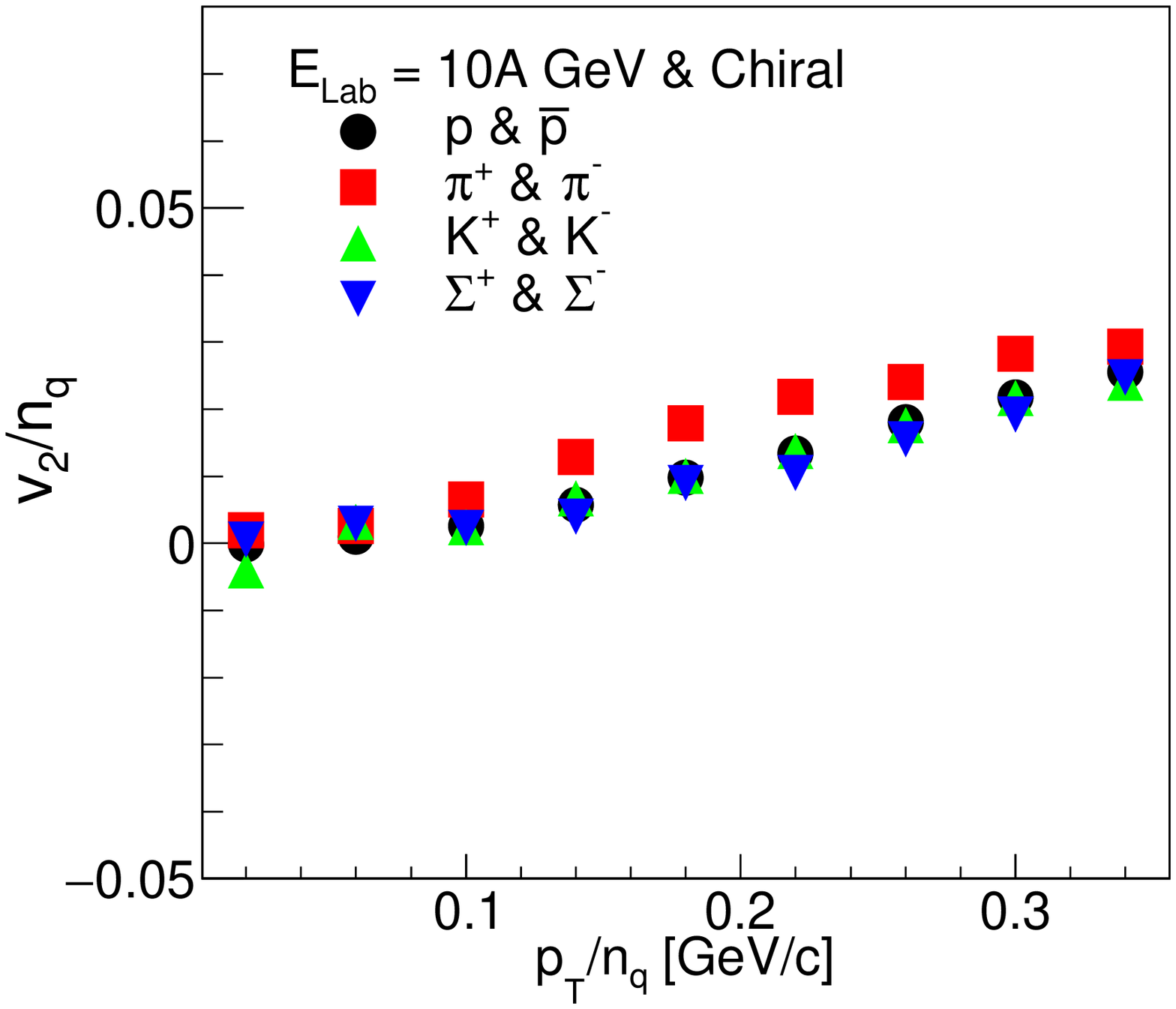}
  \includegraphics[scale=0.2]{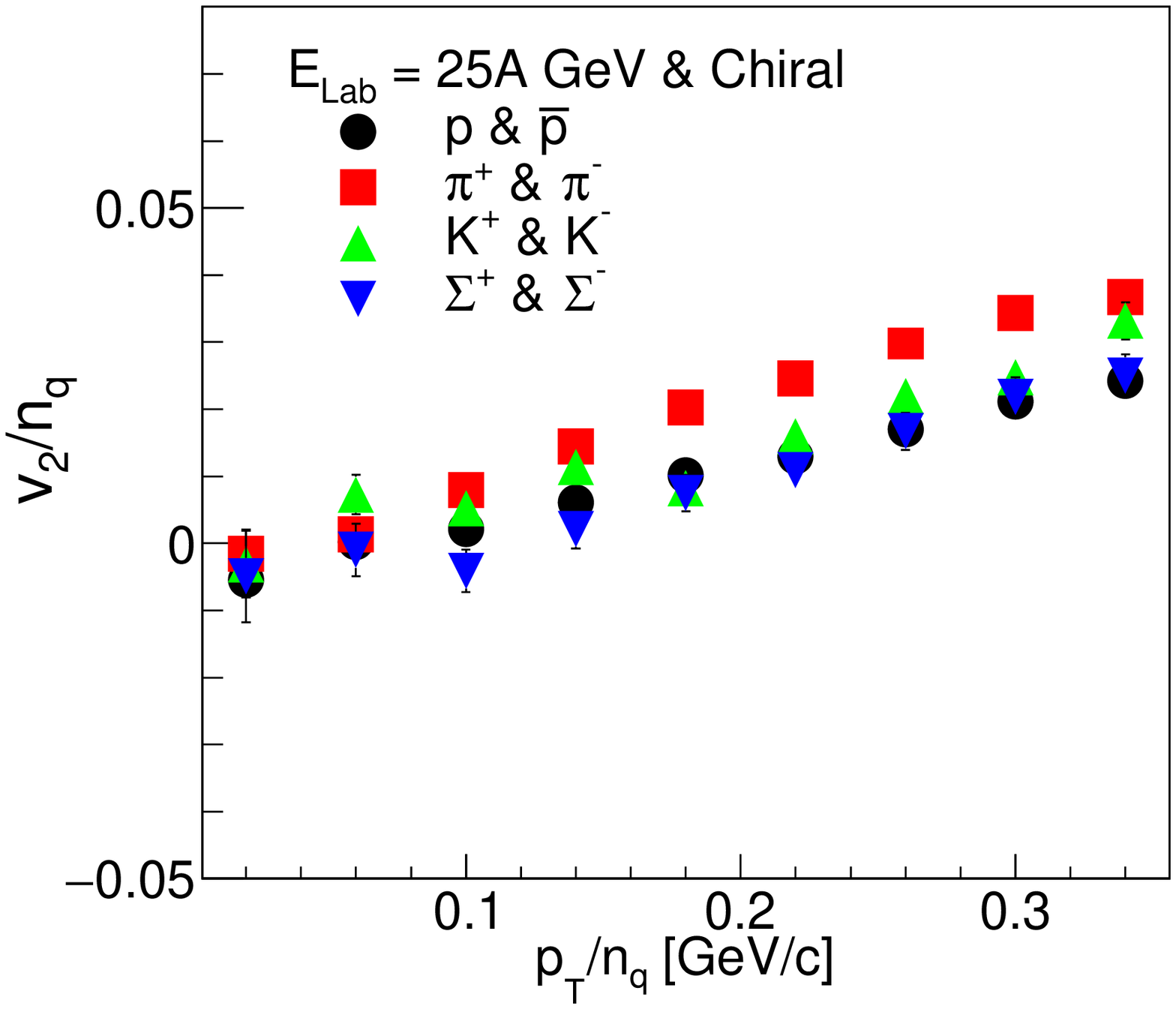}

  \caption{$v_{2}/n_{q}$ vs ${p_{\rm T}}/n_{q}$ of identified hadrons (p, $\bar{\rm p}$, $\pi^{\pm}$, $\rm K^{\pm}$ and $\Sigma^{\pm}$) using UrQMD in hydro mode using Chiral EoS for 6A, 8A, 10A and 25A GeV}
\label{scale_chi}
\end{figure}

\begin{figure}[h]
  \includegraphics[scale=0.2]{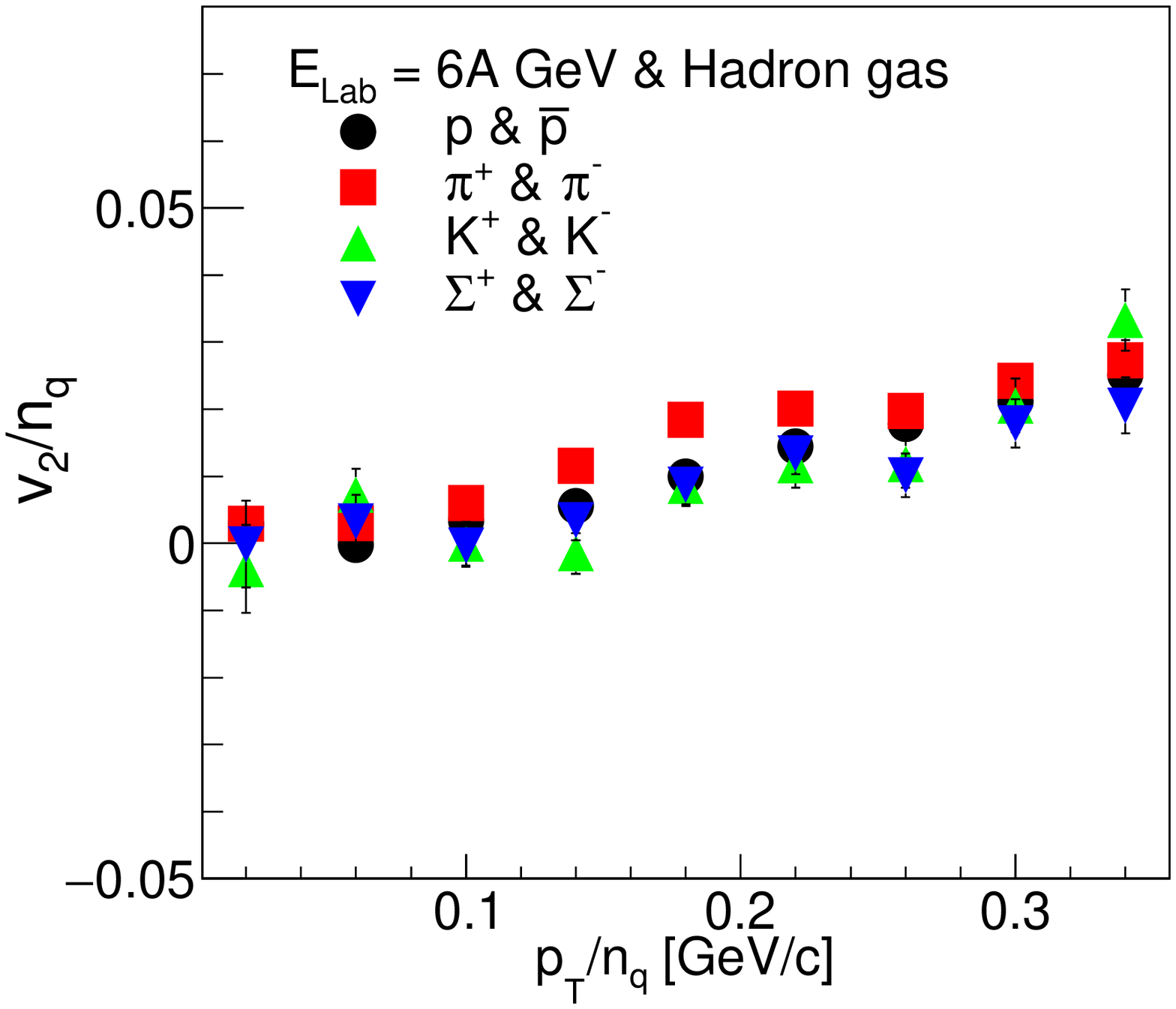}
  \includegraphics[scale=0.2]{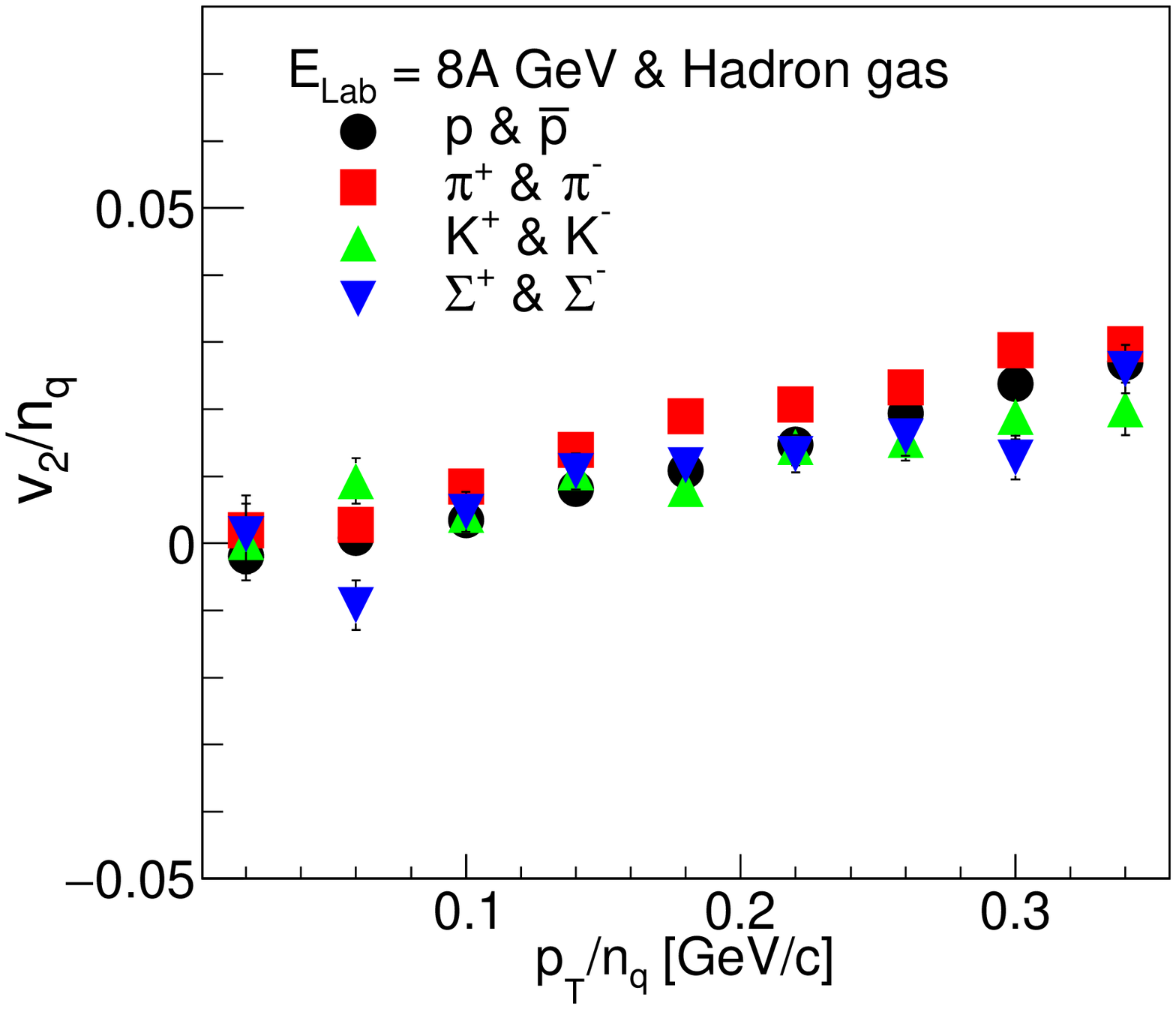}\\
  \includegraphics[scale=0.2]{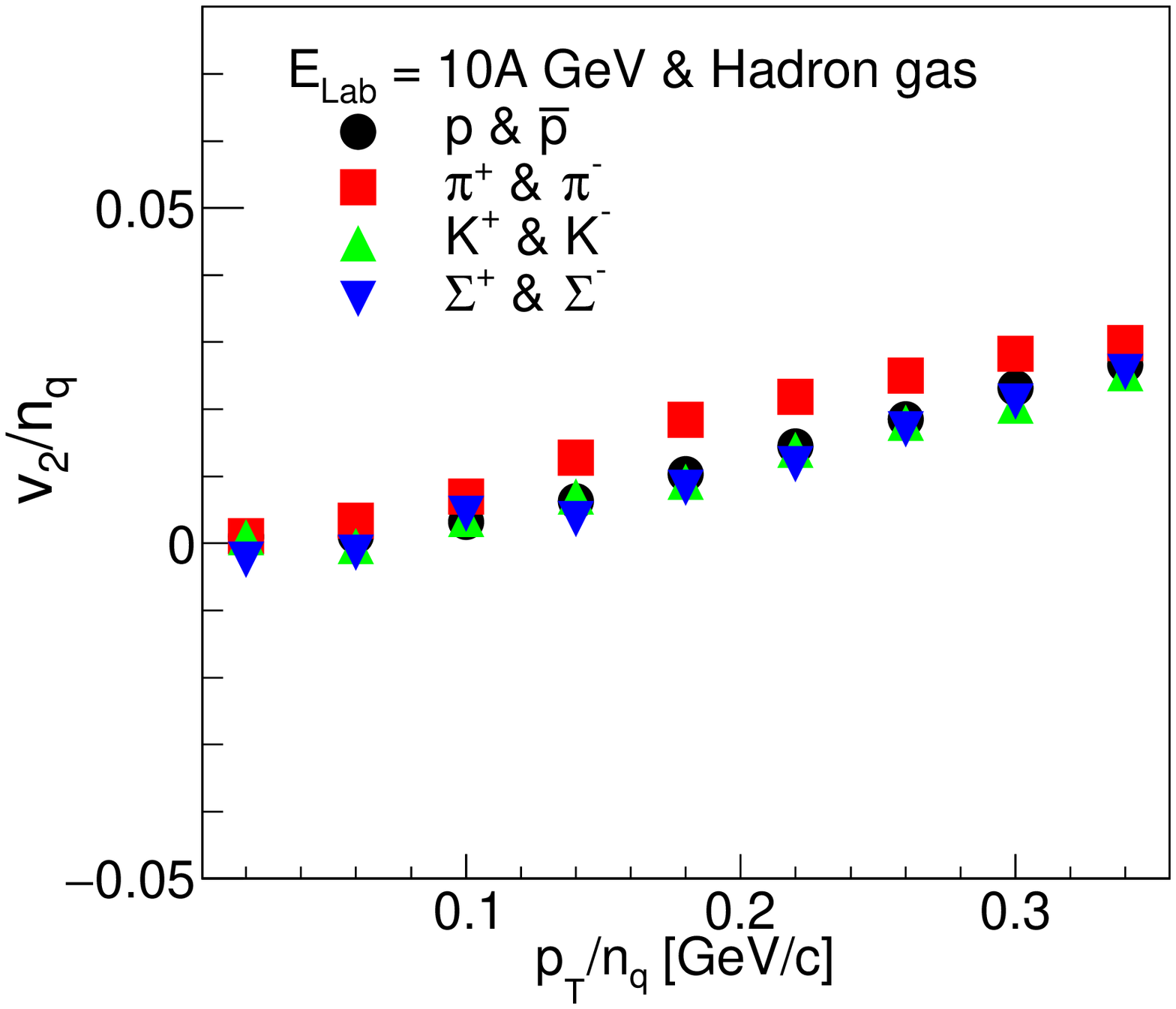}
  \includegraphics[scale=0.2]{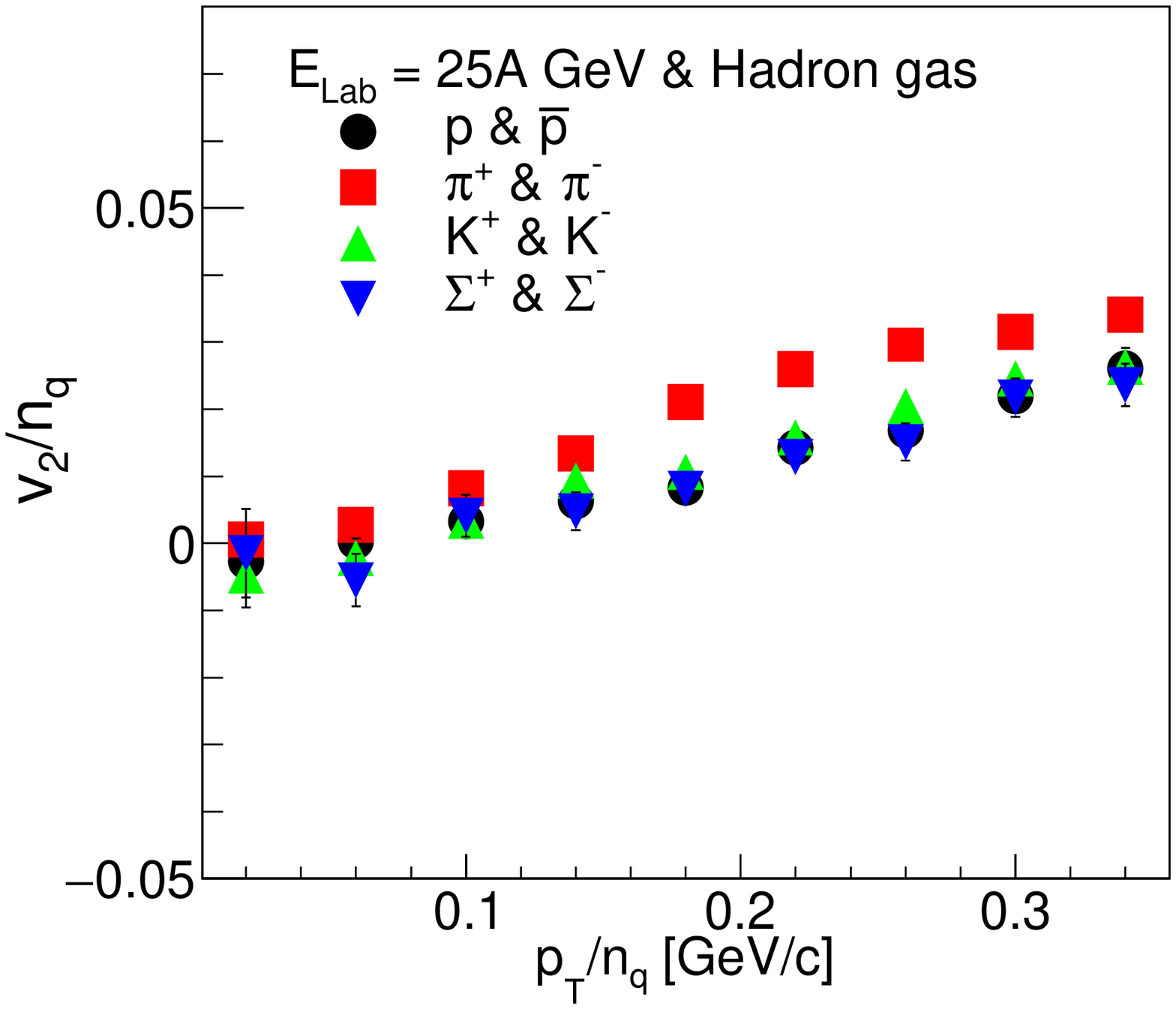}

  \caption{$v_{2}/n_{q}$ vs ${p_{\rm T}}/n_{q}$ of identified hadrons (p, $\bar{\rm p}$, $\pi^{\pm}$, $\rm K^{\pm}$ and $\Sigma^{\pm}$) using UrQMD in hydro mode using Hadron gas EoS for 6A, 8A, 10A and 25A GeV}
  \label{scale_hg}
\end{figure}

 Fig. \ref{v2pt_all_ch} shows the differential elliptic flow $v_2(p_T)$ of charged hadrons, at all four investigated energies. An approximate linear rise of elliptic flow $v_{2}$ with respect to $p_{\rm T}$ is observed for all the three cases under consideration. The magnitude of the $v_2$ depends on evolution dynamics of the produced medium, which resulted in an enhancement of $v_2$ in case of hydrodynamic scenario compared to the pure transport scenario. From the obtained results we find that at very low $p_{\rm T}$ ($p_{\rm T}$ $<$ 0.5 GeV/c), $v_{2}$ is indistinguishable in all the three different cases of evolution. This behaviour might indicate that most of the elliptic flow at such low transverse momentum is already built up during the initial scatterings in UrQMD, but by looking at the higher transverse momentum region, it can be seen that most of the elliptic flow is built up during the hydrodynamical evolution. However, the estimated $v_{2}$ does not seem to differentiate between the two EoS employed in the hydro mode over the whole $\rm p_T$ range under investigation. The enhancement in $v_{2}$ in hybrid mode compared to cascade mode can be attributed to the smaller mean free path in the previous case, generating larger pressure gradients due to the assumption of mean-field approximation in the former case. Similar $v_2$ values for two different EoS with and without explicit phase transition is due to the short duration of the hydrodynamic evolution at these low collision energies. 

 When we look at the $v_2$ of identified hadrons, in Figs. \ref{v2pt_cas_id}, \ref{v2pt_ch_id} and \ref{v2pt_hg_id}, at low $ p_{\rm T}$ ($ p_{\rm T}$ $<$ 1 GeV/c), mass ordering is observed in all the three cases and all the four energies, as expected from hydrodynamical calculations \cite{Huovinen:2001cy}.
 In the cascade mode, for all the energies under considerations, protons and pions show mass ordering up to $ p_{\rm T}$ $<$ 1 GeV/c and an inverse mass ordering for $ p_{\rm T}$ $>$ 1 GeV/c. Also in the hybrid mode with Hadron gas equation of state, mass ordering is visible up to $ p_{\rm T}$ $<$ 1 GeV/c and inverse mass ordering for $ p_{\rm T}$ $>$ 1 GeV/c is shown by mesons (pions and kaons). It should be noted that this mass ordering at low $\rm p_T$ and its violation at higher $ p_{\rm T}$ is in agreement with the measurements at RHIC \cite{Esumi:2002vy}. But in contrast, for the Chiral EoS which mimics a locally equilibrated partonic medium, no reverse ordering is not observed at $ p_{\rm T}$ $>$ 1 GeV/c.

 \begin{figure}

  \includegraphics[scale=0.2]{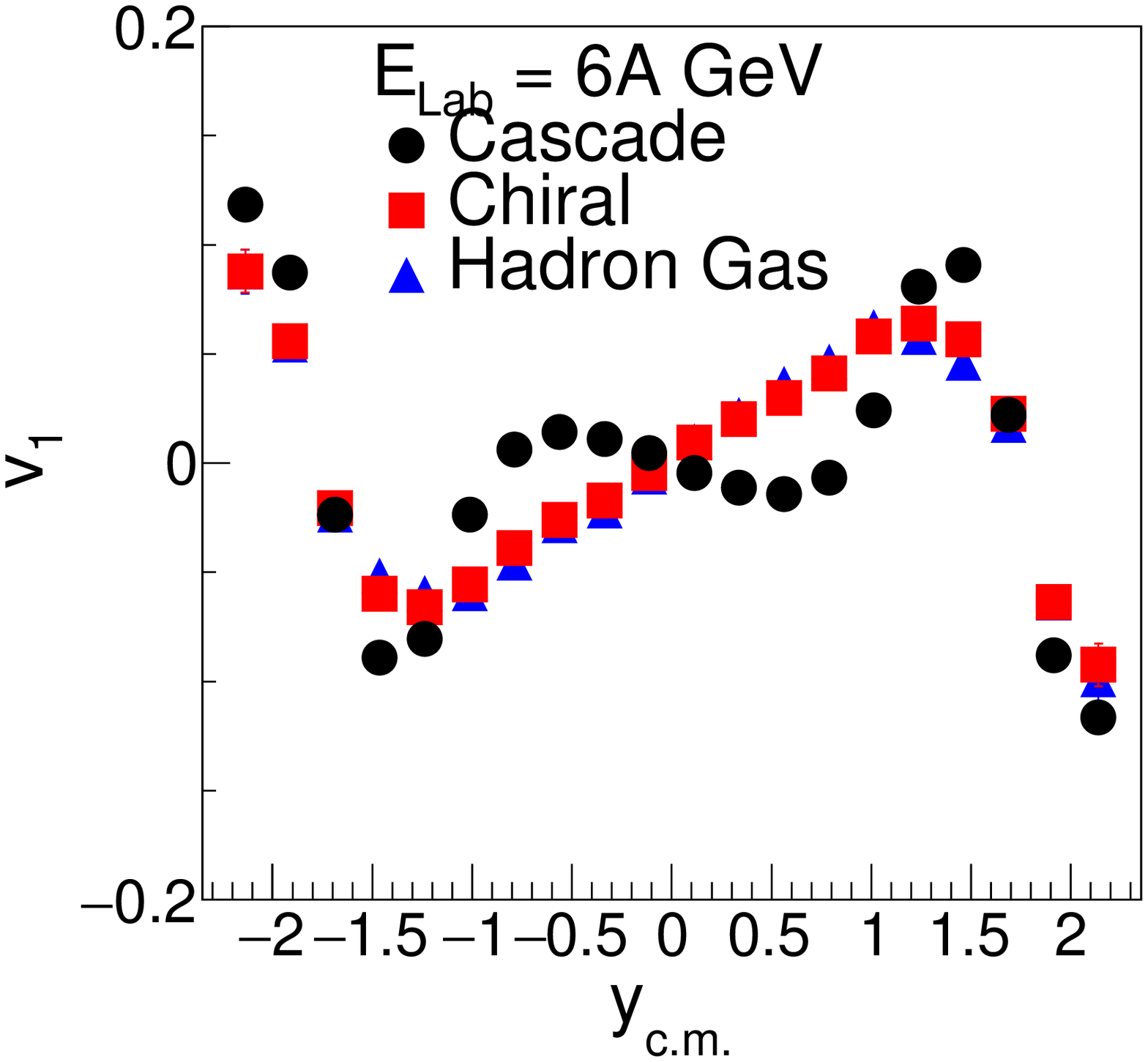}
  \includegraphics[scale=0.2]{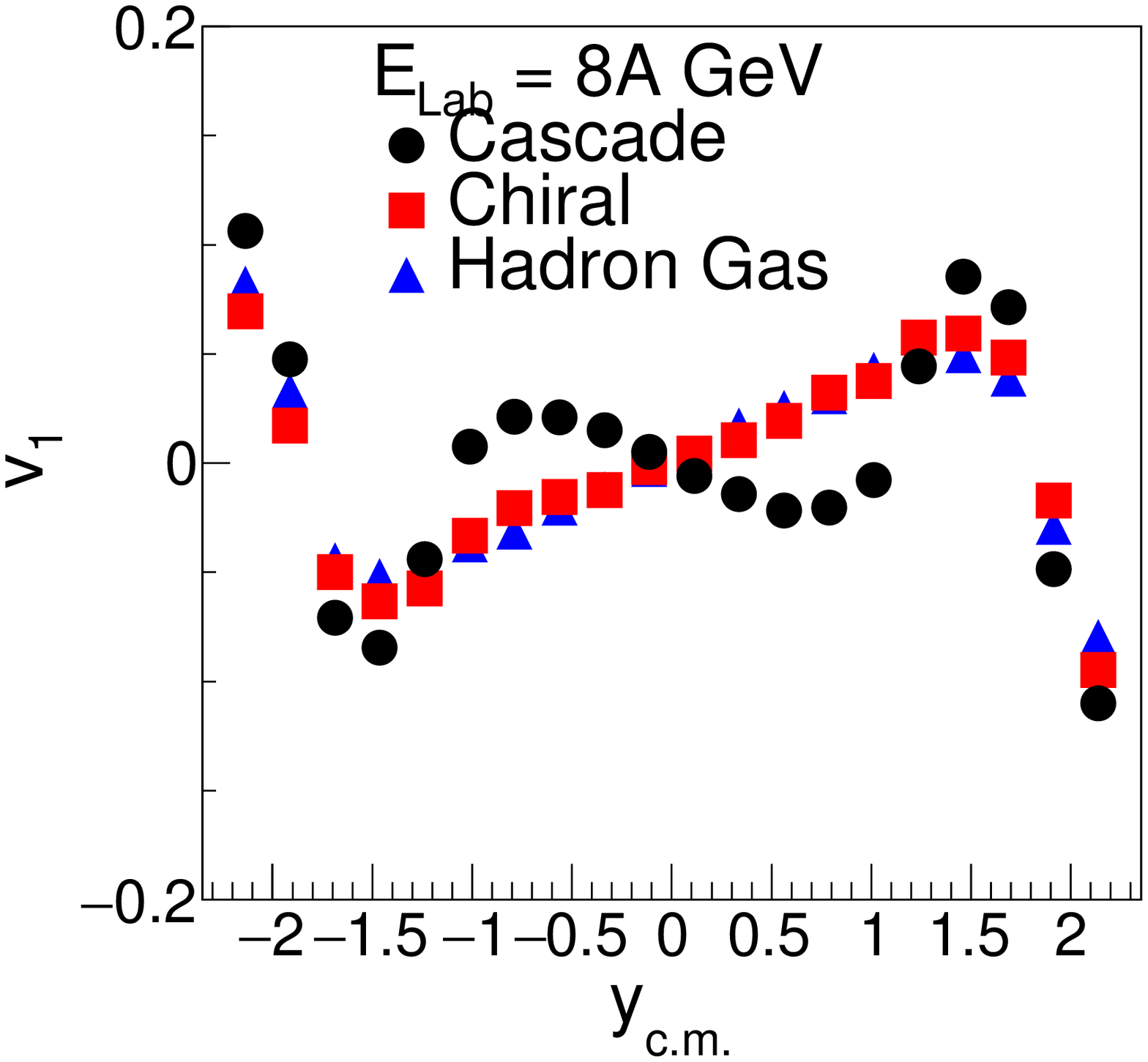}\\
  \includegraphics[scale=0.2]{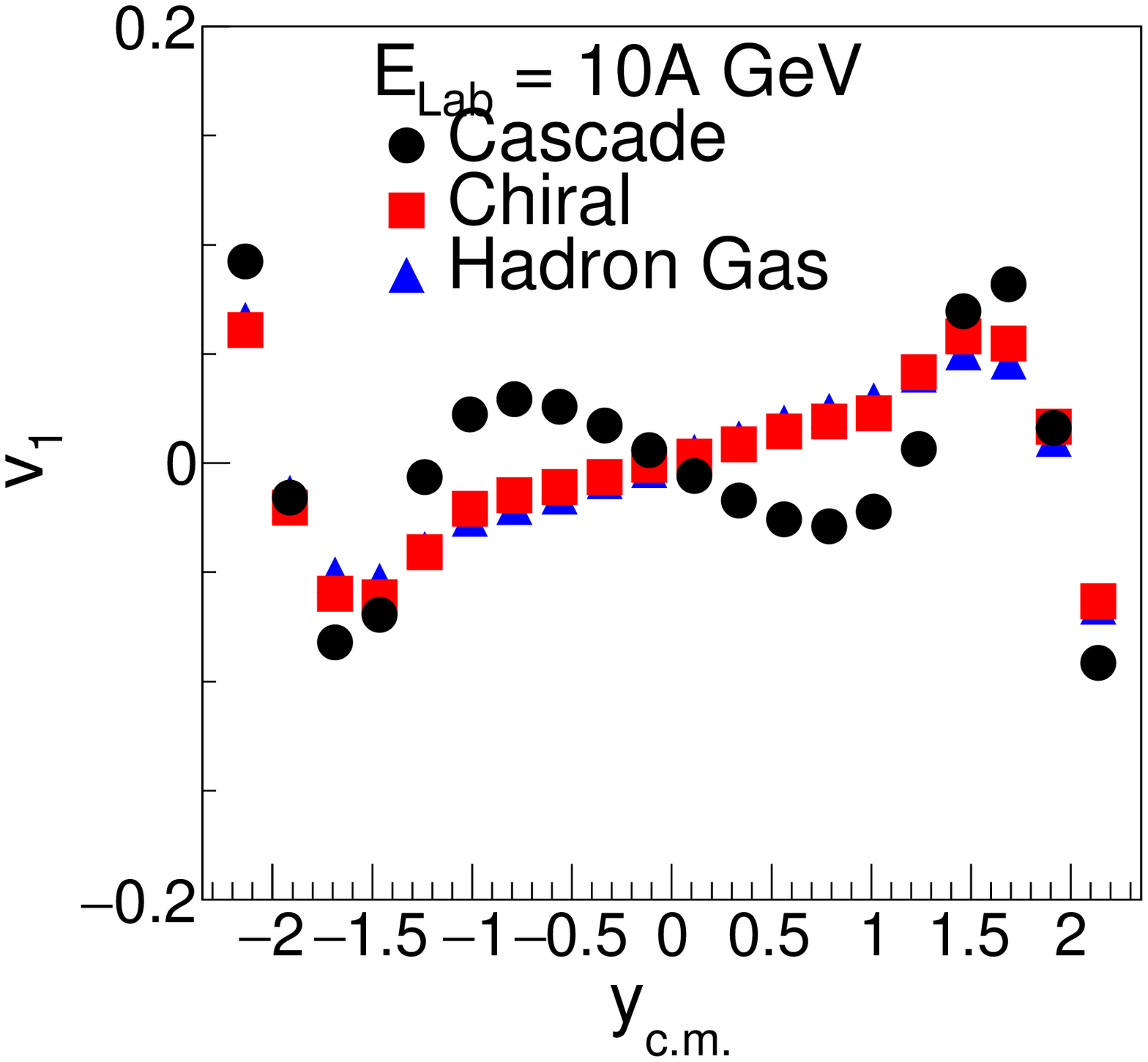}
  \includegraphics[scale=0.2]{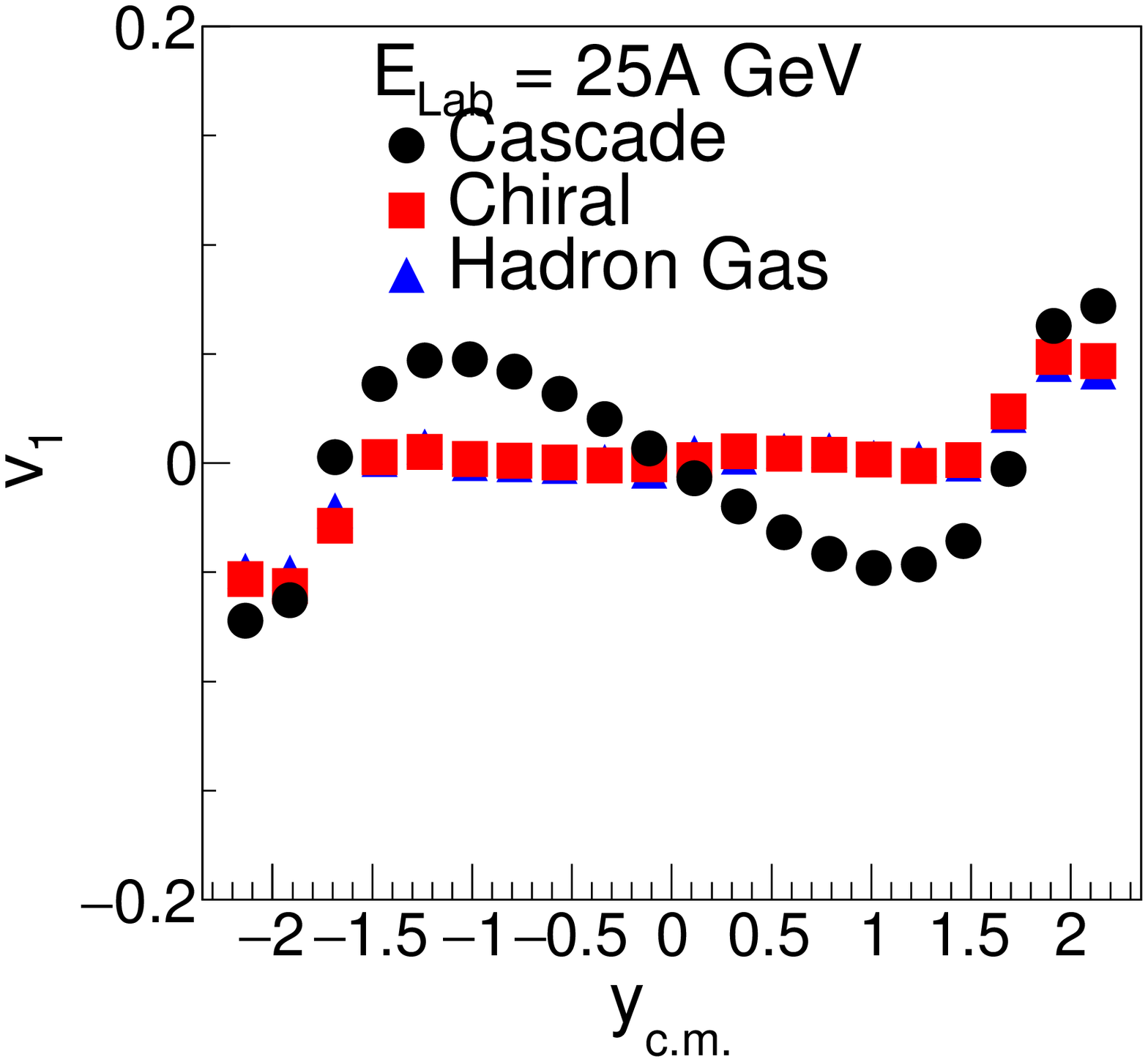}

 \caption{ $v_{1}$ vs $y_{c.m.}$ for charged hadrons using UrQMD for different EoS for 6A, 8A, 10A and 25A GeV }
 \label{v1_rap}
\end{figure}  

 \begin{figure}[t]
 \includegraphics[scale=0.2]{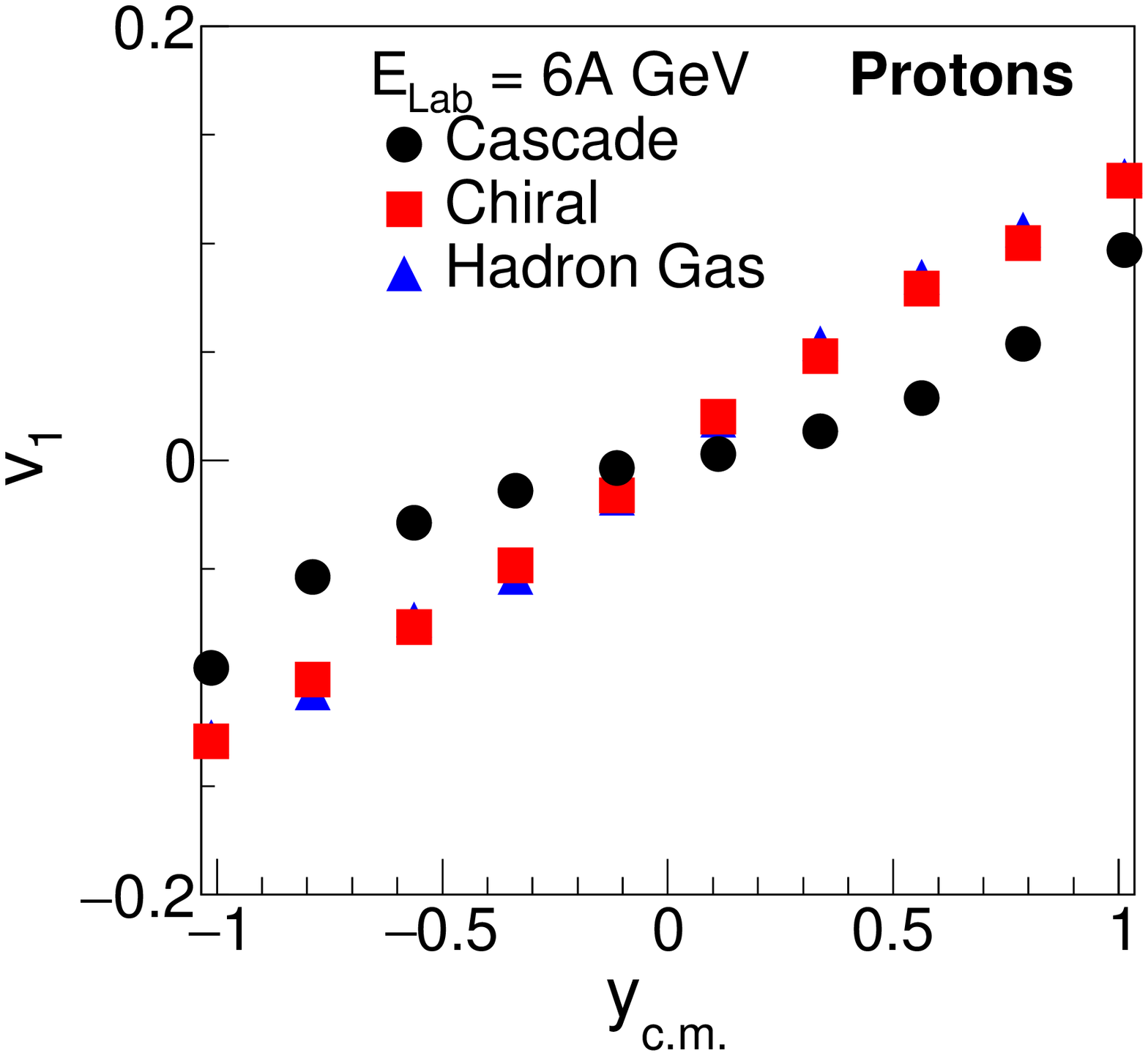}
 \includegraphics[scale=0.2]{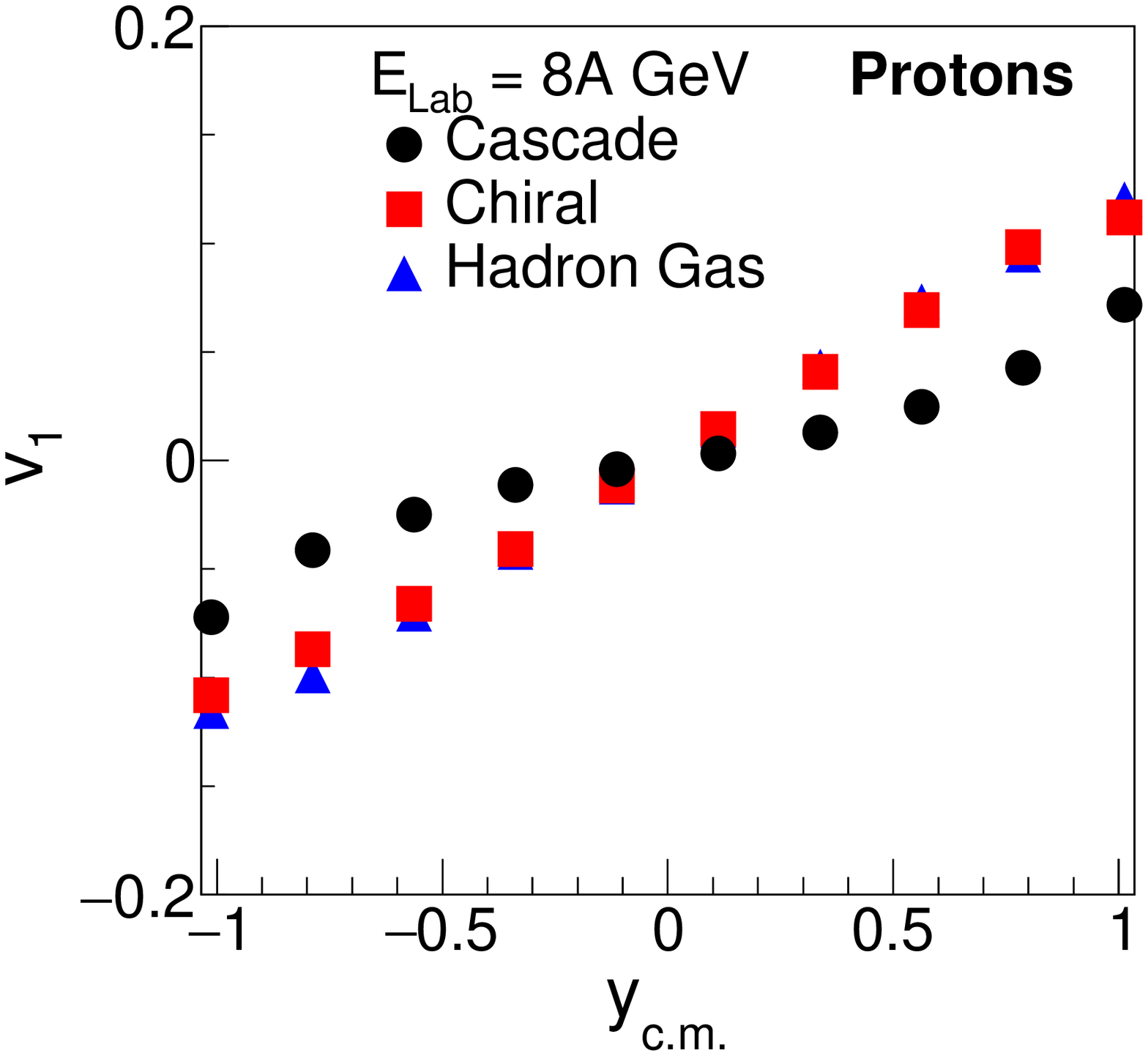}\\
 \includegraphics[scale=0.2]{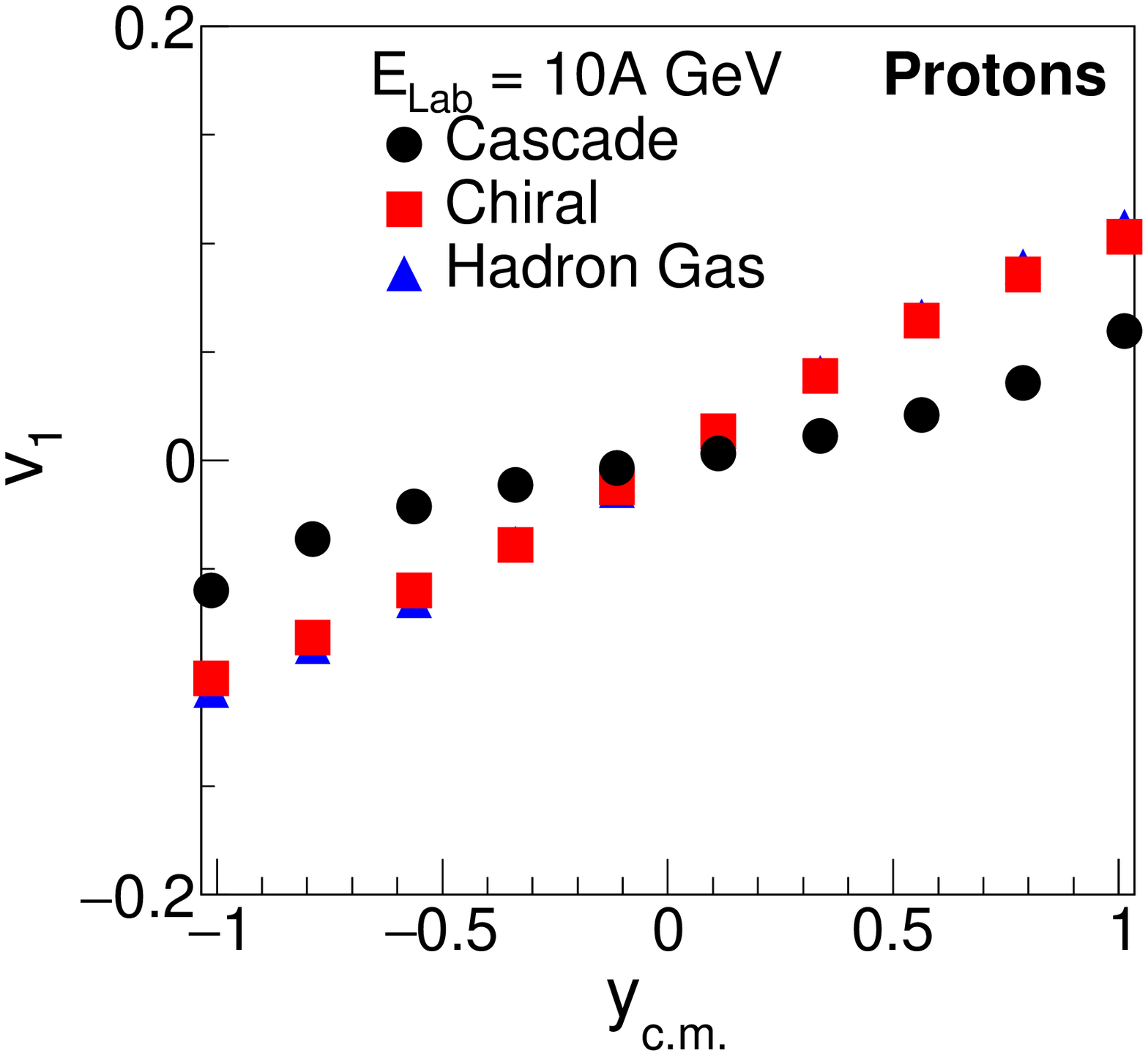}
 \includegraphics[scale=0.2]{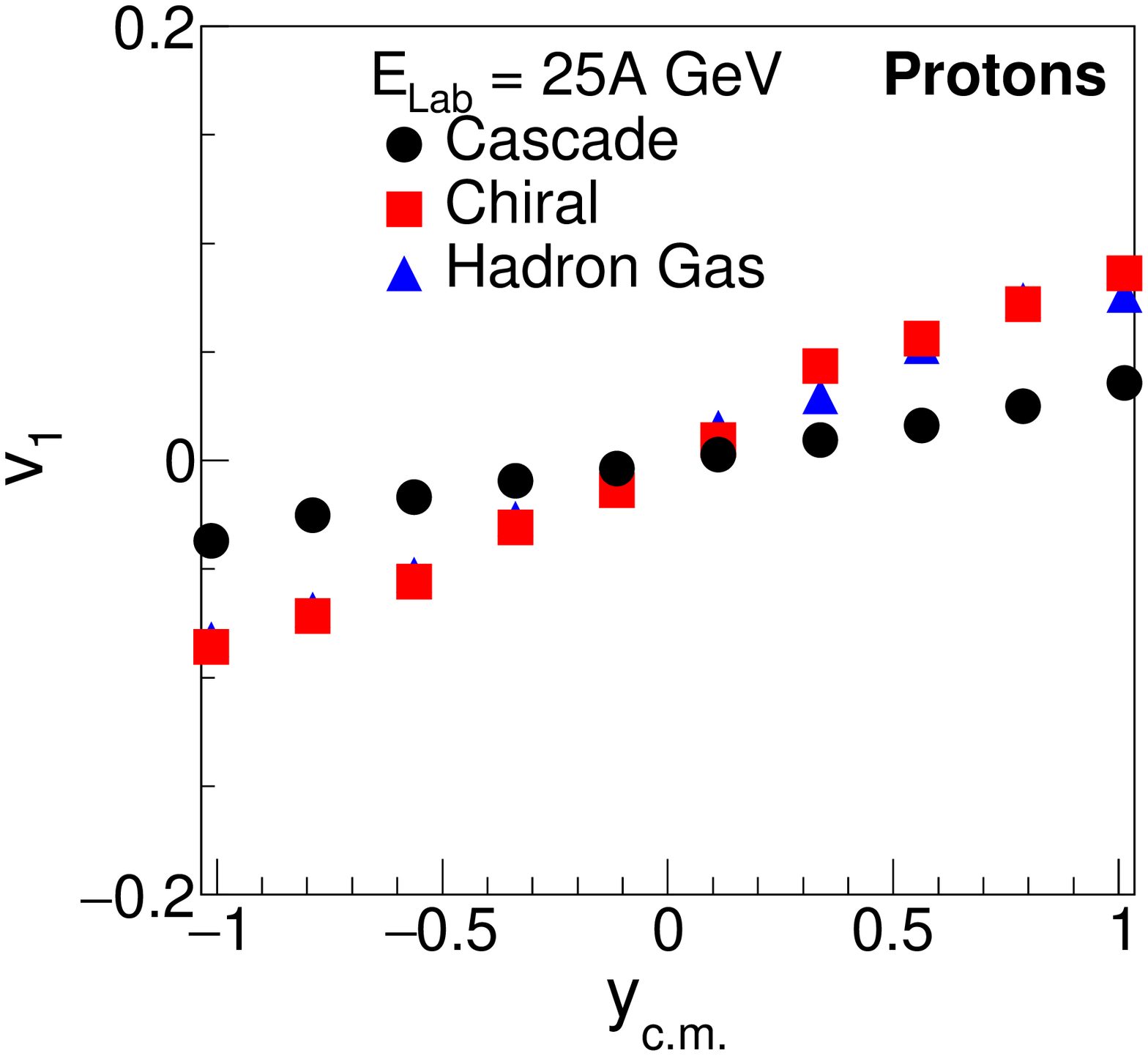}
 \caption{$v_{1}$ vs $y_{c.m.}$ for protons using UrQMD for different EoS for 6A, 8A, 10A and 25A GeV}
\label{fig1_protons}
\end{figure}
\subsection*{B. Constituent quark number scaling}   
     
The observation and mass ordering and its violation for $v_{2}(p_{T})$ of identified hadrons naturally motivates one to investigate the effect of constituent quark number scaling (NCQ) of elliptic flow ($v_{2}$/$n_{q}$) as a function of the scaled transverse momentum ($p_{\rm T}$/$n_{q}$) \cite{Adler:2001nb,Adler:2002ct,Adler:2002pb}.
NCQ scaling is a natural outcome of the hadronization models based on coalescence and recombination of partons~\cite{Abir:2009sh,Fries:2003kq} and indicates that the collectivity developed in the early stage of the collisions is of partonic origin.

 Figs.[\ref{scale_cas}],[\ref{scale_chi}] and [\ref{scale_hg}] show the variation of $v_2/n_q$ with ${p_{\rm T}}/n_q$. From the figures, it is evident that $v_2$ shows reasonably good scaling with ${p_{\rm T}}/n_{q}$ at low $p_{\rm T}$ and the degree of scaling seems to be same for all three scenarios and all energies under consideration. The scaling becomes more prominent when observed in terms of transverse kinetic energy $KE_T (= m_T - m_0)$, a variable that takes care of relativistic effects. 
 This indicates that such scaling behaviour observed in terms of $p_{\rm T}$ (or $KE_{T}$) is insensitive to the onset of partonic collectivity, rather such this is a natural outcome of the mass ordering in a boosted thermal model, an observation in line with previous calculations performed at FAIR SIS-300 energies~\cite{Bhaduri:2010wi}.

 \subsection*{C. Rapidity dependence}

 \begin{figure}[t]
 \includegraphics[scale=0.2]{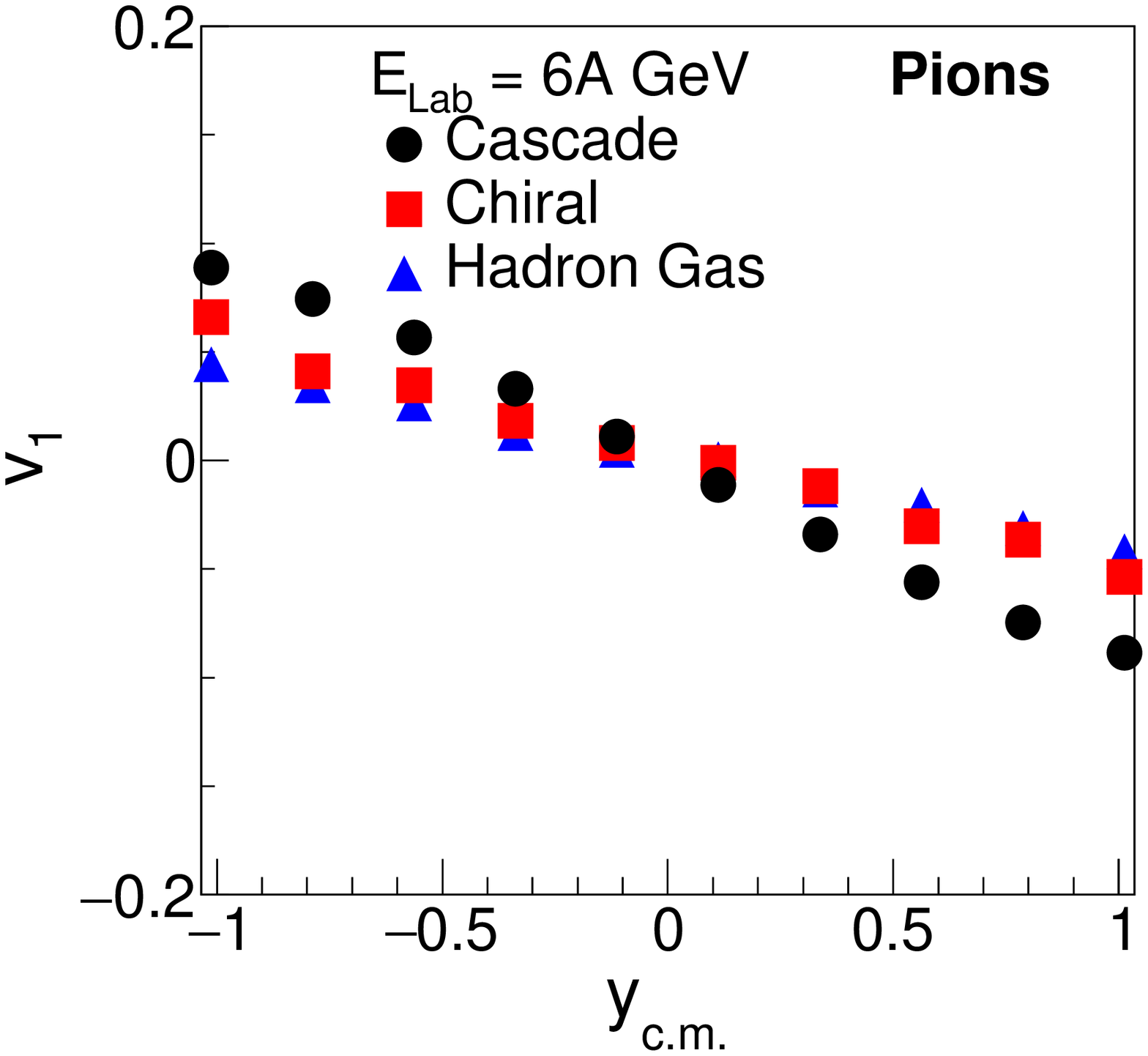}
 \includegraphics[scale=0.2]{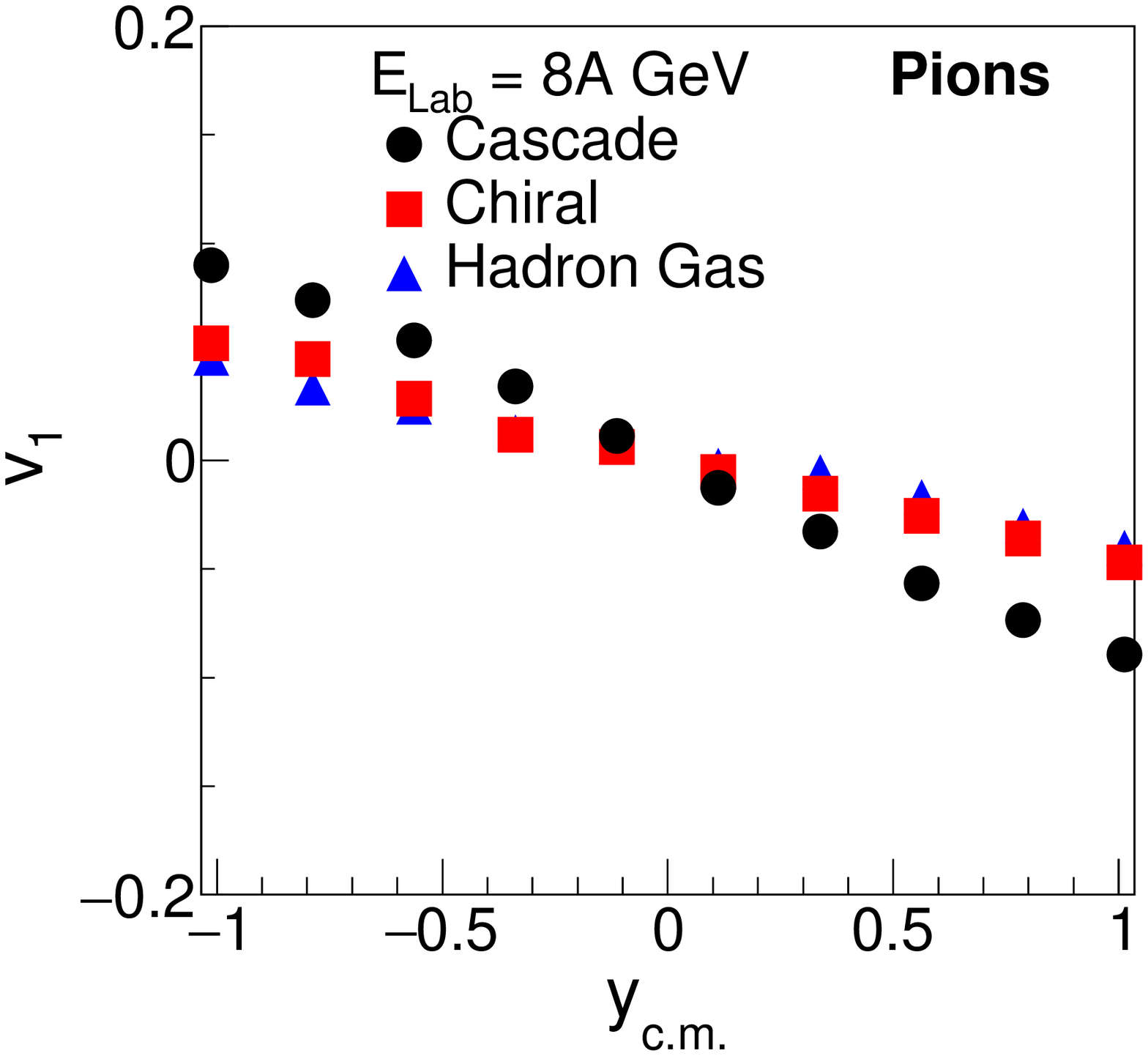}\\
 \includegraphics[scale=0.2]{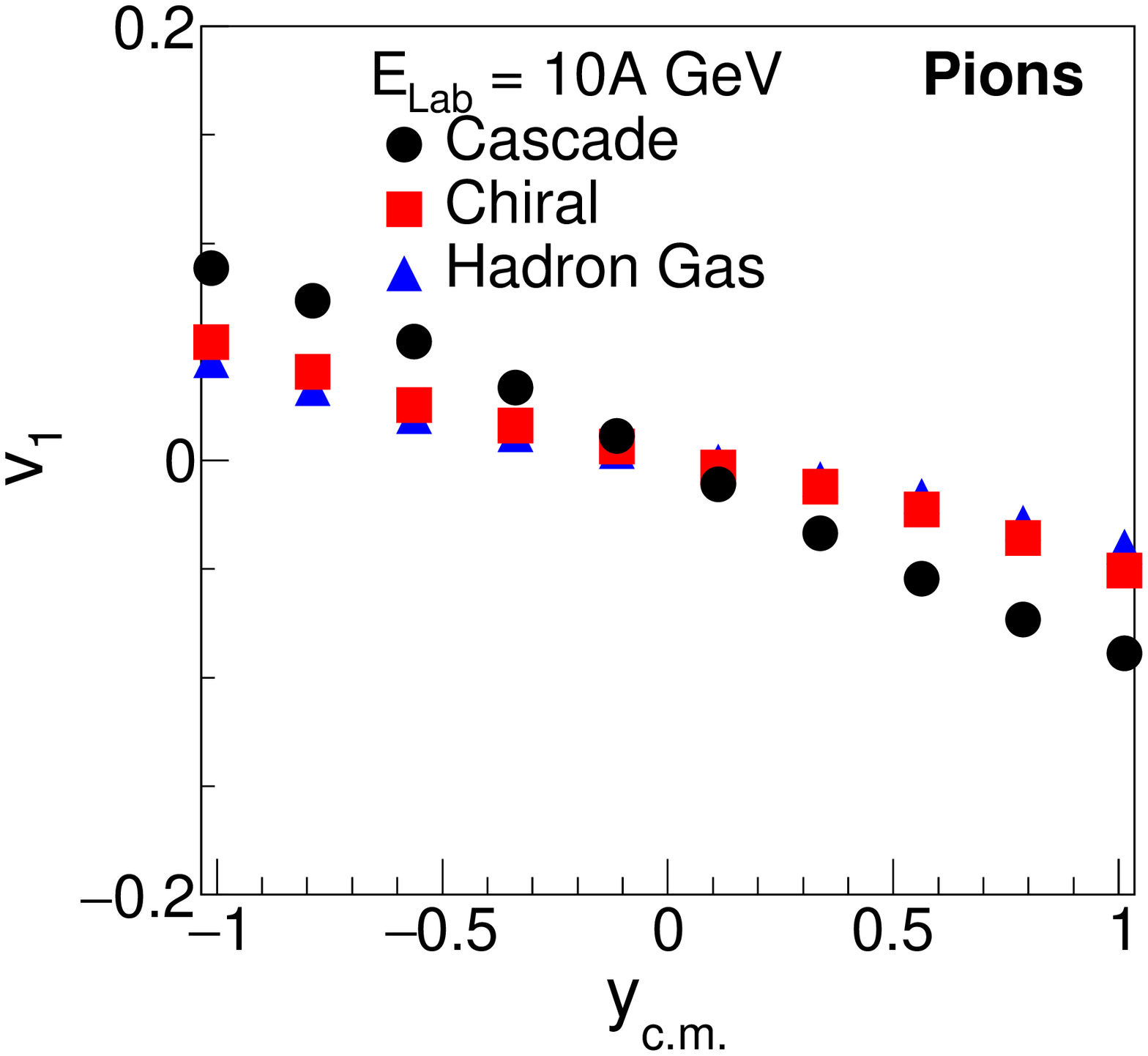}
 \includegraphics[scale=0.2]{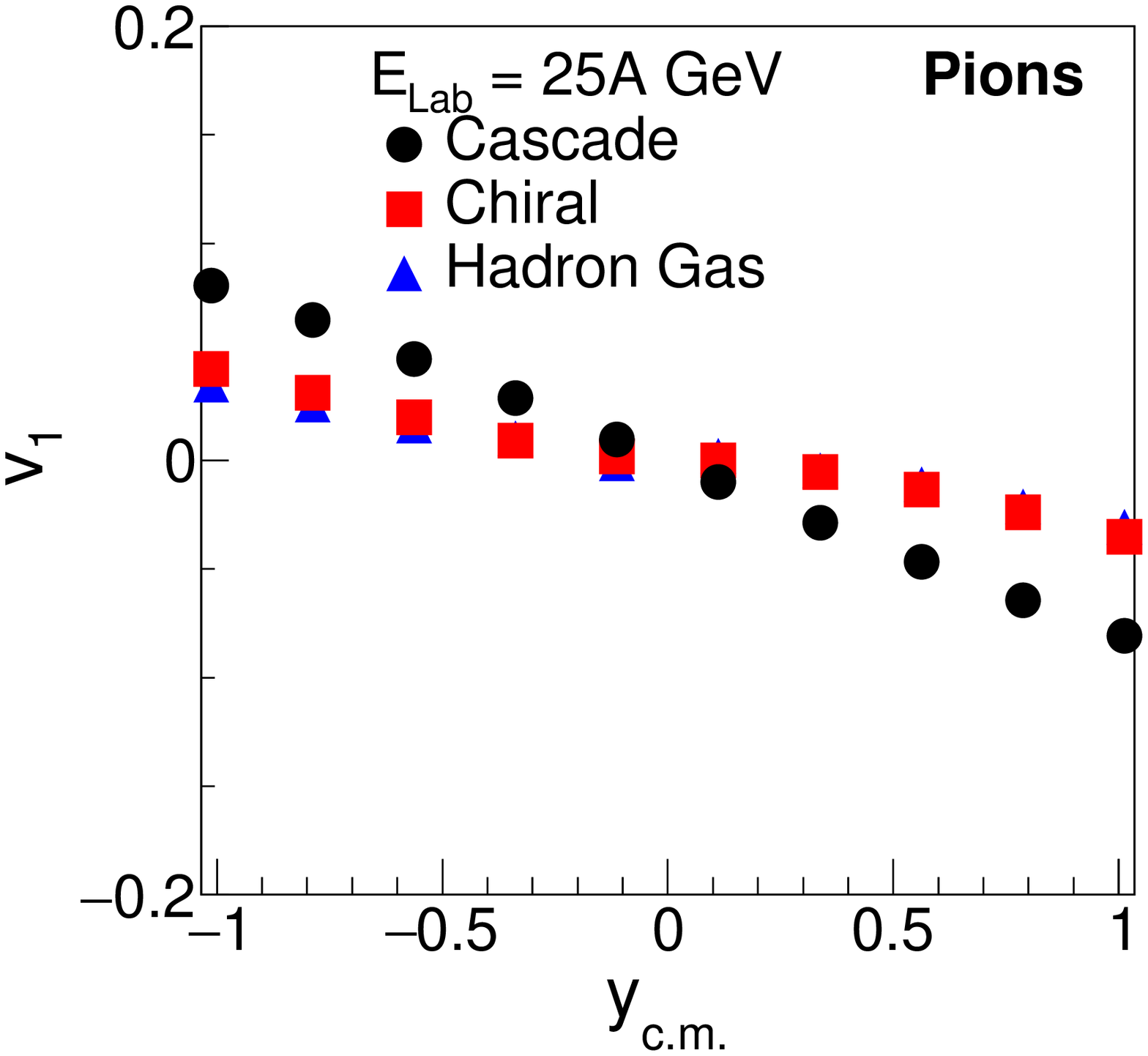}
 \caption{$v_{1}$ vs $y_{c.m.}$ for pions using UrQMD for different EoS for 6A, 8A, 10A and 25A GeV}
\label{fig1_pions}
\end{figure}

 \begin{figure}[t]
 \includegraphics[scale=0.2]{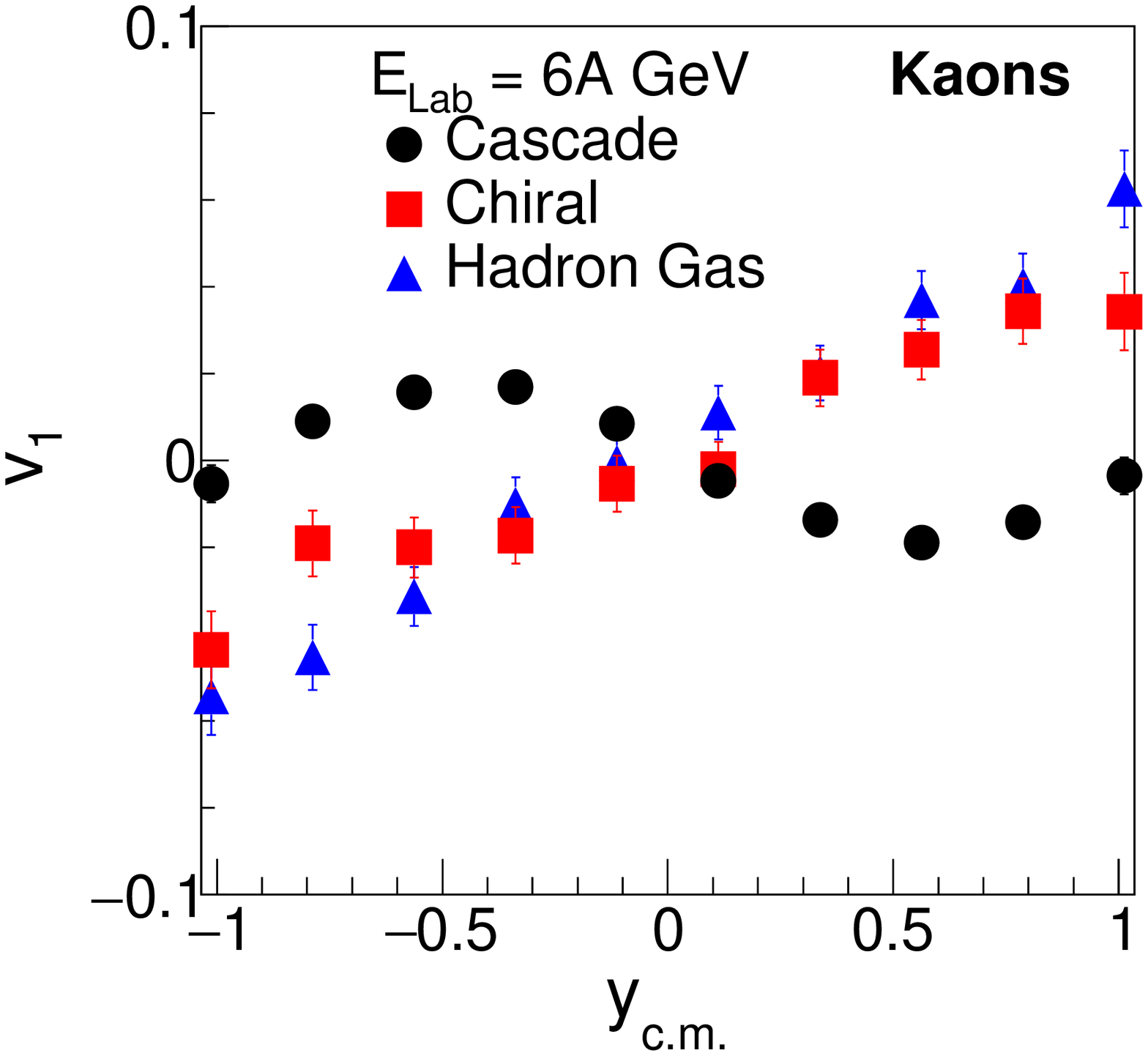}
 \includegraphics[scale=0.2]{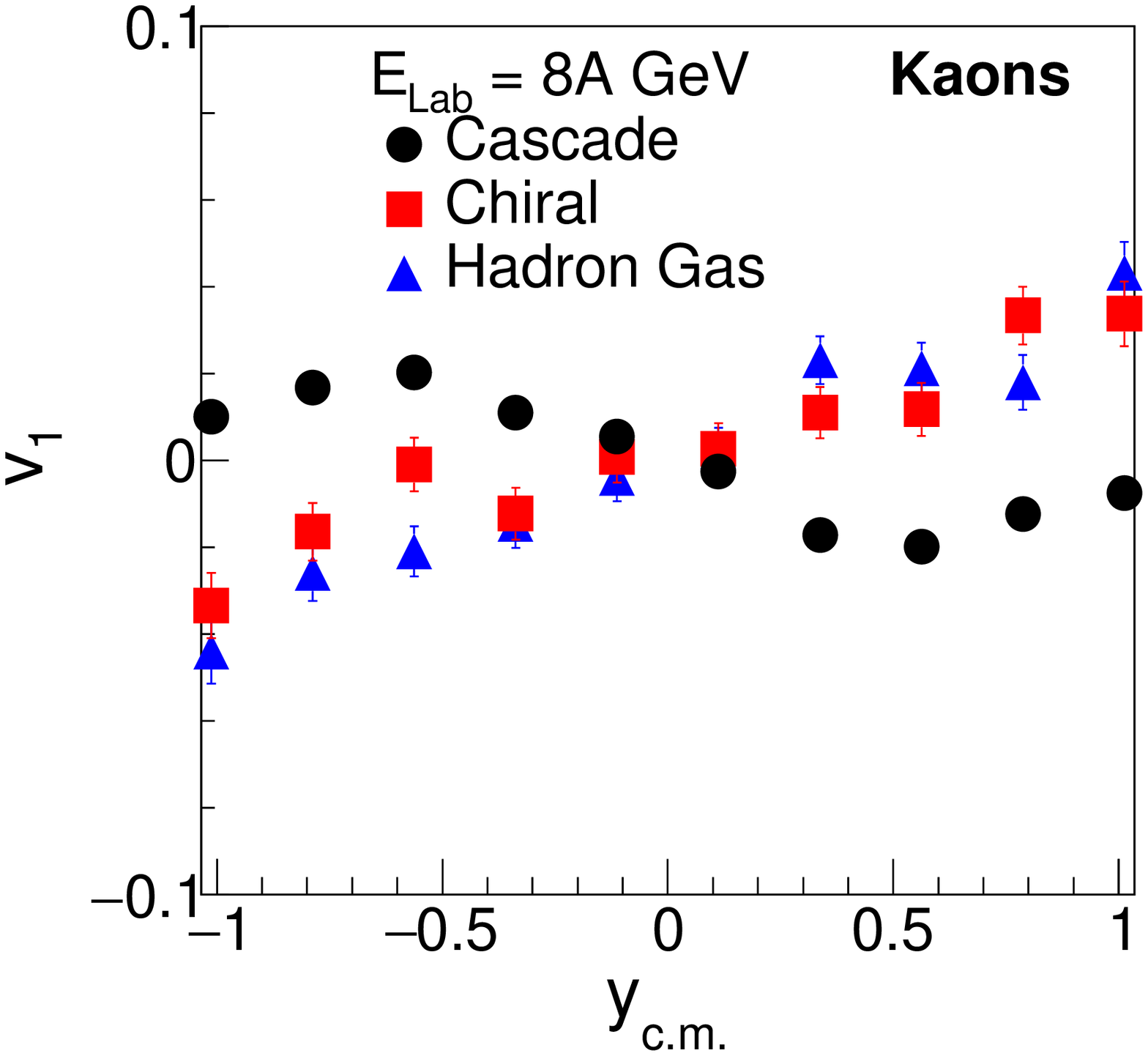}\\
 \includegraphics[scale=0.2]{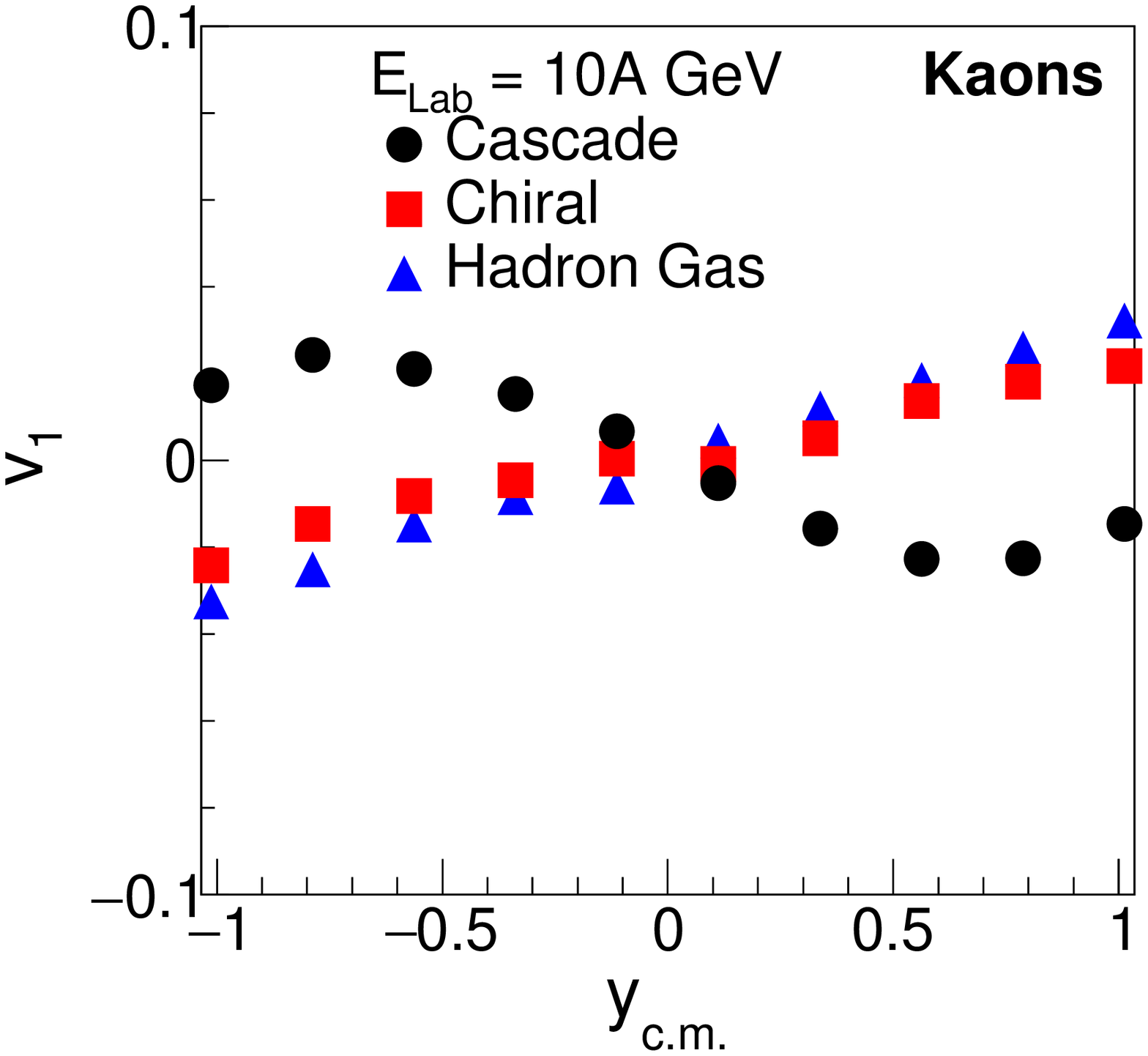}
 \includegraphics[scale=0.2]{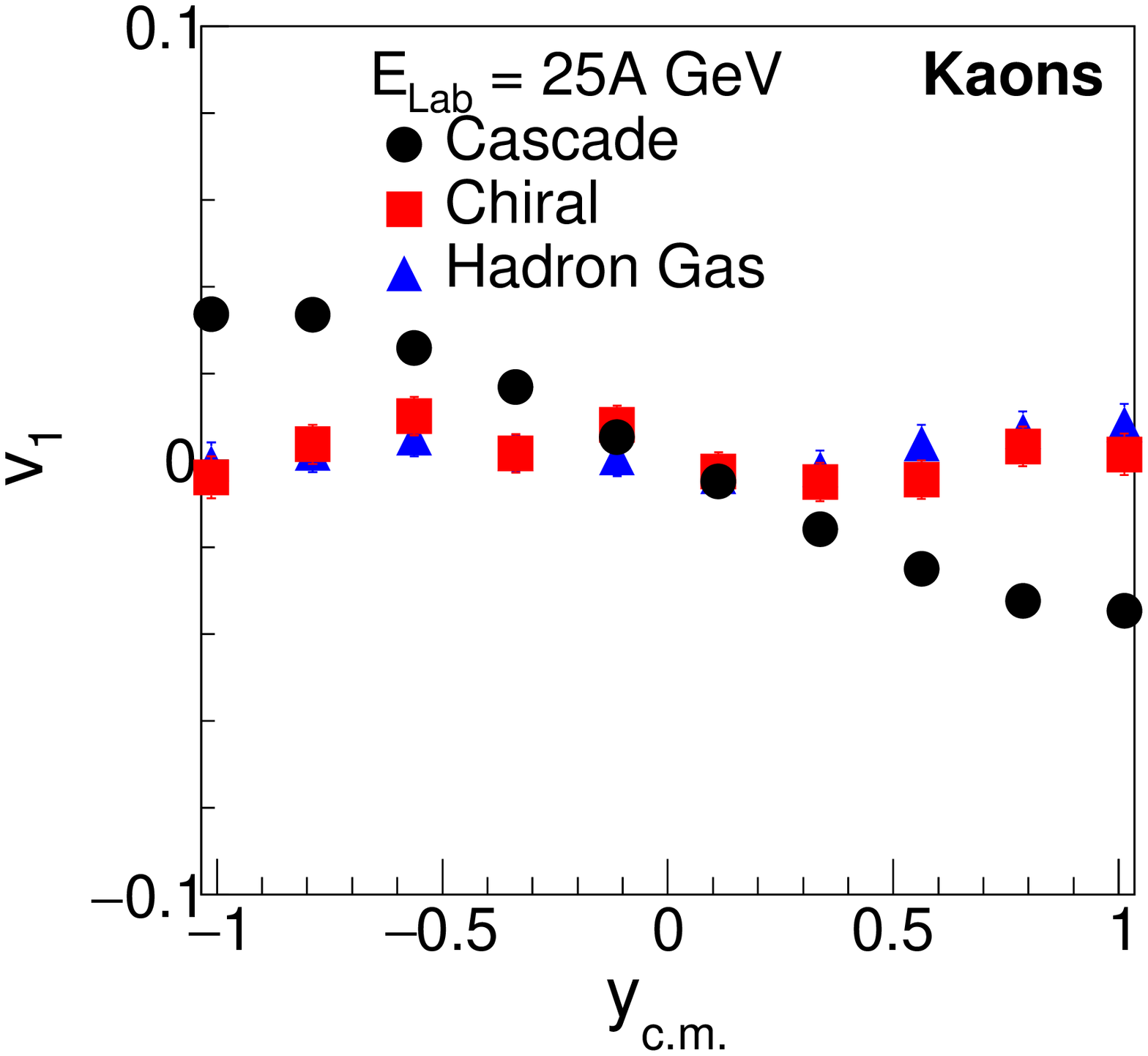}
 \caption{$v_{1}$ vs $y_{c.m.}$ for kaons using UrQMD for different EoS for 6A, 8A, 10A and 25A GeV}
\label{fig1_kaons}
\end{figure}

 \begin{figure}[t]
 \includegraphics[scale=0.2]{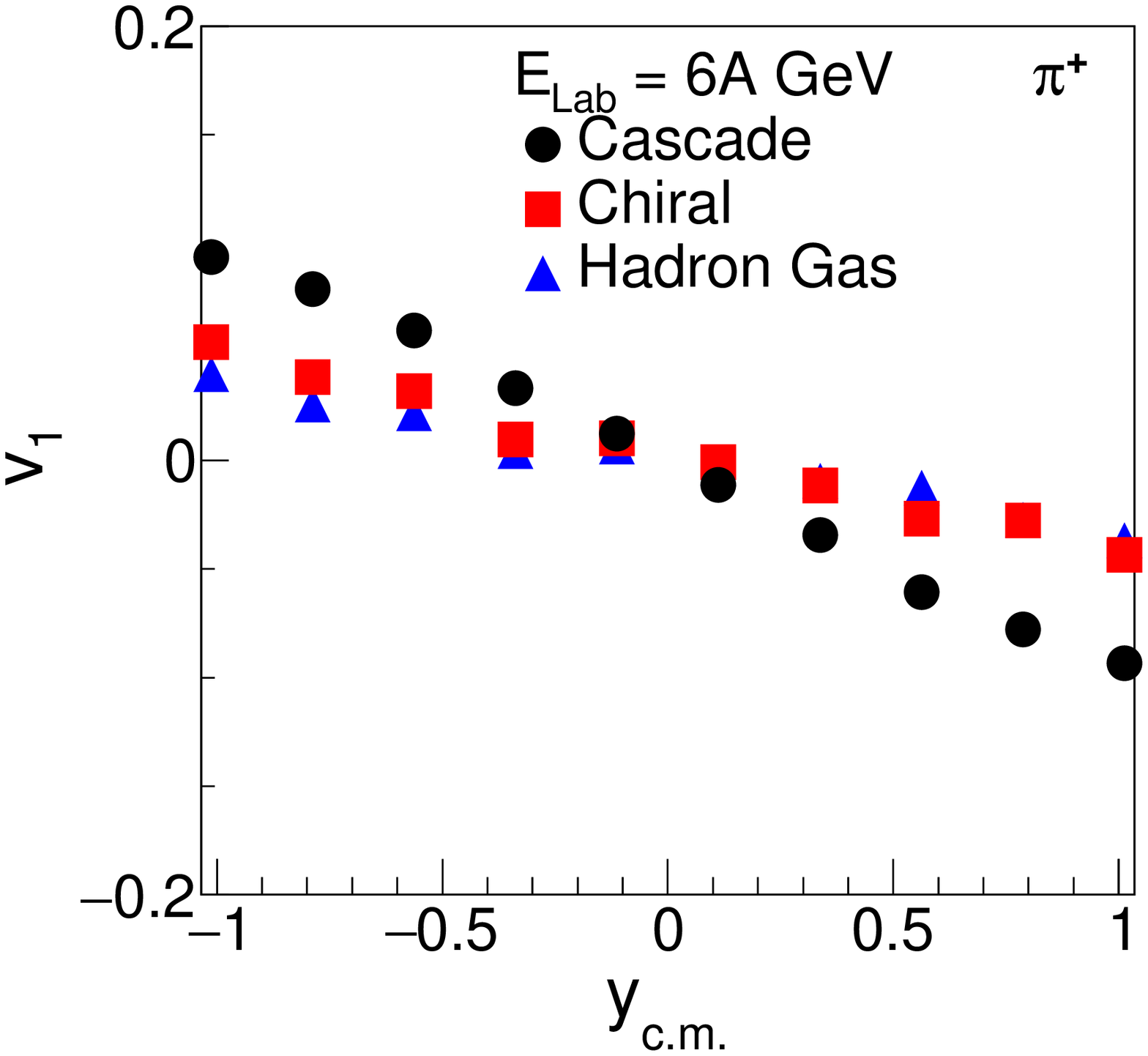}
 \includegraphics[scale=0.2]{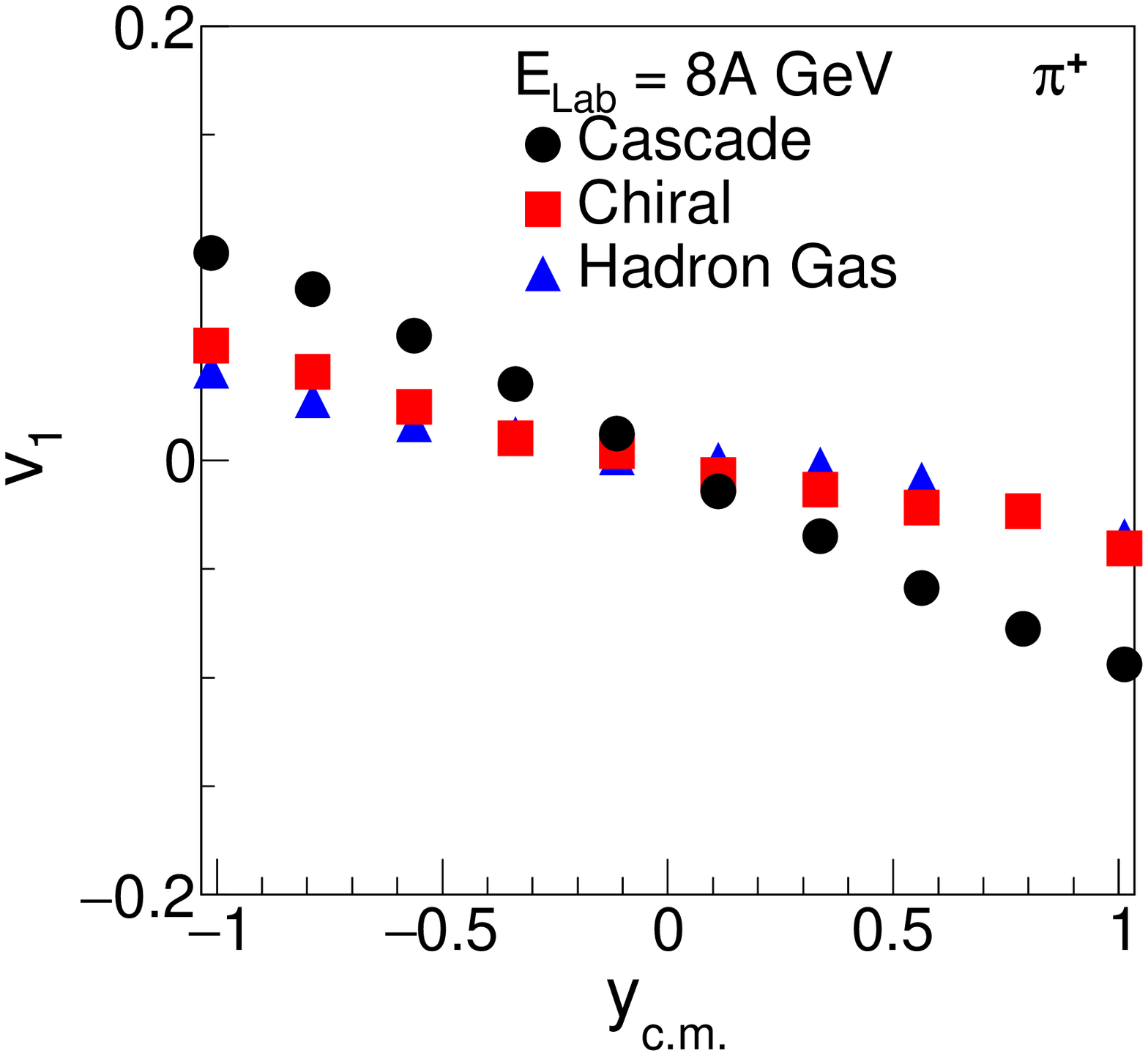}\\
 \includegraphics[scale=0.2]{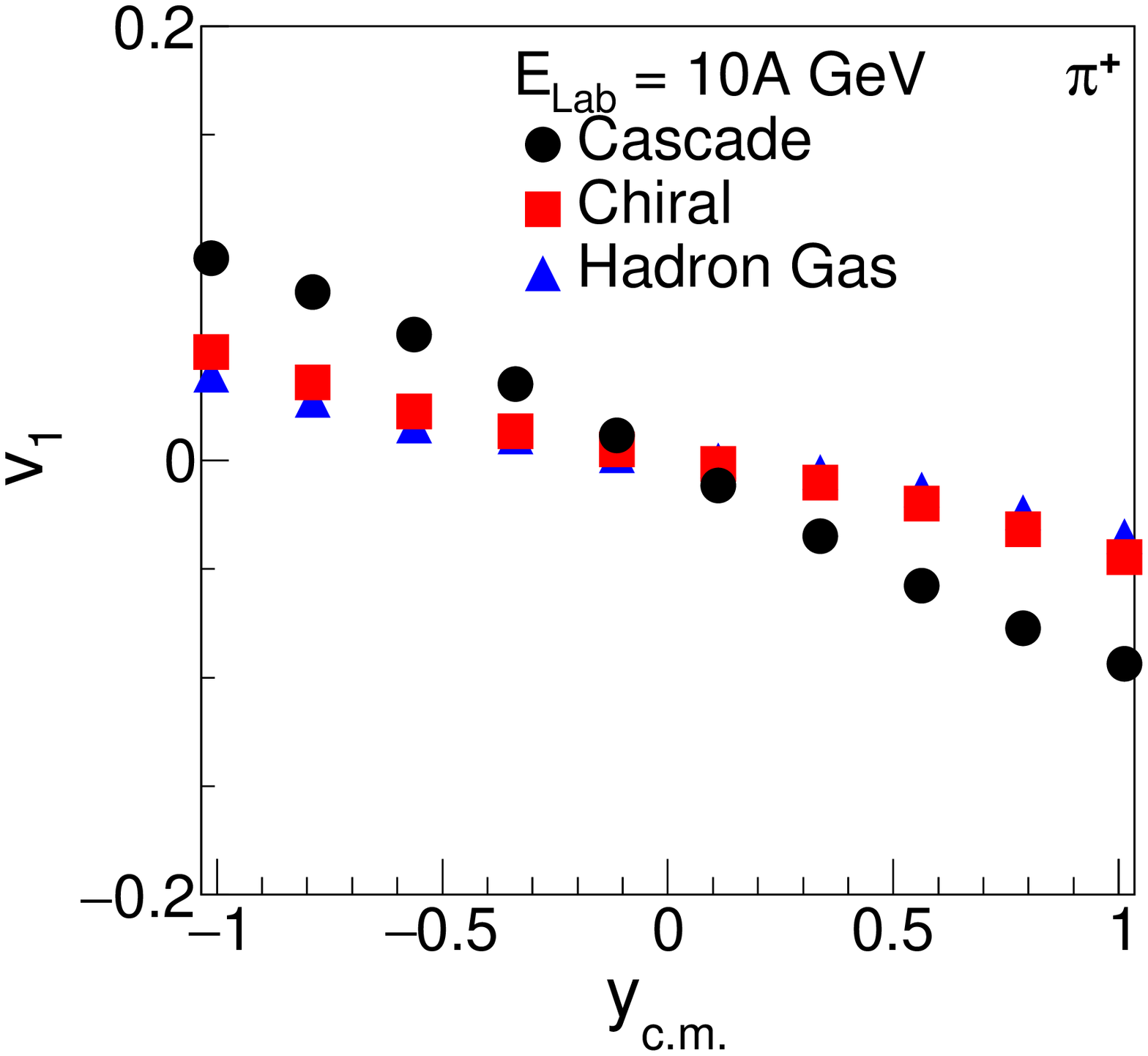}
 \includegraphics[scale=0.2]{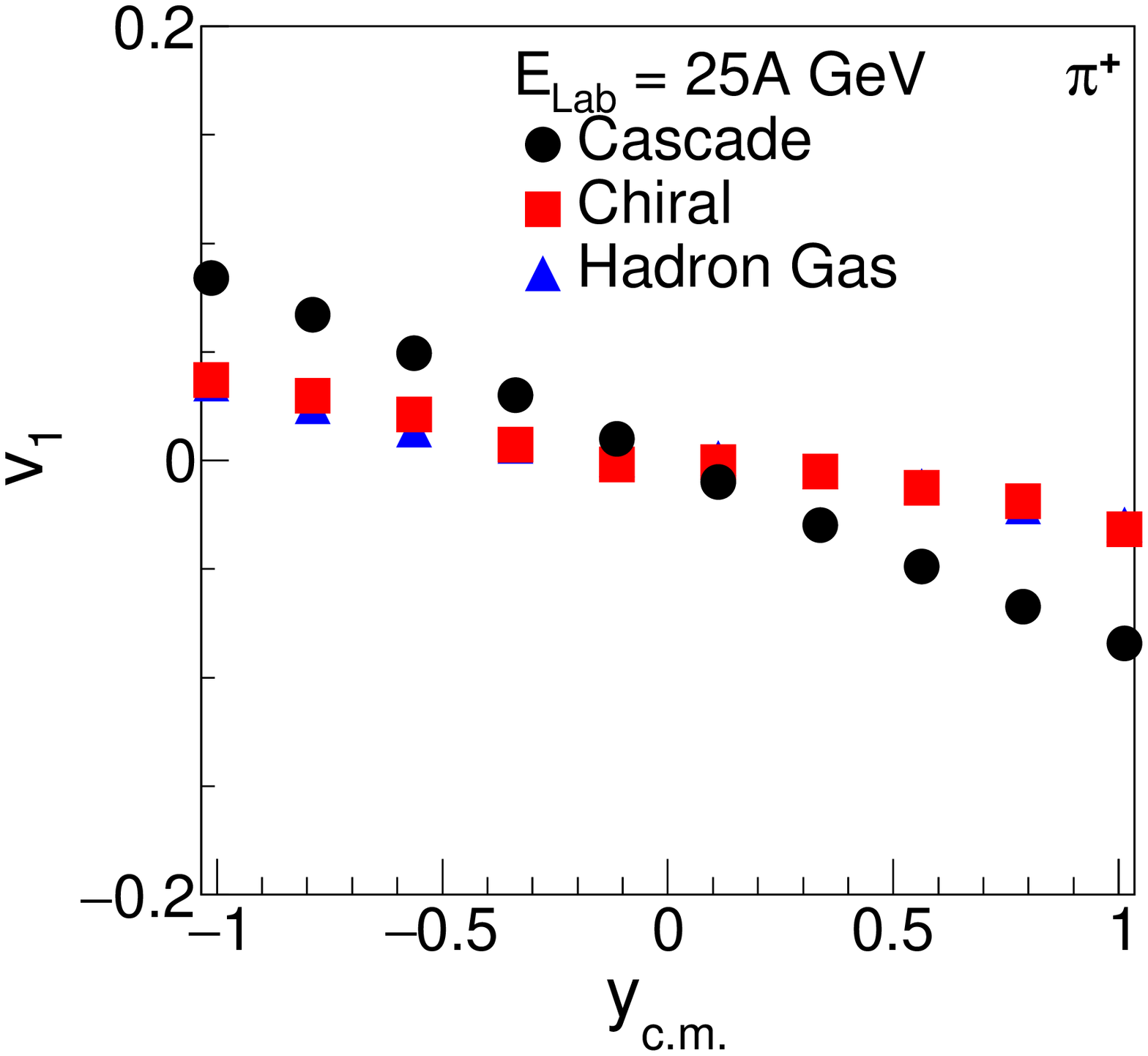}
\caption{Rapidity dependence of directed flow ($v_{1}(y_{c.m.})$) of positive pions ($\pi^{+}$) in mid-central Au+Au collisions at bombarding energies $E_b= 6A, 8A, 10A and 25A$ GeV}

\label{fig1_pip}
\end{figure}

 \begin{figure}[t]
 \includegraphics[scale=0.2]{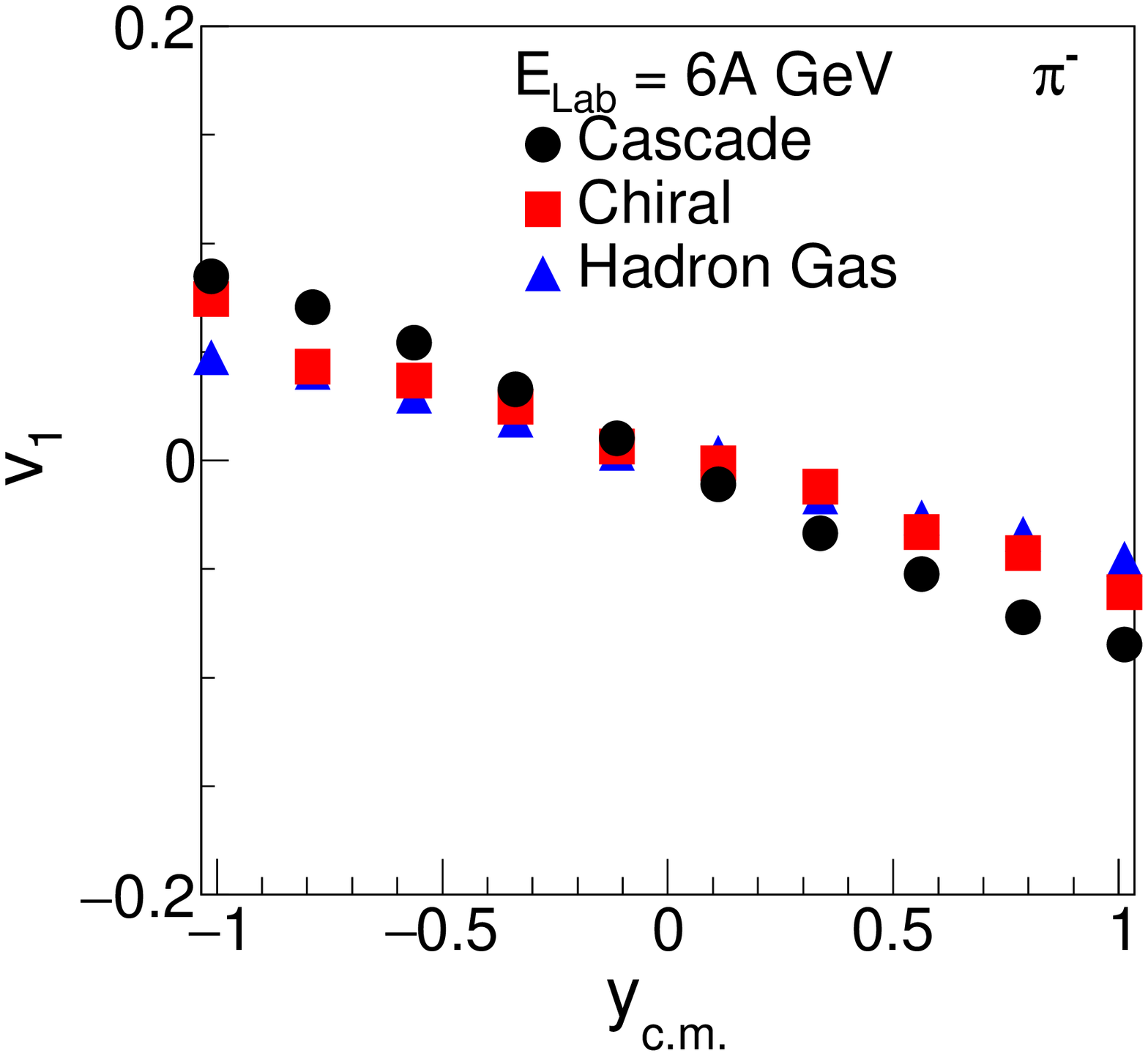}
 \includegraphics[scale=0.2]{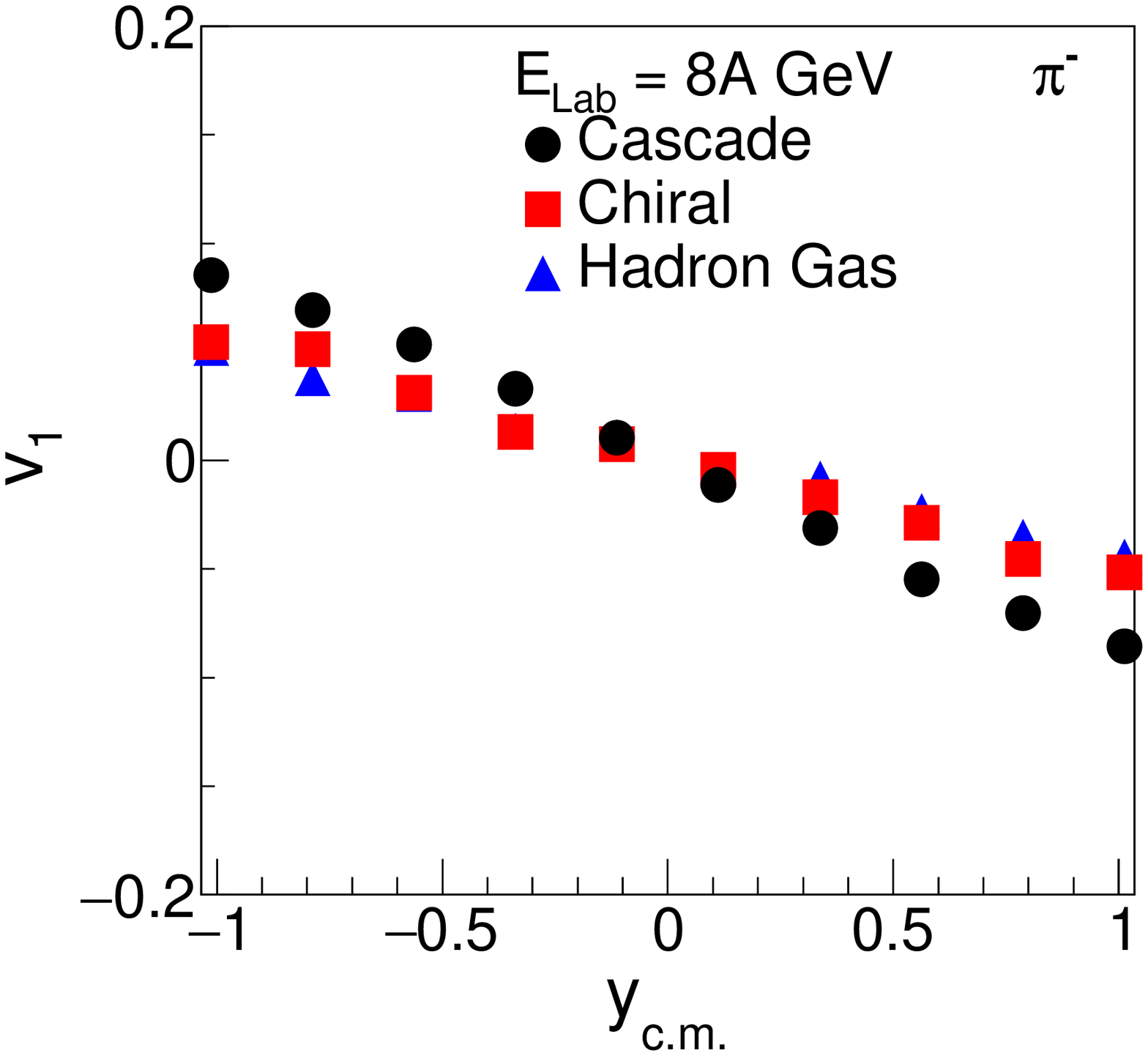}\\
 \includegraphics[scale=0.2]{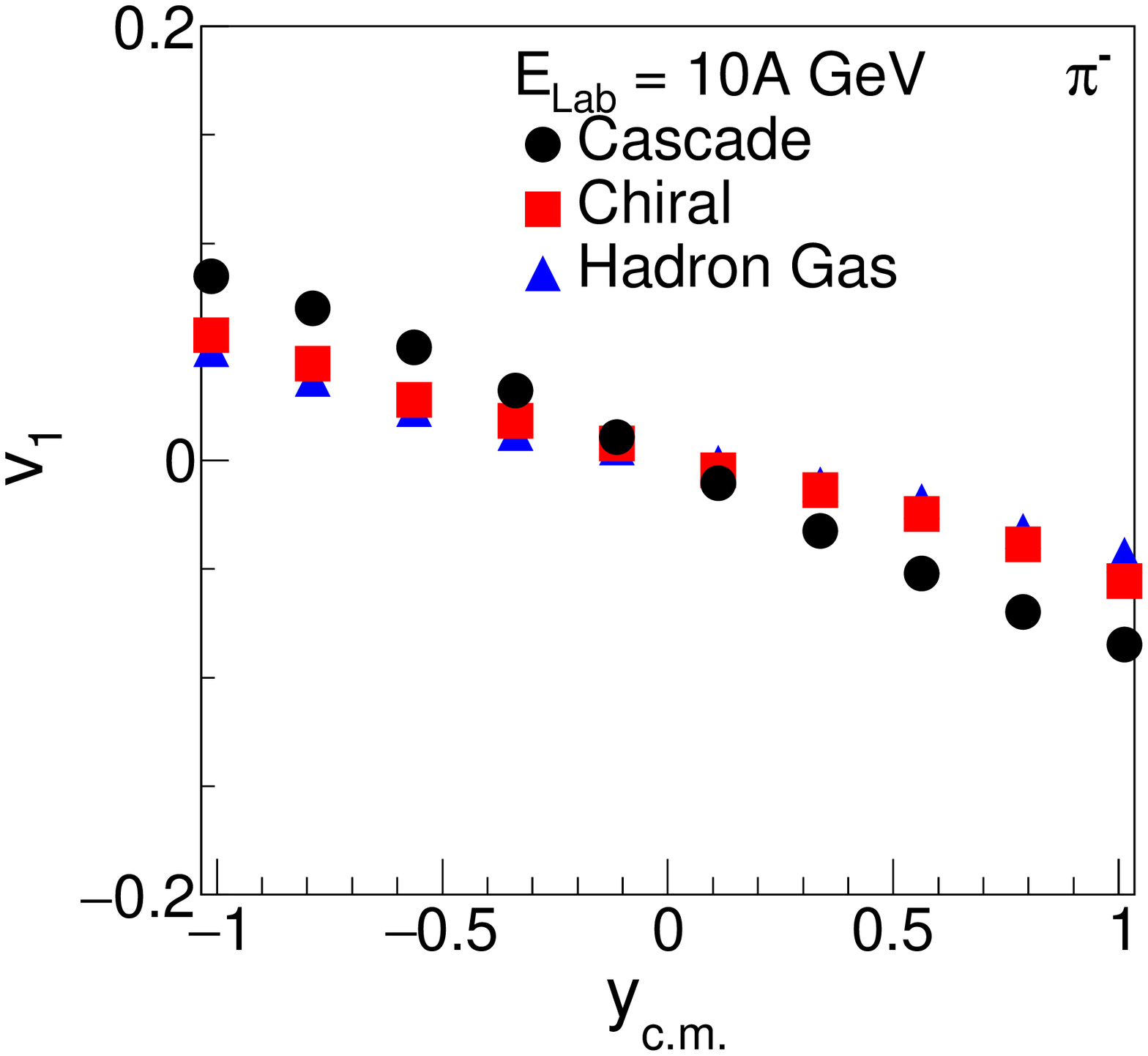}
 \includegraphics[scale=0.2]{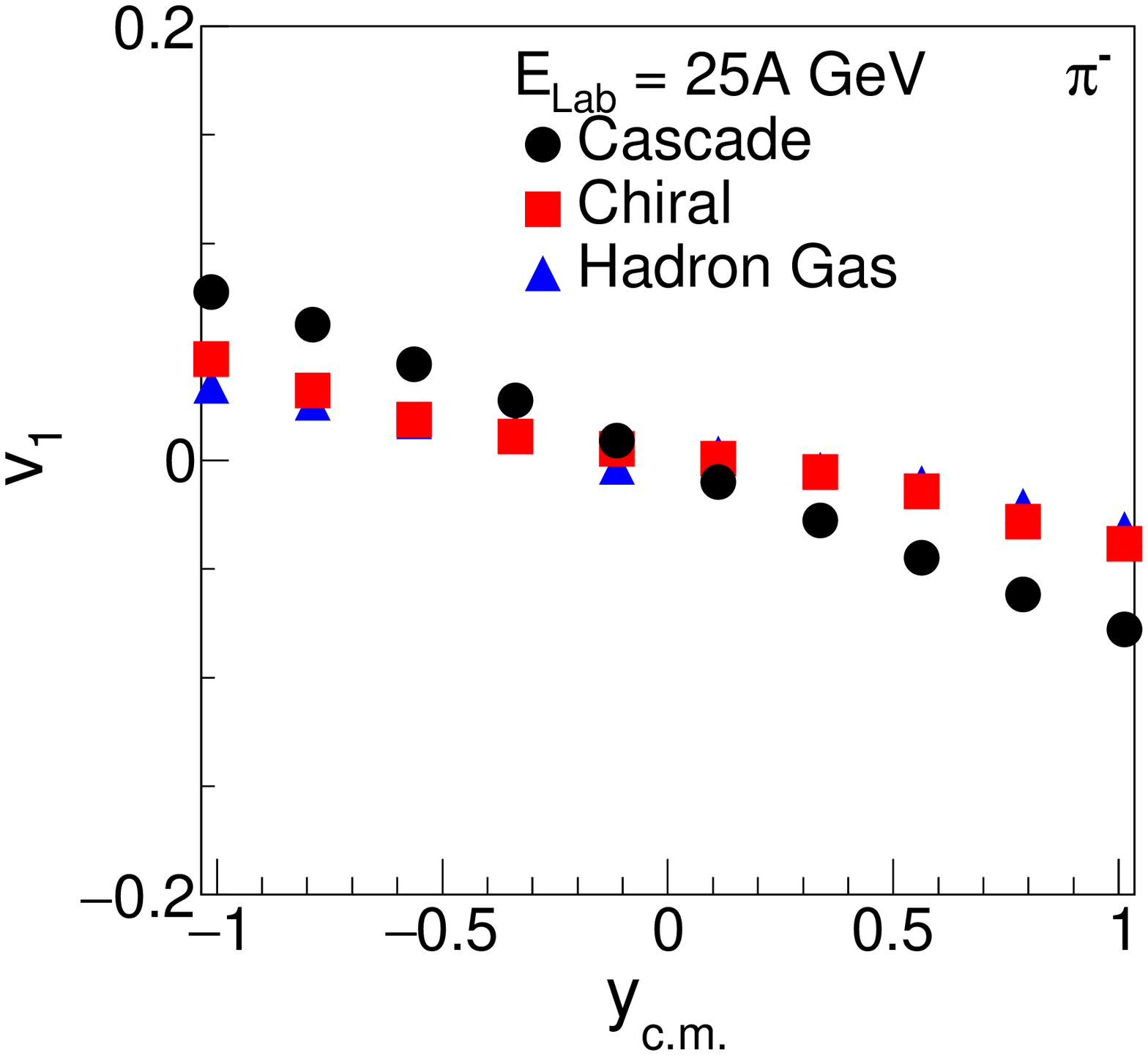}
 \caption{Rapidity dependence of directed flow ($v_{1}(y_{c.m.})$) of negative pions ($\pi^{-}$) in mid-central Au+Au collisions at bombarding energies $E_b= 6A, 8A, 10A and 25A$ GeV}
\label{fig1_pim}
\end{figure}

\begin{figure}[t]
 \includegraphics[scale=0.2]{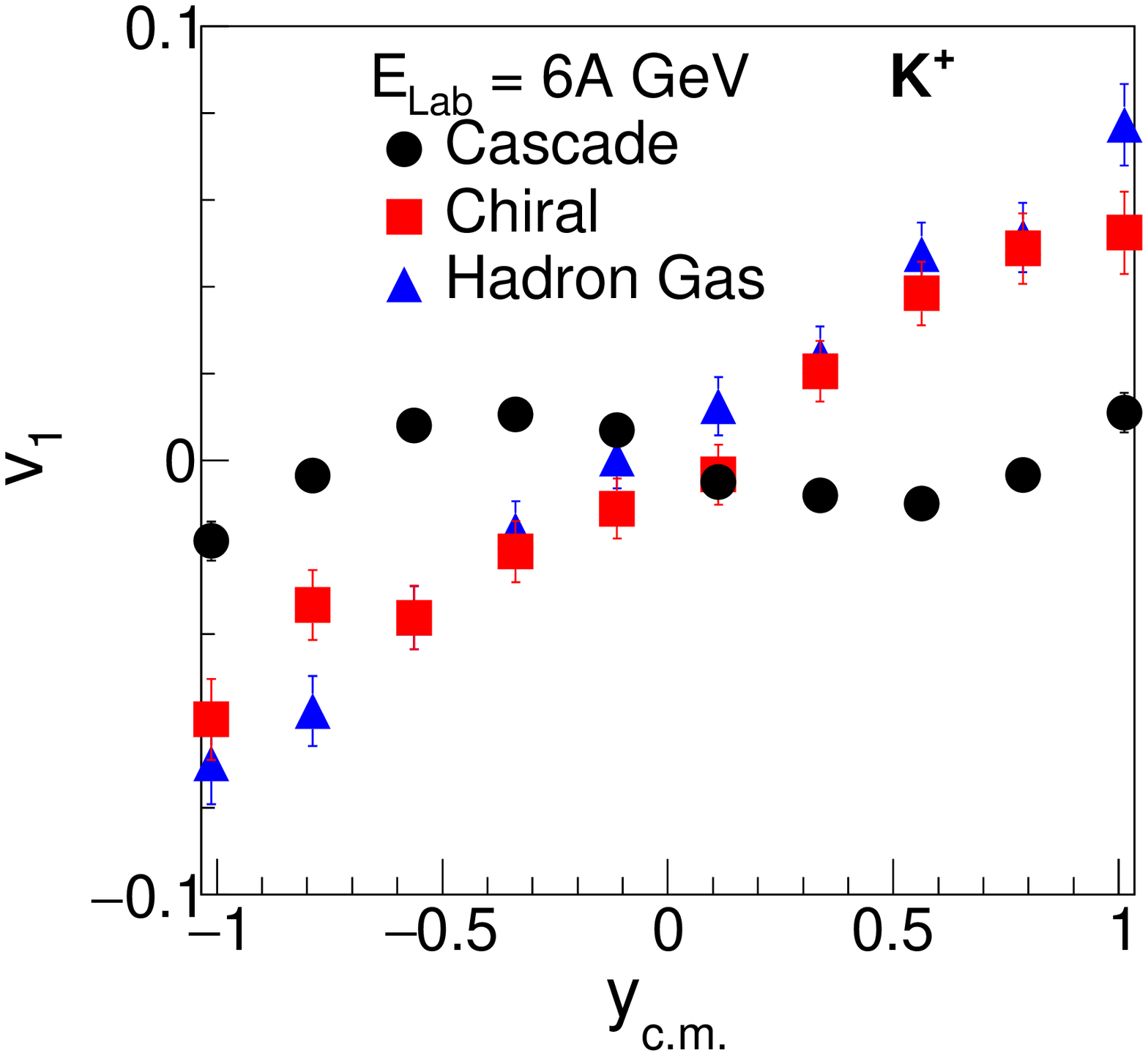}
 \includegraphics[scale=0.2]{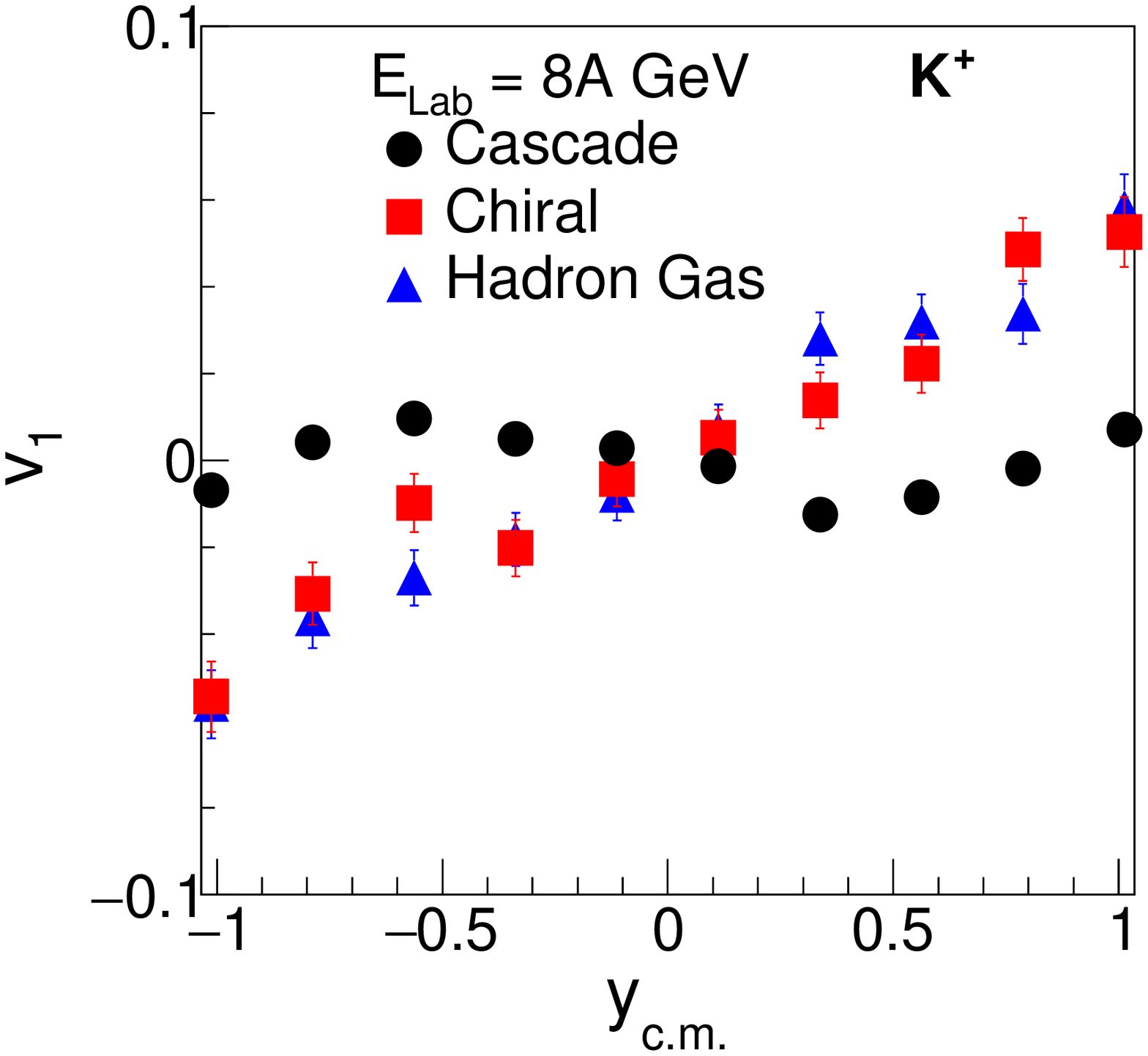}\\
 \includegraphics[scale=0.2]{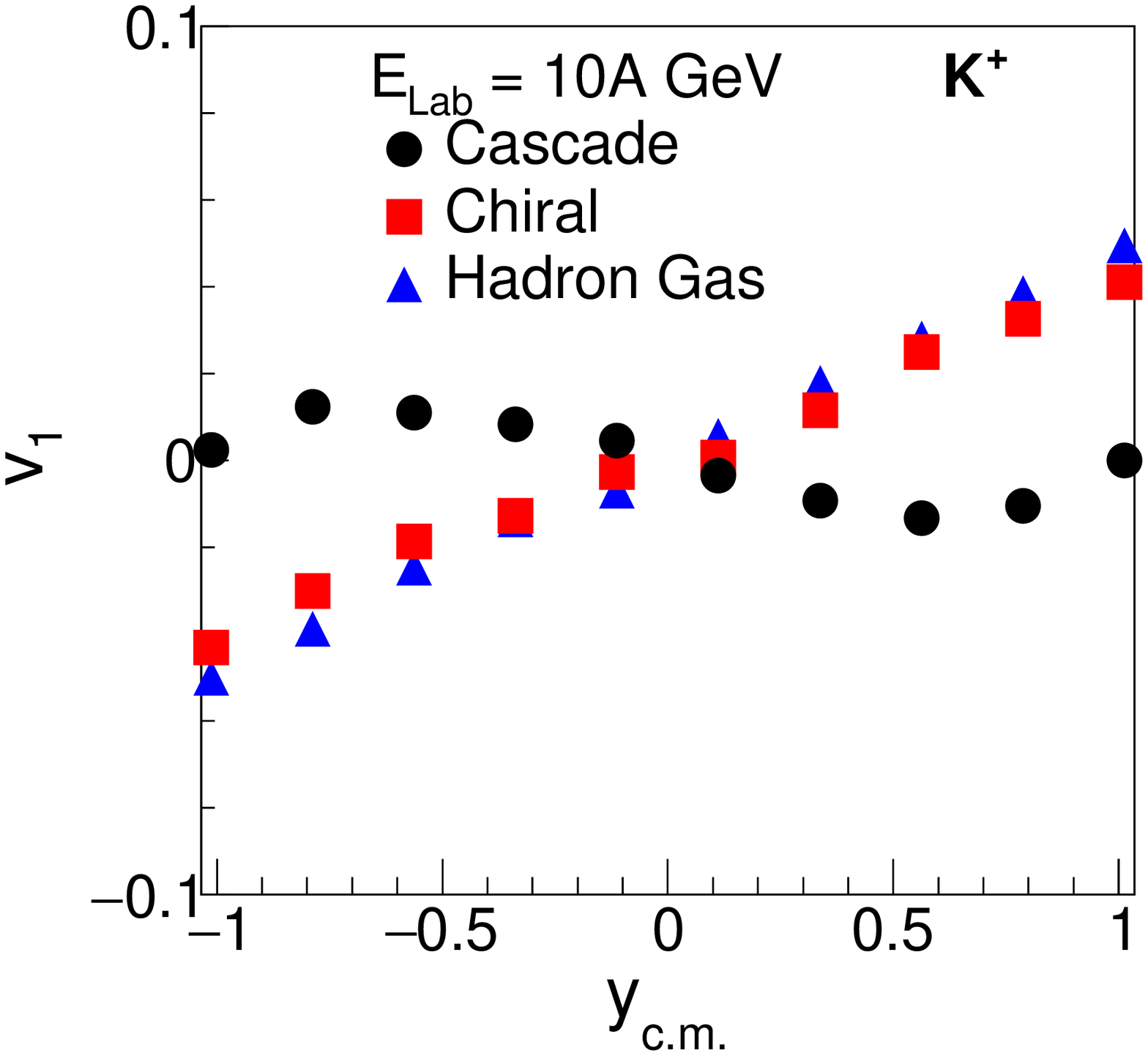}
 \includegraphics[scale=0.2]{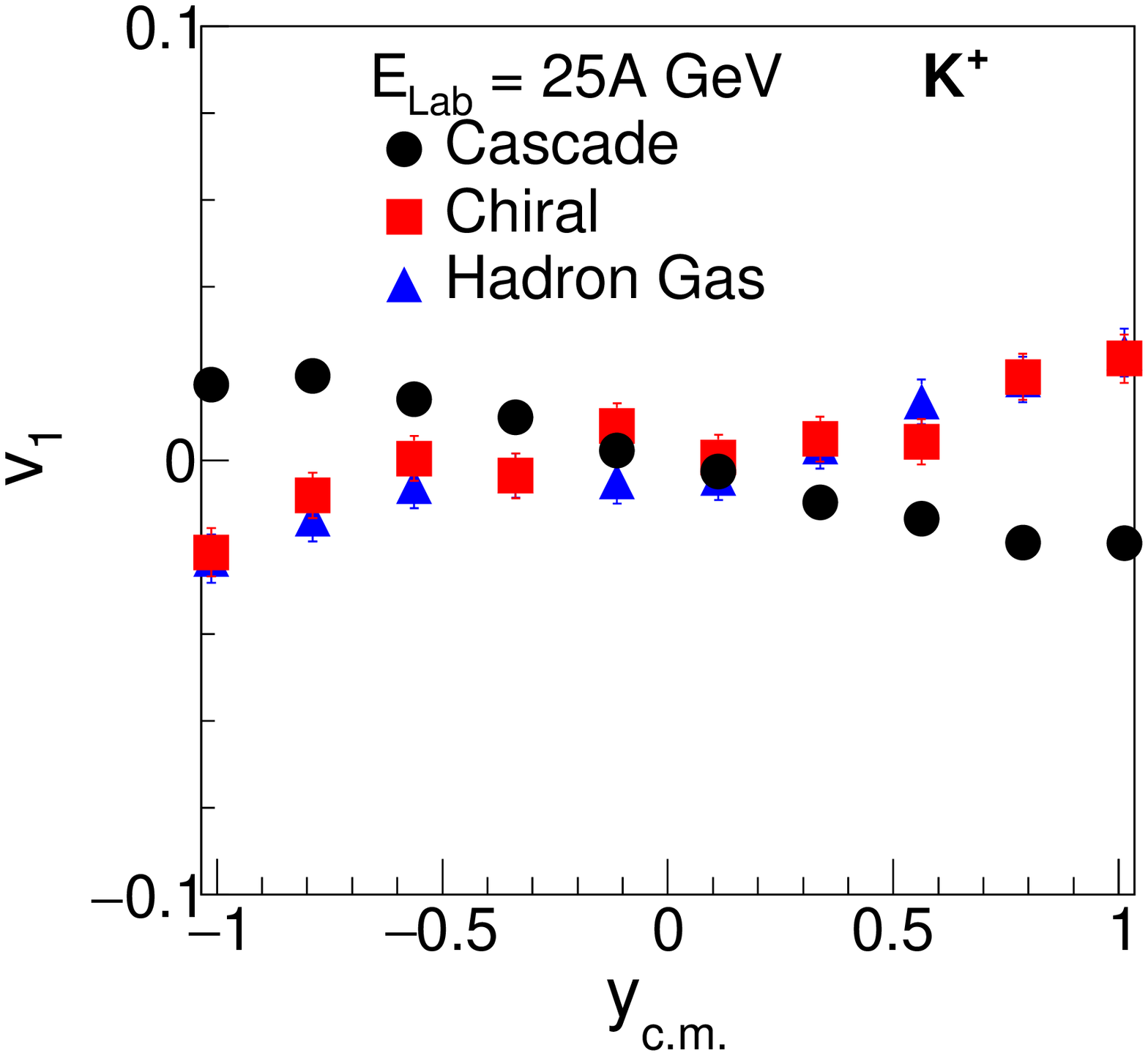}
\caption{Rapidity dependence of directed flow ($v_{1}(y_{c.m.})$) of positive kaons ($K^{+}$) in mid-central Au+Au collisions at bombarding energies $E_b= 6A, 8A, 10A and 25A$ GeV}

\label{fig1_kp}
\end{figure}

 \begin{figure}[t]
 \includegraphics[scale=0.2]{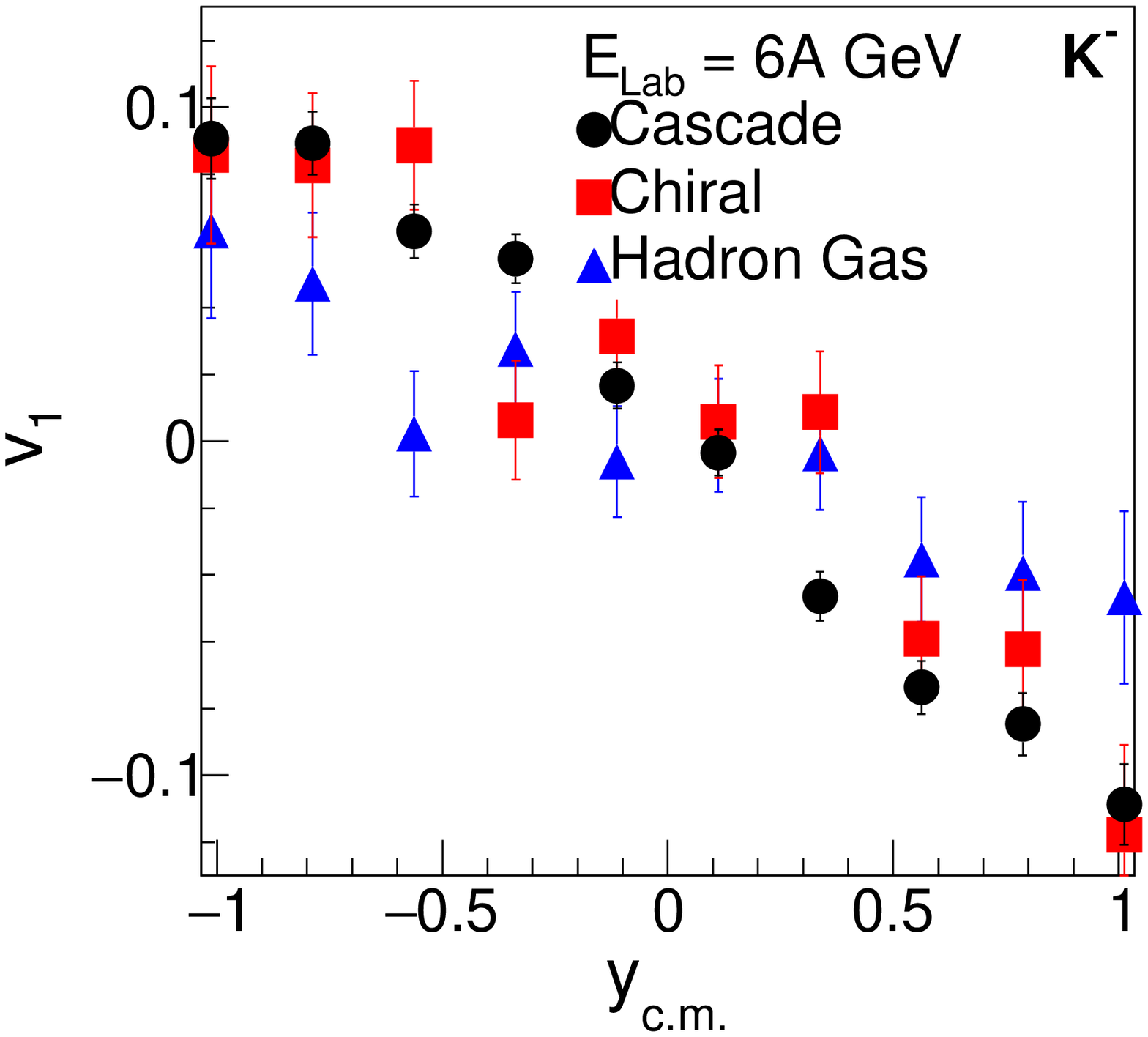}
 \includegraphics[scale=0.2]{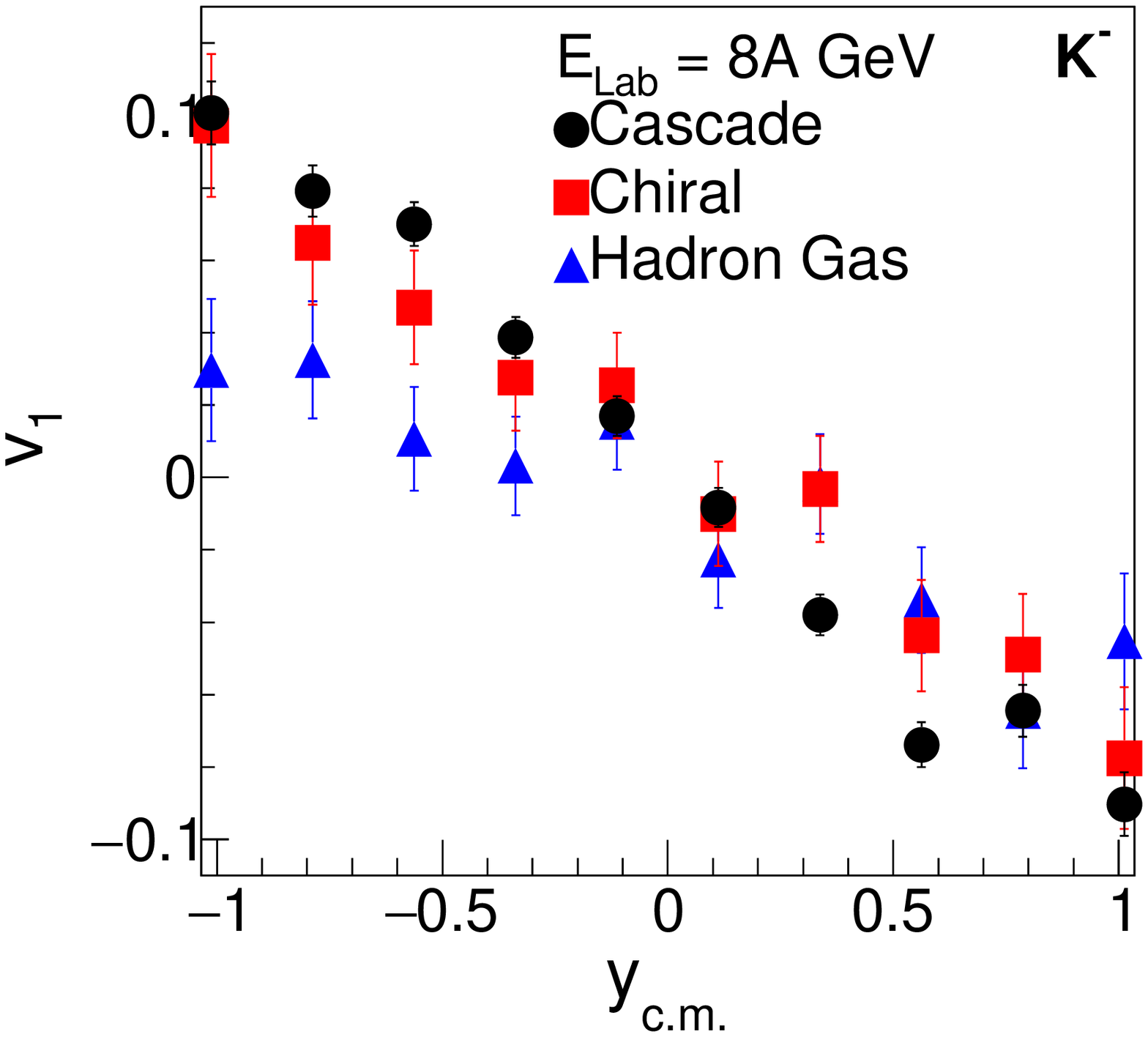}\\
 \includegraphics[scale=0.2]{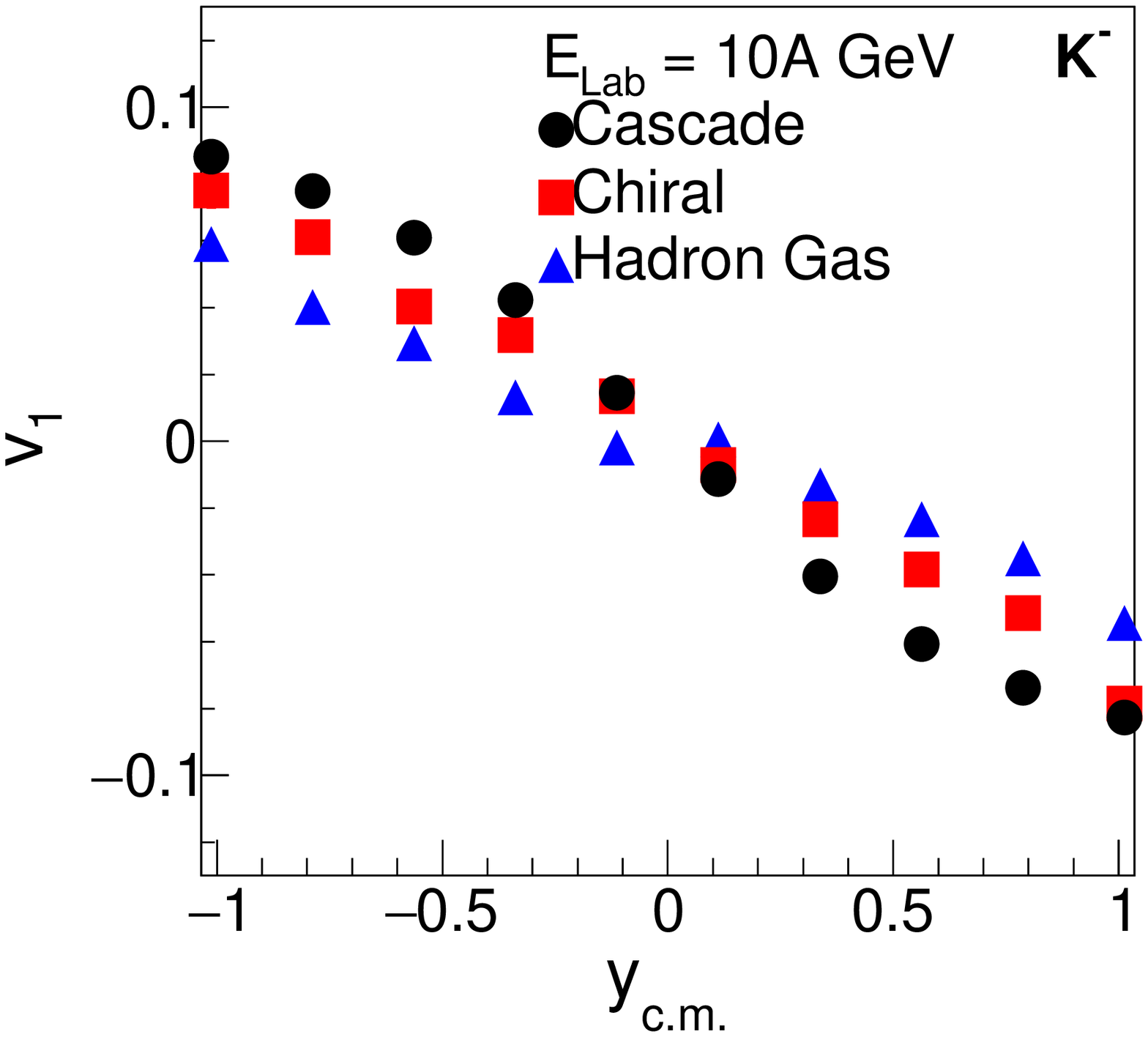}
 \includegraphics[scale=0.2]{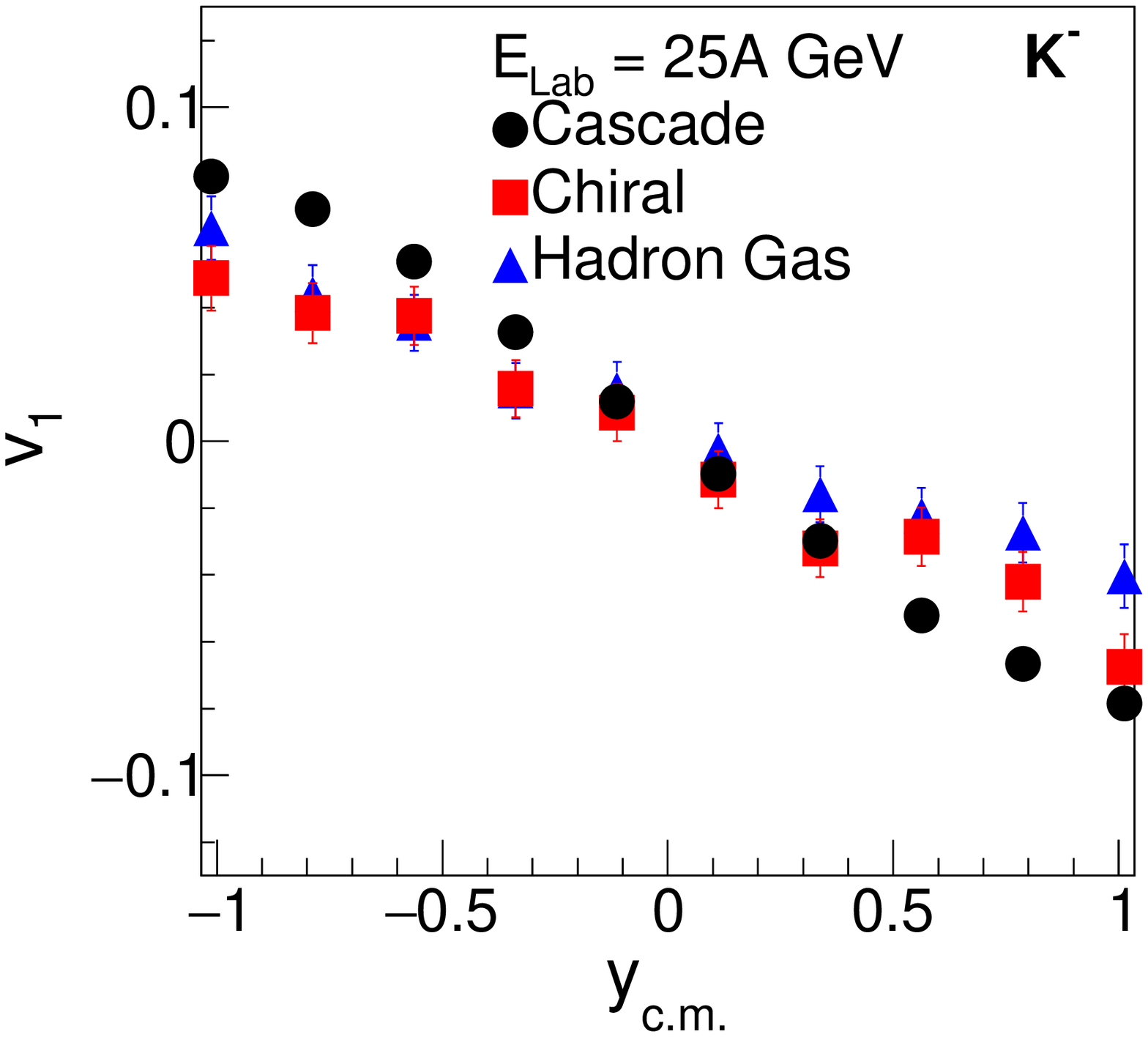}
\caption{Rapidity dependence of directed flow ($v_{1}(y_{c.m.})$) of negative kaons ($K^{-}$) in mid-central Au+Au collisions at bombarding energies $E_b= 6A, 8A, 10A and 25A$ GeV}

\label{fig1_km}
\end{figure} 

We now move on, to study the behaviour of $v_1$ and $v_2$ as a function of rapidity for different nuclear EoS. Because of its sensitivity to the longitudinal dynamics of the medium, $v_1$ is an interesting parameter to study as a function of the rapidity. The directed flow of charged hadrons as a function of the rapidity for different EoS and at different energies is shown in Fig. \ref{v1_rap}. The slope of $v_1$ at mid-rapidity ($y_{c.m} \approx 0$) shows an interesting behaviour at different energies, which is sensitive to the onset of the hydrodynamical expansion in comparison to the pure transport approach. 
 
 In the hybrid mode, the slope is positive (normal flow), and it decreases as energy goes up and almost flattens out at 25 AGeV. However, for pure transport mode, the slope is negative (antiflow) around $y_{c.m} \approx 0$, and it is showing a decreasing trend as energy goes up. Since the directed flow is expected to develop at the early stages of the collision, Fig.\ref{v1_rap} suggests otherwise that early stage does not solely responsible for the determination of final directed flow of the charged hadrons and can have the contribution from the intermediate stages of evolution. In \cite{Snellings:1999bt}, the authors have claimed that the shape of $v_{1}(y_{c.m.})$ around mid-rapidity shows sensitivity to the space-momentum correlation along with the correlation between position of the nucleons and amount of stopping which further depends on the underlying equation of state and, in turn, affects the slope around mid-rapidity.  
 
 Furthermore, to understand the species dependent effect of directed flow for different EoS, we studied the directed flow of the protons (and anti-protons), pions ($\pi^{\pm}$) and kaons ($\rm K^{\pm}$) as shown in Figs. \ref{fig1_protons}, \ref{fig1_pions} and \ref{fig1_kaons} respectively. The slope of the directed flow in case of protons is always positive, whereas the slope of the directed flow of pions is always negative for three cases of the EoS around mid-rapidity. It is higher in case of hybrid mode and smaller in case of the pure transport mode. For kaons, normal flow is observed in the case of hydro mode and anti-flow in case of pure transport mode. 
We have also studied the effect of the nuclear EoS on the directed flow of the individual charged hadrons, namely $\pi^{+}$, $\pi^{-}$, $\rm K^{+}$, $\rm K^{-}$. In Figs. \ref{fig1_kp} and \ref{fig1_km}, it seems that $\rm K^{+}$ and $\rm K^{-}$ are treated differently in the presence of hydrodynamic expansion which is visible by looking at the slope of $v_{1}(y)$ for $\rm K^{+}$ and $\rm K^{-}$ at mid-rapidity ($y_{c.m.}  \approx  0$). $v_{1}$ for $\rm K^{+}$ is similar to proton flow (normal flow) and $\rm K^{-}$ flow is anti-correlated (anti-flow) in hybrid mode. Our results are in disagreement with observations reported in~\cite{Song:1998id,Pal:2000yc}. Experimentally, it is found that $\rm K^{+}$ shows anti-flow and $\rm K^{-}$ shows normal flow as nucleons due to different potentials they experience while propagating through medium derived from the effective chiral models \cite{Kaplan:1986yq}. But on the other hand, such behaviour is in agreement with the UrQMD results published recently ~\cite{Bravina:2019efl}. The authors have studied the influence of the inclusion of mean-field potentials on the directed flow of hadrons. This interesting feature of kaon flow in UrQMD definitely needs further investigation.
 
\begin{figure}[t]
 \includegraphics[scale=0.45]{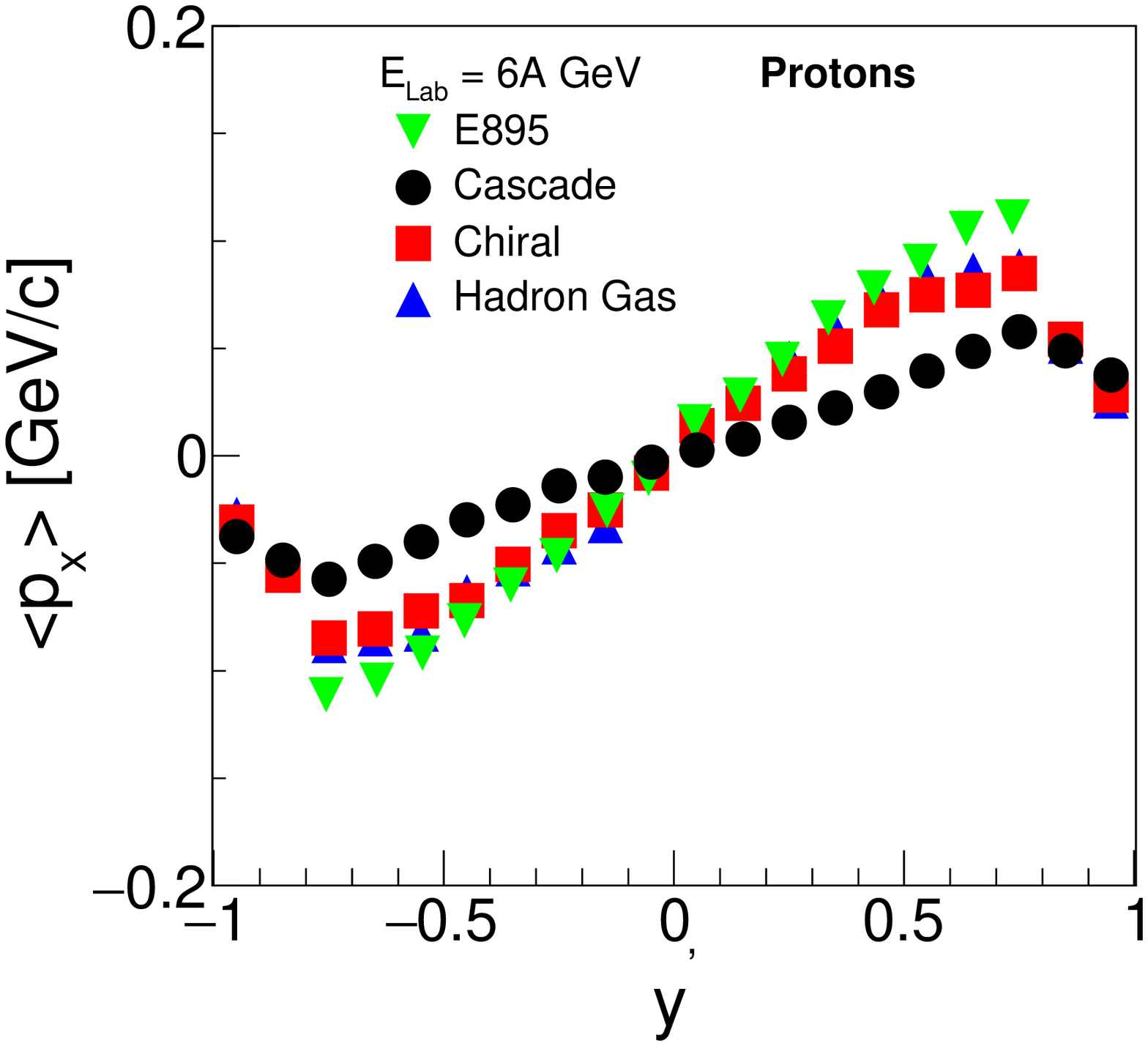}
 \includegraphics[scale=0.45]{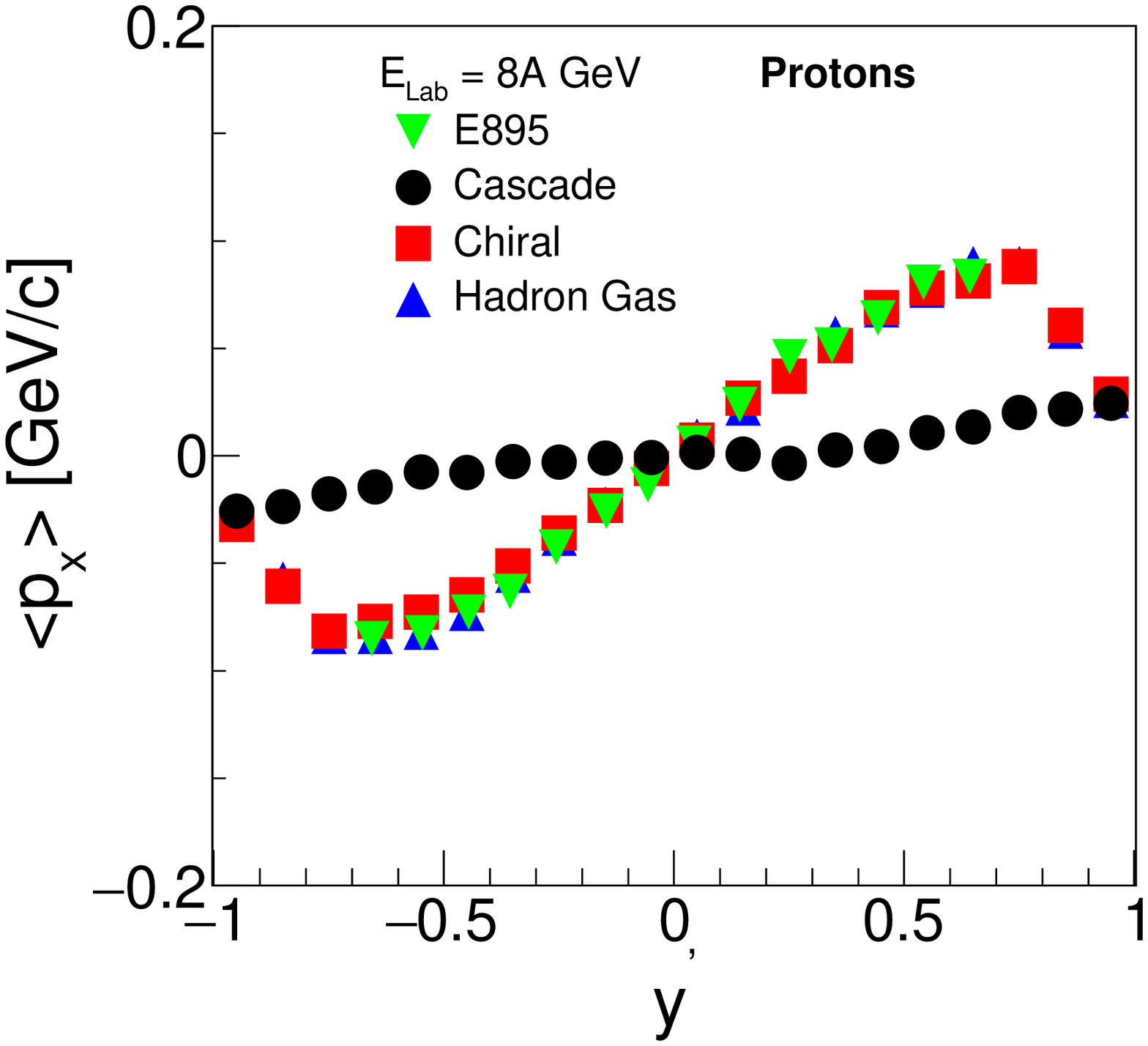}
 
 \caption{Comparison of $\rm \langle p_{x} \rangle$ vs normalized rapidity ($y^{'}$) for protons from E895 experiment \cite{Liu:2000am} at AGS energies with UrQMD for different EoS for 6A and 8A GeV}
\label{fig_data_comp}
\end{figure}

 In Fig. \ref{fig_data_comp}, we have compared the sideward flow ($\langle \rm p_{x} \rangle$) of protons as a function of normalized rapidity ($y^{'}$) from E895 \cite{Liu:2000am} at 6A and 8A GeV in mid-central ($b= 5-7$ fm) Au+Au collisions with the model results, in both cascade and hybrid mode. $y^{'}$ is normalized in a such a way that the rapidity of target and projectile become -1 and +1 respectively and is defined as, $y^{'}$ $=$ $y^{lab}/y^{mid}$ - 1. $y^{lab}$ is the rapidity in laboratory frame and $y^{mid}$ is mid-rapidity between target and projectile. From the figure, it is evident that both the EoS used in the hydrodynamic scenario, which are governed by mean-field approximation, reproduce the data quite well. The slopes of $v_{1}$ using both EoS are similar to the data. As shown in Fig. 1 of \cite{Liu:2000am}, mean-field approximation came close to explain the data well compared to the cascade scenario with the former case allows generating the additional pressure in the medium.

\begin{figure}[t]
\includegraphics[scale=0.2]
{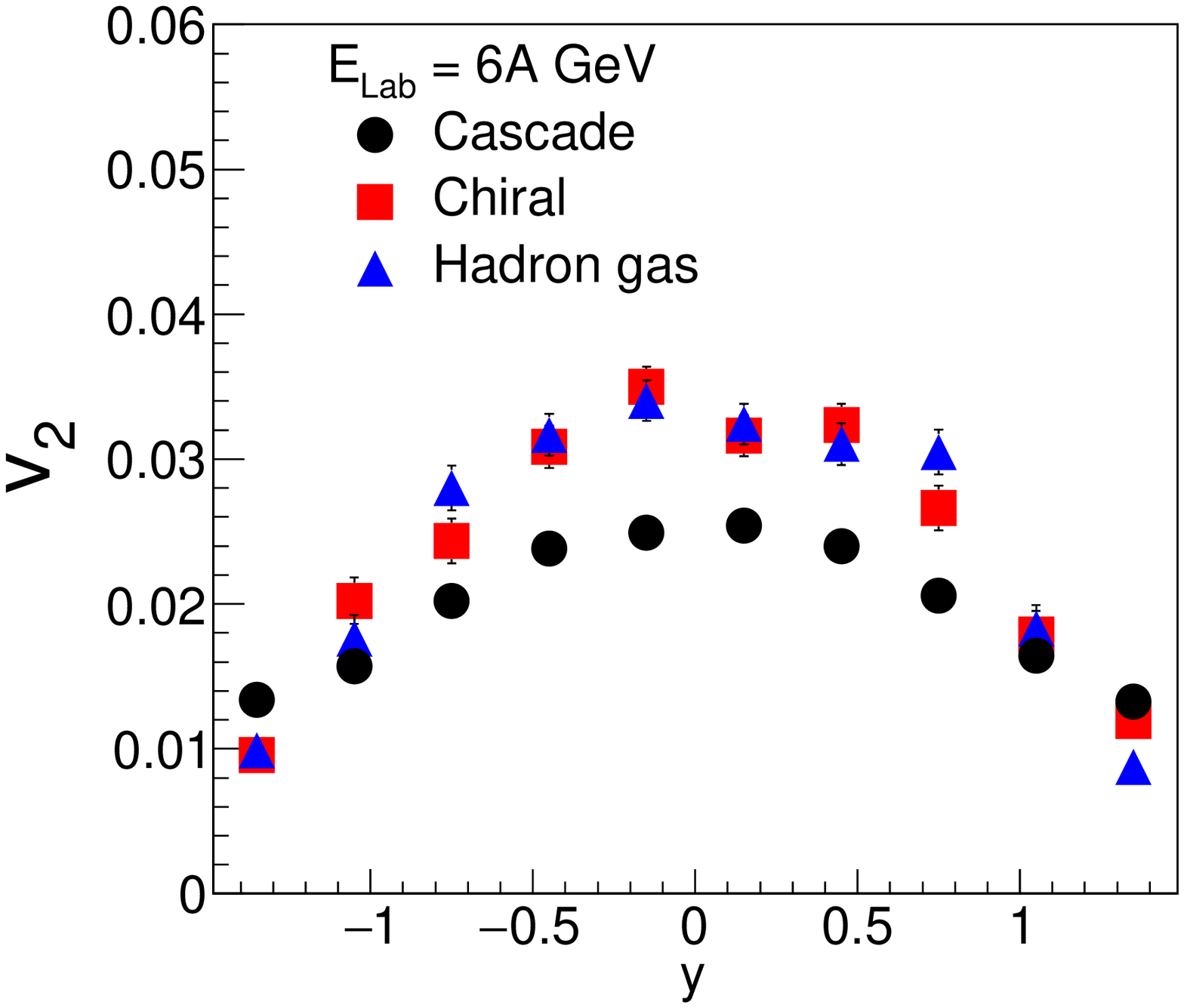}
  \includegraphics[scale=0.2]{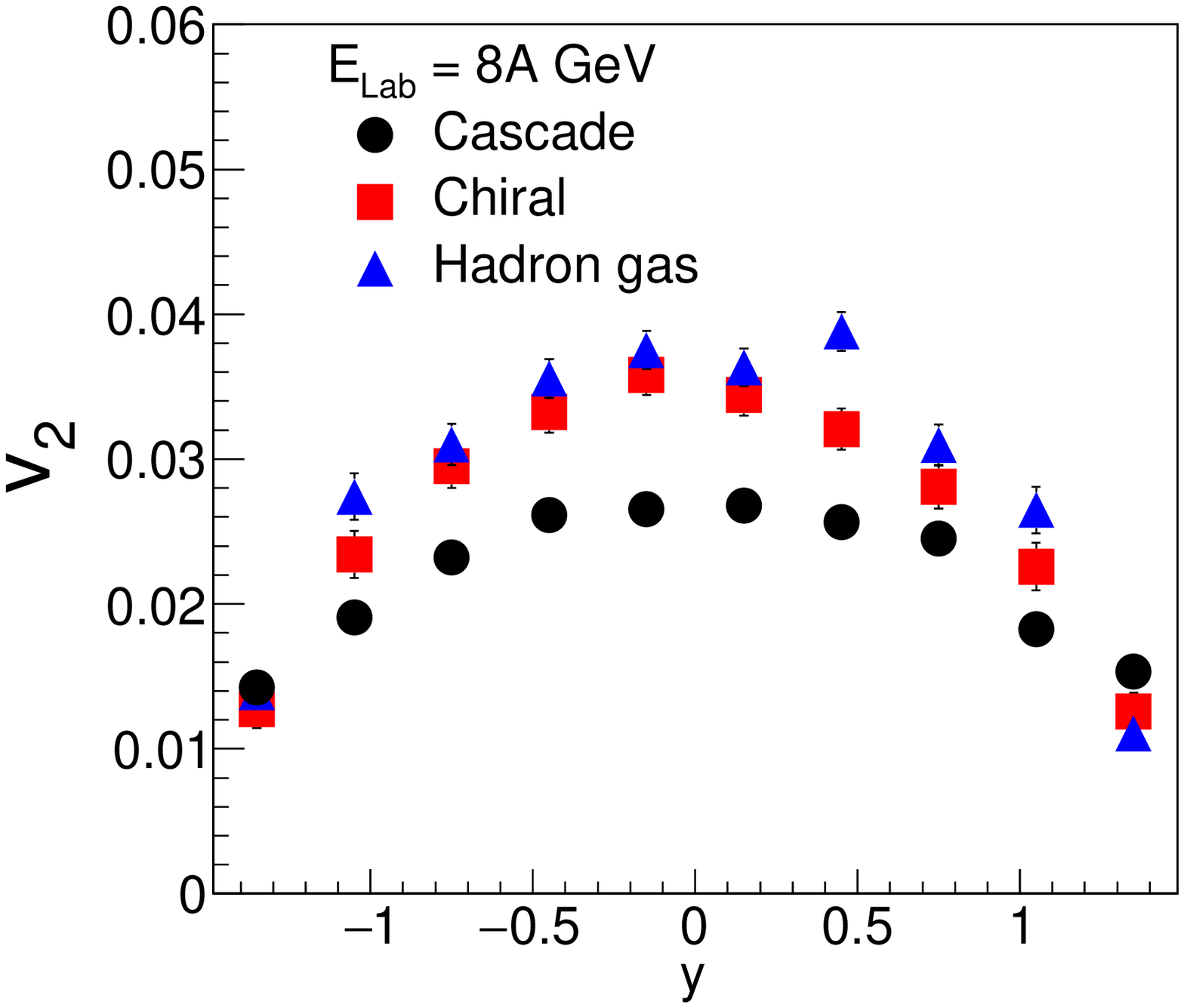}\\
  \includegraphics[scale=0.2]{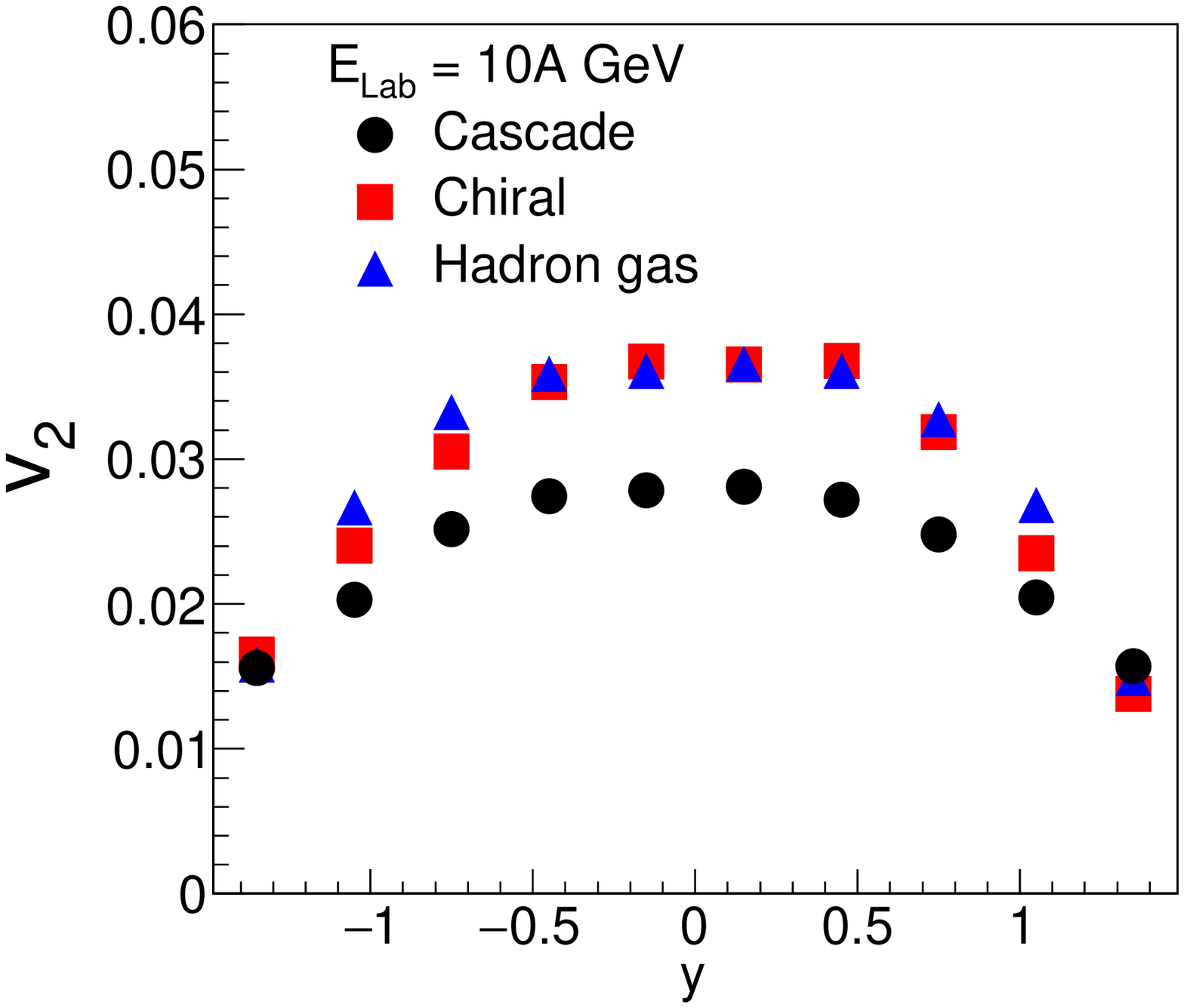}
   \includegraphics[scale=0.2]{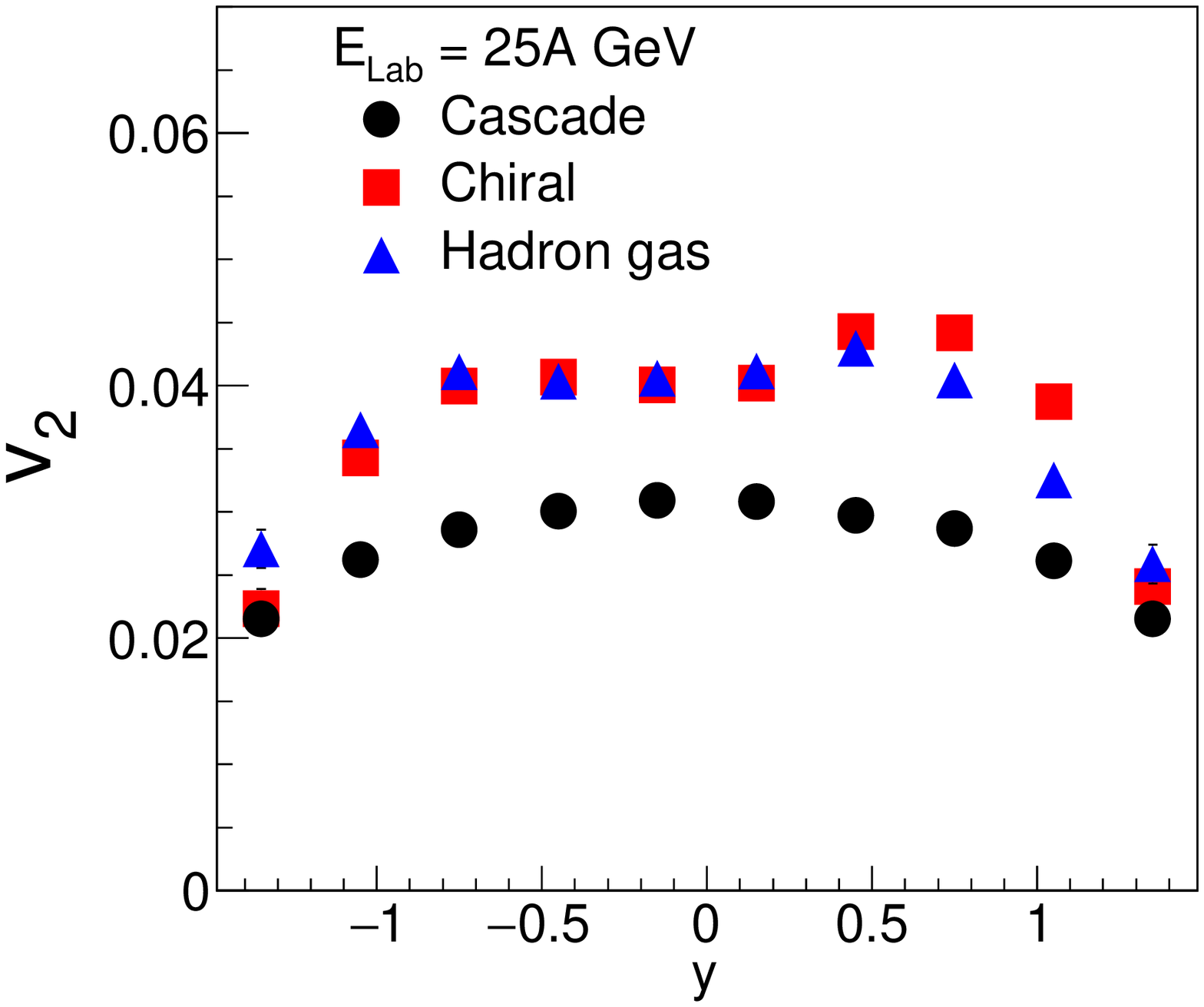}

    \caption{$v_{2}$ vs $y_{c.m.}$ for charged hadrons using UrQMD for different EoS for 6A, 8A, 10A and 25A GeV}
    \label{fig1_ch_v2y}
\end{figure}

 Elliptic flow is also studied as a function of rapidity, as shown in Fig \ref{fig1_ch_v2y}. $v_{2}$ is highest at mid-rapidity and decreases for forward rapidities. Like mentioned above, here also elliptic flow is higher for the hydrodynamic scenario suggesting the generation of magnified pressure gradients in the medium.

Before we move forward, it should be noted that flow of protons at energies as low as 6 and 8 A GeV may be sensitive to the light nuclei production. However this feature is not included in either (cascade or hybrid) public versions of the UrQMD model, which we use for simulations.

\begin{figure}
\includegraphics[scale=0.45]{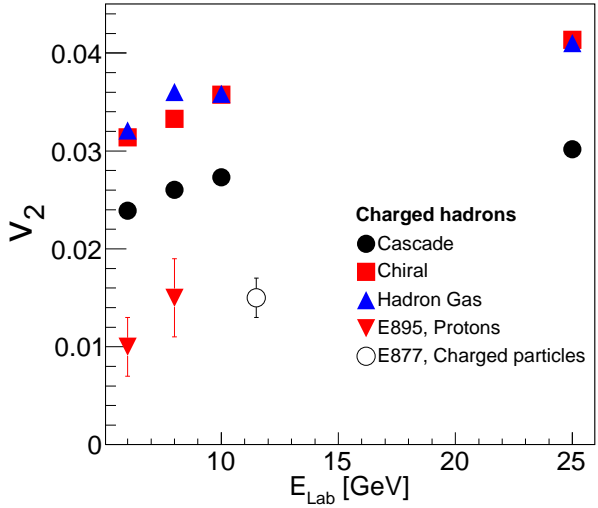}
\caption{$v_{2}$ as a function of beam energy ($\rm E_{\rm Lab}$)  for different EoS at midrapidity. It is compared with the $v_{2}$ of the protons and the charged particles at E895 and E877~\cite{Chung:2001qr,Petersen:2006vm} respectively.}
\label{v2_energy}
\end{figure}

\begin{figure}[t]
\includegraphics[scale=0.45]{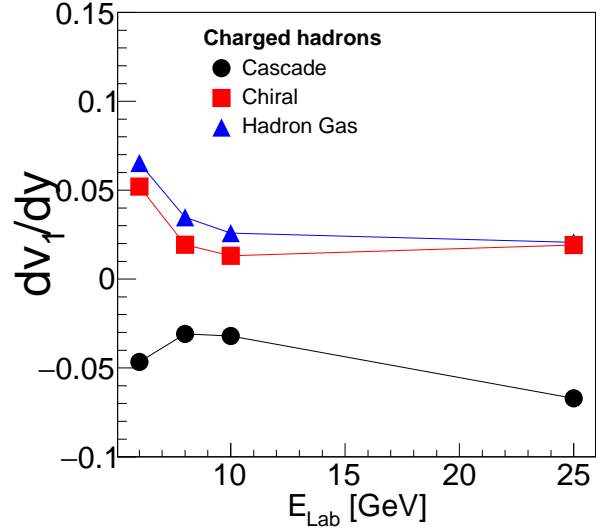}

\caption{Slope ($\frac{dv_{1}}{dy}$) at $y_{c.m.}$ $\approx$ 0 by fitting with polynomials as a function of beam energy ($\rm E_{\rm Lab}$) for different EoS at midrapidity.}
\label{dv1dy_energy}
\end{figure}

\subsection*{D. Energy dependence}

 \begin{figure}[t]
\includegraphics[scale=0.45]{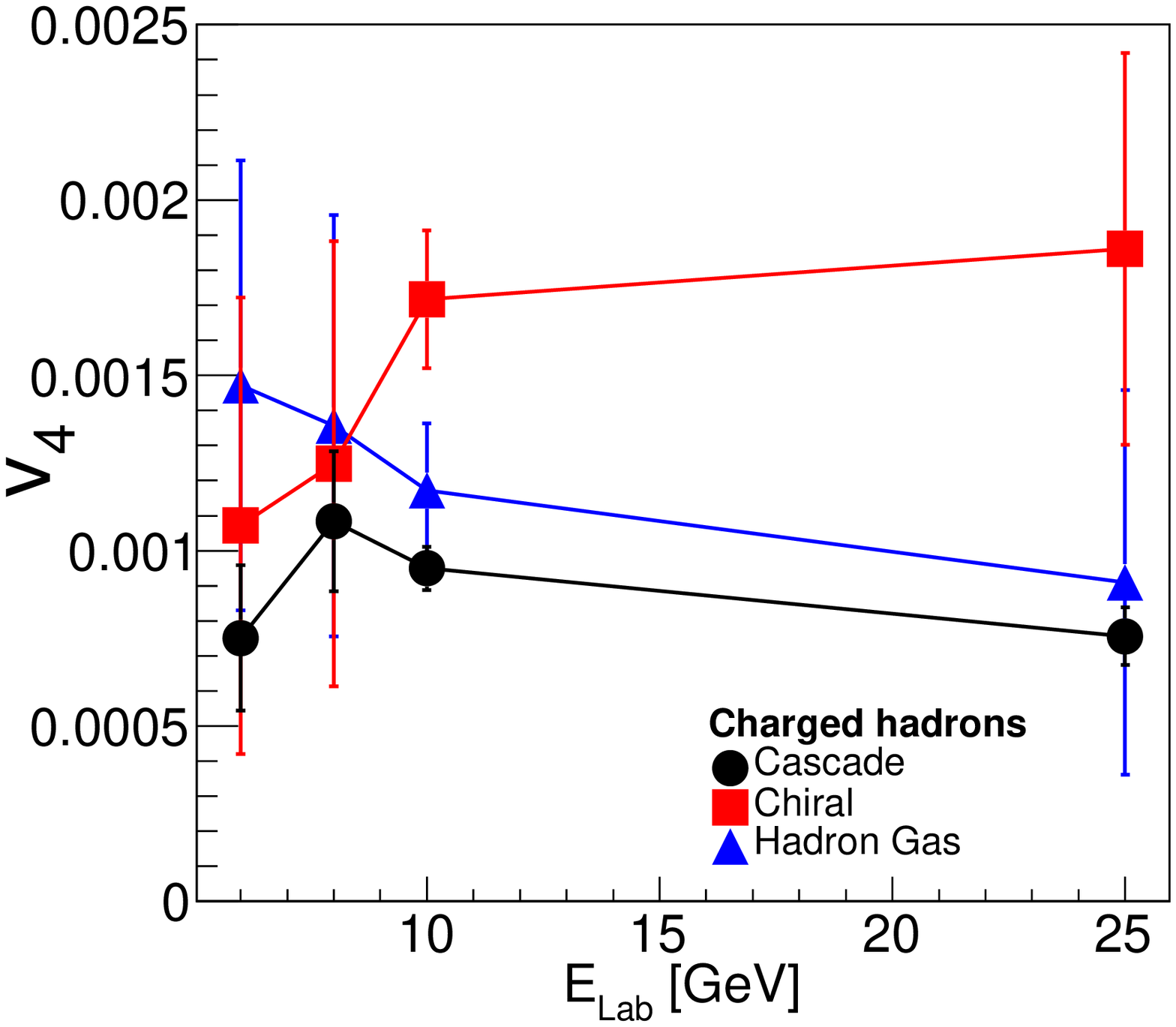}
\caption{$v_{4}$ as a function of beam energy ($\rm E_{\rm Lab}$) for different EoS at midrapidity.}
\label{v4_energy}
\end{figure}

 \begin{figure}[h]
\includegraphics[scale=0.45]{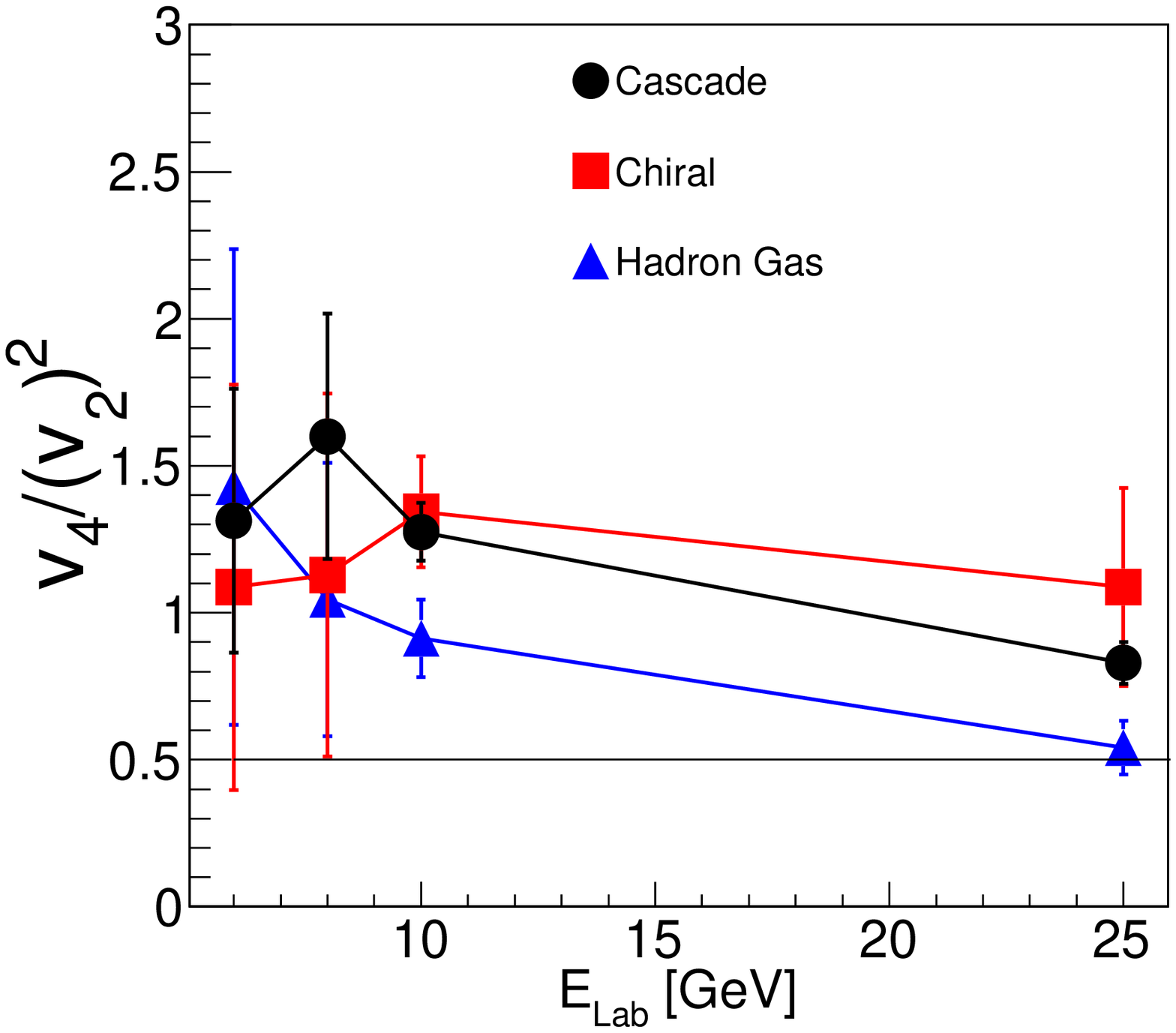}
\caption{$v_{4}/(v_{2})^{2}$ as a function of beam energy ($\rm E_{\rm Lab}$) for different EoS at midrapidity.}
\label{v4v2sq_energy}
\end{figure}
Finally, we investigate the beam energy ($\rm E_{\rm Lab}$) dependence of the anisotropic flow parameters $v_{1}$, $v_{2}$, $v_{4}$ and $v_{4}/(v_{2})^{2}$. In Fig.~\ref{v2_energy}, beam energy ($\rm E_{\rm Lab}$) dependence of integrated $v_{2}$ is shown at mid-rapidity for all three versions of the model. As expected, $v_{2}$ increases with increase in energy. In the hybrid mode, the integrated $v_{2}$ for both the EoS is about 30$\%$ larger compared to pure transport mode. Our calculated results are also compared with the data available from E877 and E895 experiments~\cite{Chung:2001qr,Petersen:2006vm}. Our results for all three configurations of UrQMD model, clearly overestimate the data, in the investigated energy range. This observation in line with the previous calculations~\cite{Petersen:2006vm}, where the energy excitation function of the charged particle $v_2$ has been compared to data over a wide energy range, from ($\rm E_{\rm Lab} = 90 A$ MeV to $\sqrt{s_{NN}} = 200$ GeV, using pure UrQMD (v2.2) model.

The slope of the directed flow as a function of beam energy is sensitive to the underlying EoS and can provide insights about the dynamics of the QCD medium. The slope ($\frac{dv_{1}}{dy}$) is estimated for charged hadrons, at mid-rapidity within the interval $|{y_{c.m.}|} \le 0.75$, as displayed in Fig. \ref{dv1dy_energy}. The slope demonstrates opposite trends in the cascade and hydrodynamic mode of evolution. Note that the slope is negative in case of pure transport and positive in case of Hadron gas and Chiral EoS. The slope also shows some sensitivity to the underlying EoS. It starts to saturate at higher energies in case of hybrid mode and decreases in the absence of hydrodynamic expansion.

The variation of $v_{4}$ as a function of beam energy ($\rm E_{\rm Lab}$) is also a very important observable to study due to its sensitivity to the nuclear EoS. In Fig. \ref{v4_energy}, we show the beam energy ($\rm E_{\rm Lab}$) dependence of the $v_{4}$ of charged hadrons in the mid-rapidity region (-0.75 $\leqslant$ $y_{c.m.}$ $\leqslant$ 0.75) for the three variants of the UrQMD model. Among the different evolution scenarios under study, $v_{4}$ seems to increase as energy goes up for Chiral EoS, and $v_{4}$ starts to increase for cascade case up to 8A GeV and then drops a bit down. In contrast, $v_{4}$ appears to have a monotonic decreasing trend as a function of beam energy in case of Hadron gas EoS. By taking into account the large statistical fluctuations, one could make a statement that the $v_{4}$ at high baryon densities bears an effect of different EoS. But any concrete comment can only be made upon reduction of these uncertainties.

According to \cite{Borghini:2005kd,Gombeaud:2009ye,Luzum:2010ae}, the generation of $v_{4}$ is governed by both the intrinsic $v_{2}$ and the $4^{th}$ order moment of collective flow. The contribution of $v_{2}$ to $v_{4}$ is simply estimated as $v_{4}$ = 0.5$(v_{2})^{2}$, within ideal fluid dynamics and in the absence of any fluctuations. Hence, with the ratio $v_{4}/(v_{2})^{2}$, one can gain some insights about the dynamics of the collision. Results available at RHIC~\cite{Adams:2003zg,Masui:2005aa,Abelev:2007qg,Huang:2008vd} show double the value of $v_{4}$, $v_{4}/(v_{2})^{2}$ $\approx$ 1. Also, note that the results from Parton-Hadron-String Dynamics (PHSD) show four times higher value, $v_{4}/(v_{2})^{2}$ $\approx$ 2 \cite{Konchakovski:2012yg}, over a range of beam energies studied in min-bias Au + Au collisions. In \cite{Nara:2018ijw}, an attempt has been made to study this ratio using JAM model. In Fig \ref{v4v2sq_energy}, we show this ratio as function of beam energy ($\rm E_{\rm Lab}$). It is higher than 0.5 and goes maximum up to about 2, within the predictions from PHSD results. Here also, the results suffer from statistical fluctuations for hybrid mode, making it difficult to make any strong conclusions, and the results can be more reliable upon reduction of these uncertainties. From Figs. \ref{v4_energy} and \ref{v4v2sq_energy}, one can note that the descending trend in both $v_{4}$ and $v_{4}/(v_{2})^{2}$ prevails for Hadron gas EoS and pure transport case. In \cite{Bhalerao:2005mm}, the authors argued about the incomplete equilibration in the medium in the context of $v_{4}/(v_{2})^{2}$. They explained the behavior of $v_{4}/(v_{2})^{2}$ as a function $K^{-1}$ which is the typical number of collisions per particle, where $K$ is the Knudsen number, a dimensionless parameter to characterize the degree of thermalization, and is related to the beam energy and system size. For $K^{-1}$ $\gg$ 1, local equilibrium is expected to be achieved. Incomplete thermalization leads to specific deviations from the ideal hydrodynamic behaviour. If the ratio $v_{4}/(v_{2})^{2}$ $>$ 0.5, the medium is not expected to be fully equilibrated which also can be seen in Fig \ref{v4v2sq_energy}. However this would prevent the use of ideal hydrodynamic model to describe the  medium evolution in such low energy collisions.  A viscous hydrodynamic expansion might be a more reliable tool, but this is beyond the scope of the present work. A conclusive picture can only be drawn once data on flow measurements will be available from the future experiments at FAIR and NICA.

\section*{IV Summary}
  In this paper, we made an attempt to address a long-standing issue of probing the equation-of-state of the strongly-interacting matter, from the measurement of collective flow observables in heavy-ion collisions. We focus on the flow parameters $v_1$, $v_2$ and $v_4$ at mid-rapidity in semi-central Au+Au collisions, in the beam energy range $6 - 25 A$ GeV, where the future FAIR and NICA accelerators would be operated. The UrQMD transport approach coupled with the ideal hydrodynamic expansion for different nuclear equations of state is employed for this purpose. We start with the examination of the elliptic flow parameter, $v_2$ of charged and identified hadrons as a function of transverse momentum and rapidity. We have noticed that $v_2$ is always higher in the hydrodynamic scenario when compared with the transport mode of the UrQMD model but fails to differentiate between the partonic and hadronic equations of state. 
The observed insensitivity can be attributed to the small life time of the hydrodynamic phase in such low eenrgy collisions. We have also studied the constituent quark number scaling of elliptic flow for all the energies and nuclear EoS. From the results in hand, $v_{2}$ shows reasonably good scaling. 

Furthermore, an attempt has been made to study the directed flow of charged and identified hadrons as a function of rapidity. For the case of charged hadrons, the slope of $v_{1}$ is sensitive to the hydrodynamical scenario and able to differentiate the pure transport mode from the hydro mode. On the other hand, similar to the observation in the case of $v_{2}$, it fails to distinguish between the two EoS and is rather insensitive to the underlying degrees of freedom in the investigated energy regime.

Along with this, efforts have been made to study the effect of different EoS on the slope of the directed flow (d$v_{1}$/dy), elliptic flow ($v_{2}$) and $v_{4}$ of charged hadrons as a function of the beam energy ($\rm E_{\rm Lab}$). Also, the ratio $v_{4}/(v_{2})^{2}$ is studied as a function of beam energy. {The ratio lies within the values 0.5 and 2 which is consistent with results obtained previously \cite{Adams:2003zg,Masui:2005aa,Abelev:2007qg,Huang:2008vd,Konchakovski:2012yg}, given the statistical fluctuations. Upon the reduction in statistical errors, more conclusive remarks can be made. These predictions will be useful once data from the experiments at NICA \cite{Kekelidze:2016wkp}, and FAIR \cite{Ablyazimov:2017guv,Sturm:2010yit} become available.

\section*{Acknowledgements}
We are grateful to Yang Wu and Declan Keane for providing us the data points from E895 experiment at AGS energies. We also thank Ralf Averbeck for critically reading the manuscript. We acknowledge the computing facility provided by the Grid Computing Facility at VECC-Kolkata, India.

\end{document}